\documentclass[printer]{aa}  

\usepackage{multicol}
\usepackage{graphicx}
\usepackage{amsmath}
\usepackage{amssymb}
\usepackage{textgreek}
\usepackage{txfonts}
\usepackage{hyperref}
\usepackage{cprotect} 
\usepackage{multirow}
\begin{document} 

   \title{PSR J1227$-$6208 and its massive white dwarf companion:\\ pulsar emission analysis, timing update and mass measurements}

   \author{Miquel Colom i Bernadich
          \inst{1}$^{\ast}$
          \and
          Vivek~Venkatraman~Krishnan\inst{1}$^{\ast}$
          \and
          David~J.~Champion\inst{1}
          \and
          Paulo~C.~C.~Freire\inst{1}
          \and
          Michael~Kramer\inst{1}
          \and
          Thomas~M.~Tauris\inst{2,1}
          \and
          Matthew~Bailes\inst{3}
          \and
          Alessandro~Ridolfi\inst{4,1} 
          \and
          Maciej~Serylak\inst{5,6}
          }

   \institute{Max-Planck-Institut für Radioastronomie, Auf dem Hügel 69, D-53121 Bonn, Germany
         \and
            Dept. of Materials and Production, Aalborg University, DK-9220~Aalborg {\O}st, Denmark
        \and
            INAF -- Osservatorio Astronomico di Cagliari, via della Scienza 5, 09047 Selargius (CA), Italy
         \and
            Centre for Astrophysics and Supercomputing, Swinburne University of Technology, P.O. Box 218, Hawthorn, Vic, 3122, Australia   
         \and
            SKA Observatory, Jodrell Bank, Lower Withington, Macclesfield, SK11 9FT, United Kingdom
         \and
            Department of Physics and Astronomy, University of the Western Cape, Bellville, Cape Town, 7535, South Africa         
            \\
        $^\ast$\email{mcbernadich@mpifr-bonn.mpg.de,vkrishnan@mpifr-bonn.mpg.de}
             }

   \date{Received ; accepted}
 
  \abstract
   {PSR J1227$-$6208 is a 34.53-ms recycled pulsar with a massive companion. This system has long been suspected to belong to the emerging class of massive recycled pulsar$-$ONeMg white dwarf systems such as PSR J2222$-$0137, PSR J1528$-$3146 and J1439$-$5501. Here we present an updated emission and timing analysis with more than 11 years of combined Parkes and MeerKAT data, including 19 hours of high-frequency data from the newly installed MeerKAT S-band receivers. We measure a scattering timescale of 1.22 ms at 1 GHz with a flat scattering index $3.33<\beta<3.62$, and a mean flux density of $0.53-0.62$~mJy at 1 GHz with a steep spectral index $2.06<\alpha<2.35$. Around 15\% of the emission is linearly and circularly polarised, but the polarisation angle does not follow the rotating vector model. Thanks to the sensitivity of MeerKAT, we successfully measure a rate of periastron advance of $\dot \omega=0.0171(11)$~deg\,yr$^{-1}$, and a Shapiro delay with an orthometric amplitude of $h_3=3.6\pm0.5$~\textmu s and an orthometric shape of $\varsigma=0.85\pm0.05$. The main source of uncertainty in our timing analysis is chromatic correlated dispersion measure noise, which we model as a power law in the Fourier space thanks to the large frequency coverage provided by the Parkes UWL receiver. Assuming general relativity and accounting for the measurements across all the implemented timing noise models, the total mass, companion mass, pulsar mass and inclination angle are constrained at $2.3<M_\textrm{t}/M_\odot<3.2$, $1.21<M_\textrm{c}/M_\odot<1.47$, $1.16<M_\textrm{p}/M_\odot<1.69$ and $77.5<i/\textrm{deg}<80.3$. We also constrain the longitude of ascending node to either $\Omega_\textrm{a}=266\pm78$~deg or $\Omega_\textrm{a}=86\pm78$~deg. We argue against a neutron star nature of the companion based on the very low orbital eccentric of the system ($e=1.15\times10^{-3}$), and instead classify the companion of PSR J1227$-$6208 as a rare, massive ONeMg white dwarf close to the Chandrasekhar limit.
   }


   \titlerunning{PSR J1227$-$6208 and its massive white dwarf companion}
   \authorrunning{M. C. i Bernadich et al.}

   \maketitle
%

\section{Introduction}

The Chandrasekhar limit of white dwarf (WD) masses is a topic of ongoing research in astrophysics and theoretical physics. Studies show that maximally-rotating rigid WDs could sustain masses up to $M_\textrm{WD}\approx1.48$~$M_\odot$ \citep{yoon2005rapidly}, but several effects have been speculated to allow them to exist beyond this limit, such as differential rotation \citep{yoon2005rapidly} or high magnetisation \citep{kundu2012mass}, and Super-Chandrasekhar masses have been inferred indirectly from some Type Ia supernovae \citep{tomaschitz2018chandrasekhar}. Additionally, extensions or modifications of general relativity (GR) must incorporate a prediction for the upper mass limit of WDs \citep[e.g.][]{gregoris2023chandrasekhar,mathew2021chandrasekhar}. To test these postulations, the empirical measurement of WD masses close to the Chandrasekhar limit is required. However, WDs with $M_\textrm{WD}\gtrsim1.3~M_\odot$ are a rarity in the Galactic field. One of their windows of study is optical photometry, which has enabled the characterisation of their masses based on emission models \cite[e.g.][]{caiazzo2021wd,hollands2020wd,kulebi2010wd,miller2023hydaes,pshirkov2020wd}. The other window of study are mass measurements of binary radio pulsars with massive WD companions obtained via pulsar timing. This technique of allows for the measurement of relativistic effects in the orbital motion and in the light propagation time; these are quantified by the post-Keplerian (PK) parameters. Under the assumption of GR, such measurements can result in the precise measurement of the WD and pulsar masses \citep{lorimer2012handbook}.

Timing measurements of pulsars with massive WD companions are relevant not only for probing the Chandrasekhar limit, but also for testing bianry evolution and fundamental physics. In recycled pulsar binaries, we can test how binary interaction affects the resulting WD and NS mass distribution \citep[for general insigts on binary evolution and pulsar recylcing, see][]{tauris2023physics}. For instance, there is a bi-modality in the WD mass distribution, with $M_\textrm{WD}\lesssim0.5$~$M_\odot$ WDs being found with fully recycled pulsars at and $M_\textrm{WD}\gtrsim0.7$~$M_\odot$ with mildly recycled pulsars \citep{mckee2020precise,shamohammadi2023manyS}. This divide is well understood thanks to models of binary evolution \citep{lazarus2014massive,tauris2012coII,tauris2023physics}, but the upper end of the WD mass distribution ($M_\textrm{WD}\gtrsim1.1$~$M_\odot$) remains relatively unsampled. The birth NS mass distribution can also be directly sampled in these kind of systems given the little mass accretion occurring during recycling \citep[e.g.][]{lazarus2014massive,cognard2017massive}. For instance, the measurement of a pulsar mass of $M_\textrm{p}=1.831(10)$~$M_\odot$ in PSR J2222$-$0137 \citep{guo2021improved} is evidence that NSs can be born massive (heavier than 1.4~$M_\odot$) instead of acquiring large amounts of extra mass via accretion \citep{cognard2017massive}. Furthermore, the NS mass distribution is a probe of the physics of matter under conditions of extreme density \citep[e.g.][]{ozel2016masses,Fonseca_2021,hu2020constraining}. Finally, in the most compact systems, timing of pulsars with WD companions also provides extremely precise tests of gravity theories \citep[e.g.,][]{Voisin_2020}, as shown by the strict constraints on dipolar gravitational wave emission from PSR~J2222$-$0137 \citep{guo2021improved}, which have ruled out the phenomenon of spontaneous scalarisation predicted by some alternative gravity theories \citep{Zhao_2022}.

As of today, only four pulsars with WD companions at the upper end of the mass distribution have been characterised. The most studied one is PSR J2222$-$0137, a 32.8-ms recycled pulsar in a circular 2.45-day orbit with a $M_\textrm{WD}=1.319(4)~M_\odot$ companion \citep{boyles2013gbt,cognard2017massive,guo2021improved}. With a parallax distance measured with long baseline interferometry of 268~pc \citep{deller2013vlbi,guo2021improved}, the lack of an optical detection implies a cold WD ($T<3000$~K) with a cooling age of at least several Gyr \citep{kaplan2014companion}. A similar case is PSR J1528$-$3146, a 60.8-ms recycled pulsar in a circular 3.18-day orbit with a $M_\textrm{WD}=1.33^{+0.08}_{-0.07}~M_\odot$ companion \citep{jacoby2006detection,jacoby2007discovery,berthereau2023j1528}. In this case, the optical detection implies a cooling age between 1.5 and 3.2~Gyr \citep{jacoby2006detection}, consistent with the pulsar characteristic age of 3.9~Gyr estimated by \cite{berthereau2023j1528}. A third system is PSR J1439$-$5501, with a 28.6-ms pulsar in a 2.12-day orbit with a $M_\textrm{WD}=1.27^{+0.14}_{-0.12}$~$M_\odot$ optically detected companion with a cooling age of 0.1$-$0.5~Gyr \citep[Jang et al., in prep.][]{faulkner2004finding,lorimer2006timing,pallanca2013optical}. The final system is PSR B2303+46, with a young 0.937-s pulsar in a highly eccentric 12.34-day orbit with a $M_\textrm{WD}=1.34^{+1.08}_{-0.15}$~$M_\odot$ companion \citep{thorsett1993massive,thorsett1996masses}. The WD companion is hot and young, implying a system age of 30~Myrs \citep{vanKerkwijk1999massive}.

In this work, we present a detailed study of PSR J1227$-$6208 (J1227$-$6208 from now on). Discovered independently by three different studies in data from the Murriyang Parkes telescope\footnote{\url{https://www.parkes.atnf.csiro.au/}} \citep{bates2015discovery,knispel2013discovery,mickaliger2012discovery}, it is a 34.5-ms mildly recycled pulsar in orbit with a massive companion. With an orbital period of $P_\textrm{b}=6.72$~days and a projected semimajor axis of $x=23.2$~light seconds (ls), it has a high mass function of $f_\textrm{M}=0.297$~$M_\odot$. Assuming a pulsar mass $M_\textrm{p}=1.35$~$M_\odot$, this leads to a minimum companion mass of $M_\textrm{c}>1.27$~$M_\odot$. Its low orbital eccentricity ($e=1.15\times10^{-3}$) makes the possibility of a NS companion unlikely, giving more weight to a massive WD hypothesis instead. However, owing to the low orbital eccentricity, and large timing uncertaintes, precise measurements of PK parameters in this system have been impossible until now.

It is for these reasons that the system was included in the Relativistic Binary program \citep[RelBin,][]{kramer2021relbin} of MeerTIME \citep{bailes2020meerKAT}, a large science project that takes advantage of the much higher timing precision of Southern pulsars made possible by the superb sensitivity of the MeerKAT telescope\footnote{\url{https://www.sarao.ac.za/science/meerkat/about-meerkat/}}\citep{jonas2016meerkat}. The RelBin project is designed to measure masses and test theories of gravity in 25 selected binary pulsar systems \citep{{kramer2021relbin}}, which includes J1227$-$6028. The updated timing analysis of J1227$-$6028 presented here includes a decade-long timing baseline of Parkes/Murriyang observations and two years of dedicated MeerKAT observations. This is the first timing experiment to include data from the newly commissioned S-band MeerKAT receivers \citep{barr2018sband}.

This paper is structured as follows. Section~\ref{backends} details the observations used in this study and the data reduction for timing. Section~\ref{emission_analysis} reports the emission study, including the modelling of the profile, scattering and spectral measurements, and the detection of polarised light. Section~\ref{timing_anlysis} reports the timing analysis, including the modelling of several timing noises, constraints on PK parameters, mass measurements and an exploration of the orbital orientation. Section~\ref{astrophysics} discusses the astrophysical implications of our measurements. In Section~\ref{prospects} we outline future prospects and lines of action for the study of this system. Finally, Section~\ref{conclusion} concludes the paper and summarizes the key aspects of our measurements and discussions.

\section{Data acquisition and reduction}\label{backends}

\begin{table*}

\caption[]{\label{data} Summary of the data sets used in this analysis. }
\centering
    \resizebox{\textwidth}{!}{%
    \begin{tabular}{lcccccccccccc}
    \hline
    \hline \\[-1.5ex]
    Receiver(s) & Freq. & Bandw. & Temp. & Gain & \# obs & Time & First obs. & Last obs. & \# bands & \# ToAs & Median err. \\[0.3ex]
     & (MHz) & (MHz) & K & K\,Jy$^{-1}$ &  & (hours) &  &  &  & (\textmu s)\\
    \hline \\[-1.5ex]
    Parkes multibeam & 1382 & 400  & $\sim$21 & $\sim$0.9 & 93 & 33 & 1 March 2012     & 8 April 2019 & 4 & 365 & 26.4 \\[0.5ex]
    MeerKAT L-band   & 1284 & 856  & 4$-$7    & $\sim$2.6 & 37 & 26 & 12 February 2019 & 25 May 2023 & 8 & 508 & 5.39 \\[0.5ex]
    Parkes UWL       & 2368 & 3328 & $\sim$22 & $\sim$0.9 & 56 & 72 & 4 May 2020       & 6 June 2023 & 8 & 444 & 19.5 \\[0.5ex]
    MeerKAT S-band   & 2406 & 875  & 4$-$7    & $\sim$2.3 & 10 & 19 & 12 May 2023      & 28 May 2023 & 4 & 228 & 5.43 \\[0.5ex]
    \hline
    \hline
    \end{tabular}}

\tablefoot{ 
The data were recorded in folding mode with four polarisation channels and 1,024 phase bins of 28.28 of \textmu s in length. The system temperatures are drawn from the \href{https://skaafrica.atlassian.net/wiki/spaces/ESDKB/pages/277315585/MeerKAT+specifications}{MeerKAT General documentation}, the \href{https://www.atnf.csiro.au/research/multibeam/lstavele/description.html}{Multibeam Receiver Description webpage} and \citep{hobbs2020uwl}. The gain is estimated from the collecting area, accounting that only 56 antennas are available in the MeerKAT S-band configuration. The last three columns indicate timing data reduction parameters: number of sub-bands used in multi-frequency timing, number of ToAs per data set, and the median ToA uncertainty.
}
\end{table*}

\begin{figure*}
	\centering
	\includegraphics[width=0.24\linewidth, trim={40 60 30 170}, clip]{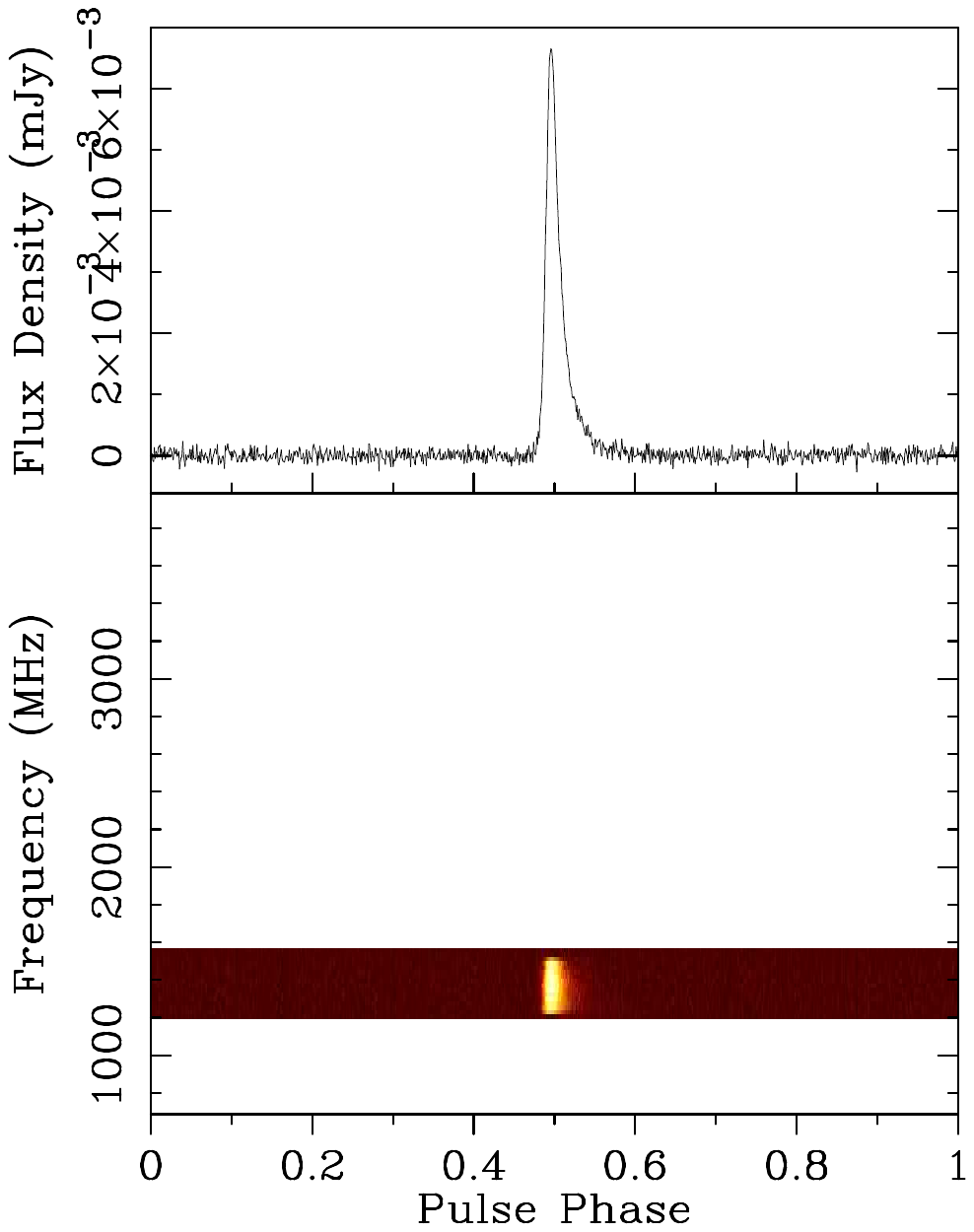}
    \includegraphics[width=0.24\linewidth, trim={40 60 30 170}, clip]{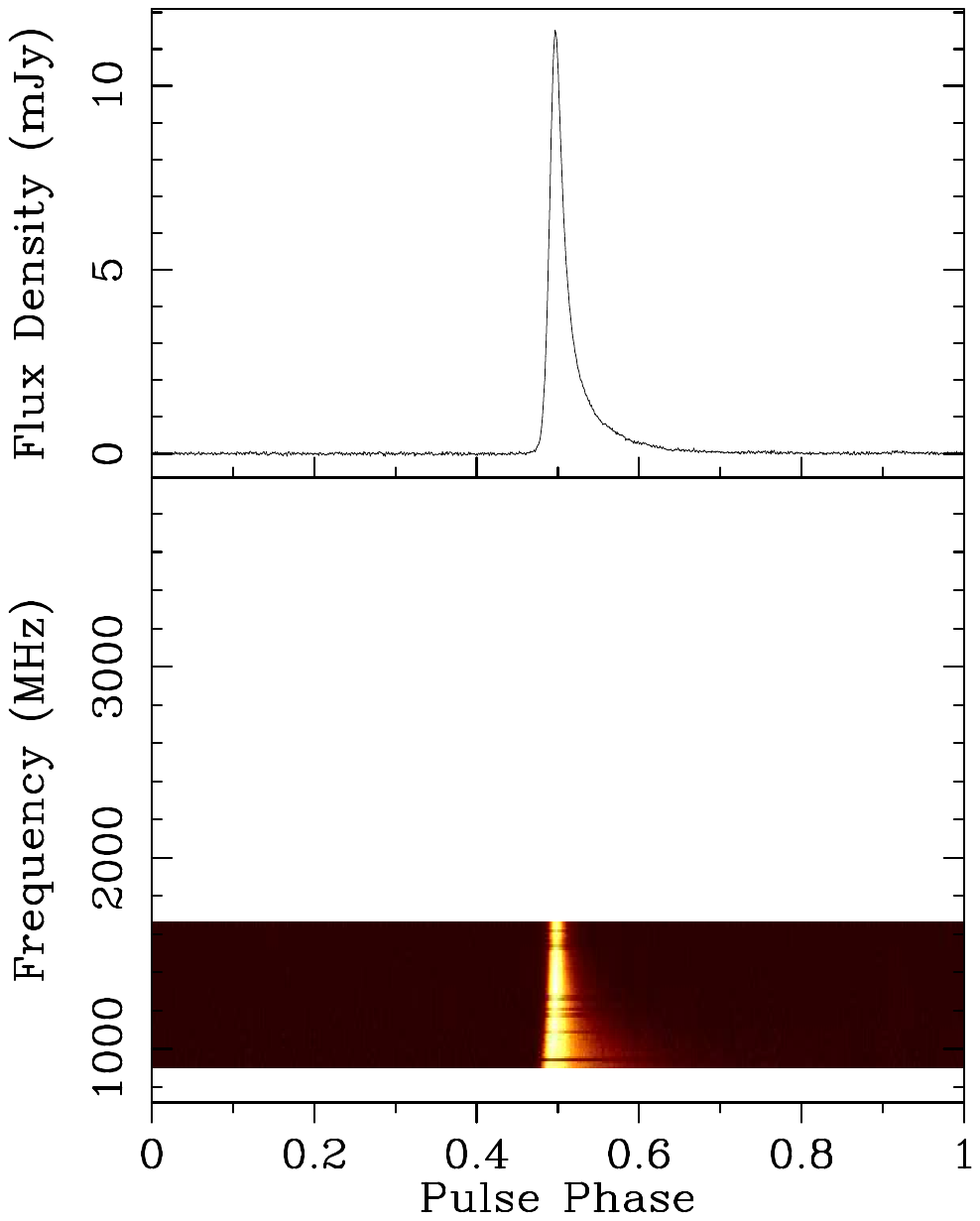}
    \includegraphics[width=0.24\linewidth, trim={40 60 30 170}, clip]{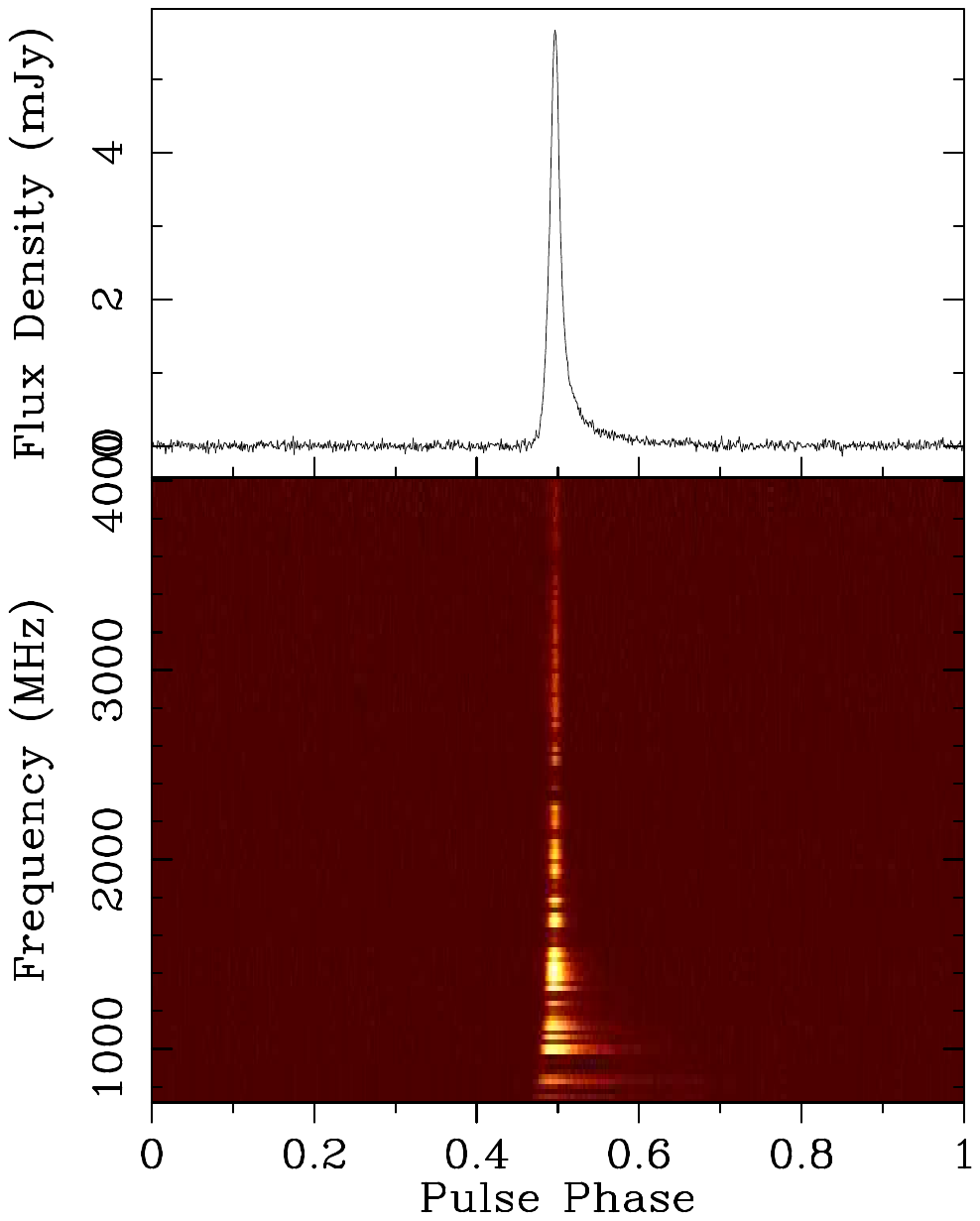}
    \includegraphics[width=0.24\linewidth, trim={40 60 30 170}, clip]{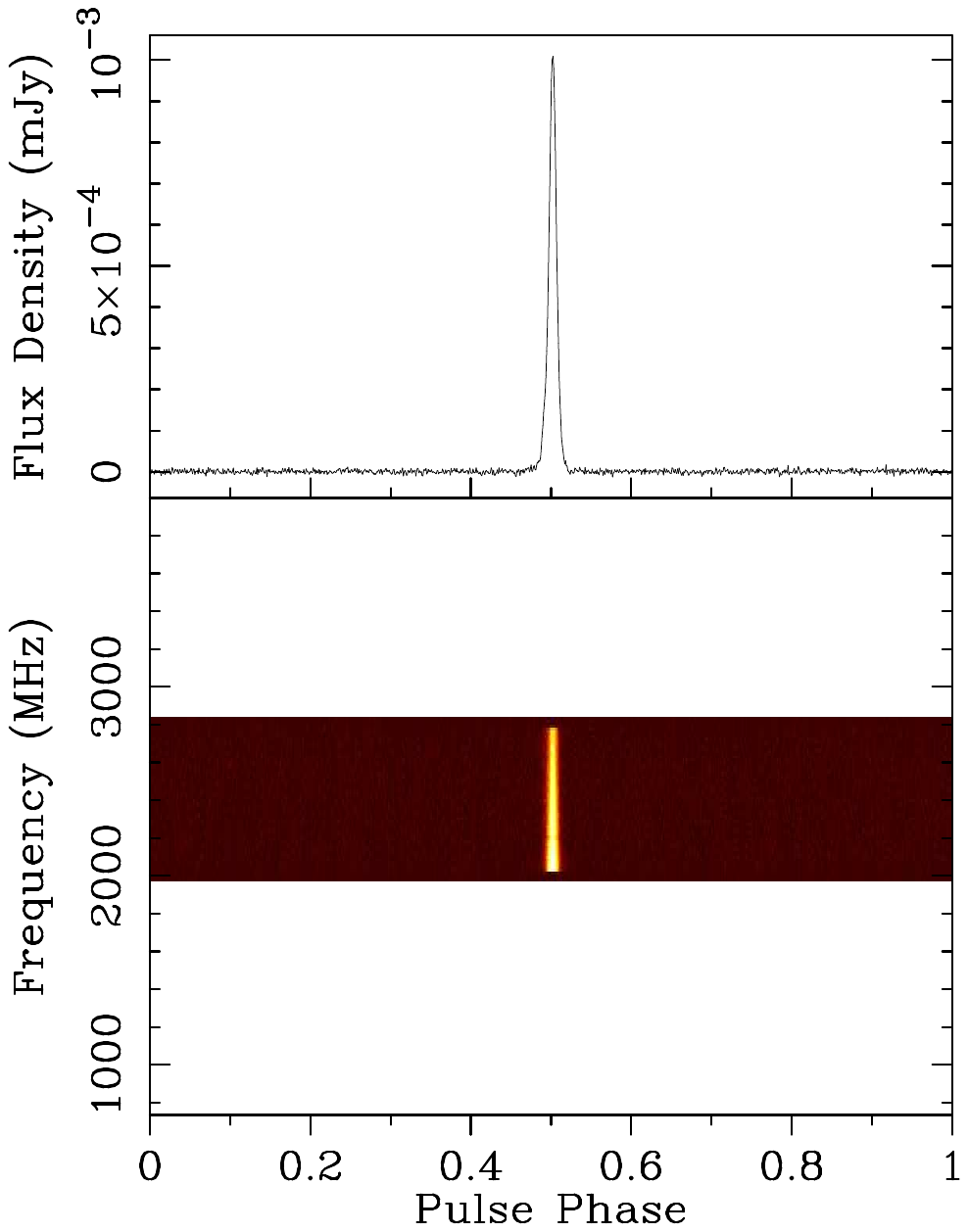}
	\caption{Time-integrated pulse profiles of J1227$-$6208 as recorded with the different receivers (from left to right: Parkes multibeam, MeerKAT L-band, Parkes UWL and MeerKAT S-band). The top plots show the intensity of the integrated emission, while the bottom plots are frequency-resolved. Only the MeerKAT L-band and Parkes UWL data sets are flux calibrated. The plots were made with \texttt{PSRCHIVE/psrplot}.}
	\label{freq_intensity_plots}
\end{figure*}

\subsection{Parkes/Murryiyang}\label{Parkes}

Table~\ref{data} shows the four data sets used in this work. Observations were performed first with the central beam of the 21-cm multibeam receiver \citep{staveley1996multibeam} and later with the ultra wide-bandwidth low-frequency receiver \citep[UWL,][]{hobbs2020uwl}. The multibeam data set provides the longest baseline for the measurement of secular variations of the Keplerian parameters, while the UWL observations are particularly useful for constraining the dispersion measure (DM) evolution due to their overlap with the MeerKAT observations and their very large bandwidth. UWL observations also include three dense orbital campaigns on the dates of 4$-$10 October 2020, 20$-$26 July 2022 and 3$-$9 May 2023, with 7 observations each, accumulating 5.74, 10.41 and 17.63 hours each. These campaigns have two aims: aiding in the measurement of the time delay of
the pulses as they propagate through the gravitational field of the companion \citep[Shapiro delay,][]{shapiro1964test} and constraining DM evolution within a single orbit. In line with this objective, the first of the orbital campaign was coordinated with the MeerKAT orbital campaign with the L-band receivers (Section~\ref{MeerKAT}) so that observations alternate each other.

The multibeam receiver data were recorded by the Center for Astronomy Signal Processing and Electronics Research (CASPER) Parkes Swinburne Recorder \cite[CASPSR,][]{sarkissian2011backends} backend. The data were folded with 512 frequency channels, four polarisation channels, 1,024 phase bins and with coherent de-dispersion at $\textrm{DM}\approx363$~cm$^{-3}$\,pc. Each observation was accompanied by a noise-diode observation for polarisation calibration. Calibration was performed on each file with the \texttt{pac} command from the \texttt{PSRCHIVE}\footnote{\url{https://psrchive.sourceforge.net/}} software package \citep{hotan2004psrchive}. The archives were manually excised of radio-frequency-interference (RFI) with the \texttt{PSRCHIVE/pazi} interface, and the 80 bottom channels and 32 top channels were zero-weighted with the \texttt{PSRCHIVE/paz} command to remove Gaussian noise caused by the loss of sensitivity at the edge of the bandpass. The bands were then scrunched to four frequency sub-bands, a single intensity channel and a single subintegration per observation with the \texttt{PSRCHIVE/pam} command for the production of frequency-resolved times of arrival (ToAs). To derive timing templates for each band, we obtained a frequency-resolved standard profile resulting from the time integration of all multibeam CAPSR observations (with the exclusion of three heavily RFI-affected observations) into a single rotational phase cycle. The analytic timing templates were then produced for each sub-band by fitting a combination of von Mises functions with \texttt{PSRCHIVE/paas} program. The ToAs were produced with the \texttt{PSRCHIVE/pat} command using the FDM algorithm with the two-dimensional, frequency-resolved timing template, where each subintegration resulted a ToA from each of the four sub-bands.

The UWL receiver data were recorded by the Medusa cluster \citep{hobbs2020uwl} with 3,328 frequency channels, four polarisation channels, 1,024 phase bins and with coherent de-dispersion at $\textrm{DM}\approx363$~cm$^{-3}$\,pc. In addition, the data were also recorded with real-time folding based on an early pulsar ephemeris. The data were processed using the \texttt{psrpype} processing pipeline\footnote{\url{https://github.com/vivekvenkris/psrpype}}. This pipeline carries out flux and polarization calibration, and also automatically removes RFI using \texttt{clfd}\footnote{\url{https://github.com/v-morello/clfd}}. It results in data that is cleansed of RFI, calibrated, and broken down into various time, frequency, and polarization resolutions. The RFI excision step also excises a standard set of frequencies, on top of whichever part of the data that the algorithm considers to be contaminated. These frequency sets were decided based on the knowledge of the known transmitter frequencies that routinely affect the data. To boost the quality of ToAs, the data was scrunched into full intensity, eight frequency channels and subintegrations of two hours in length with the \texttt{PSRCHIVE/pam} command. The ToAs were produced with the \texttt{PSRCHIVE/pat} command using the FDM algorithm. Analogously to the multibeam CASPSR data set, the timing template was modelled with \texttt{PSRCHIVE/paas} from the integration of all of the files into a 2-dimensional frequency-resolved standard profile, with the exclusion of three severely RFI-affected observations.

\subsection{MeerKAT}\label{MeerKAT}

\begin{figure*}
	\centering
	\includegraphics[width=0.32\linewidth, clip]{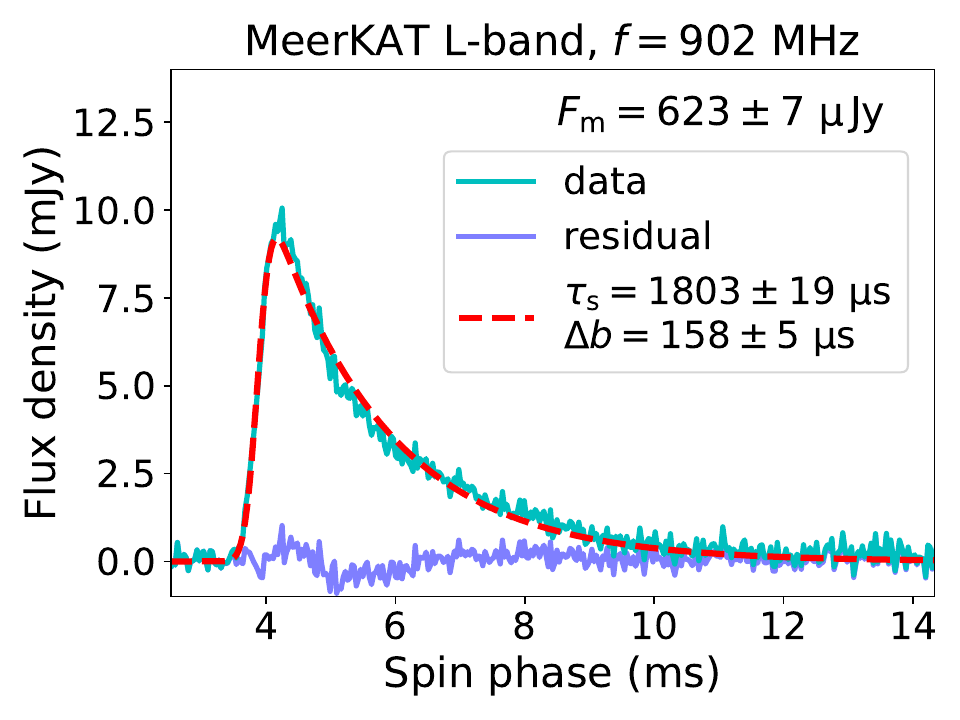}
    \includegraphics[width=0.32\linewidth, clip]{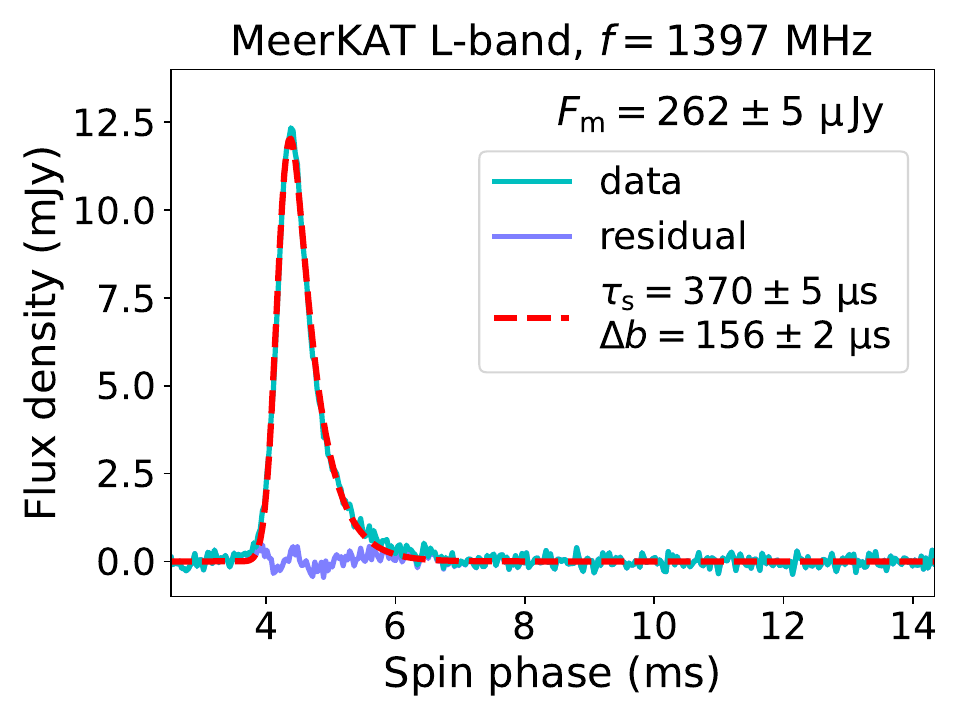}
    \includegraphics[width=0.32\linewidth, clip]{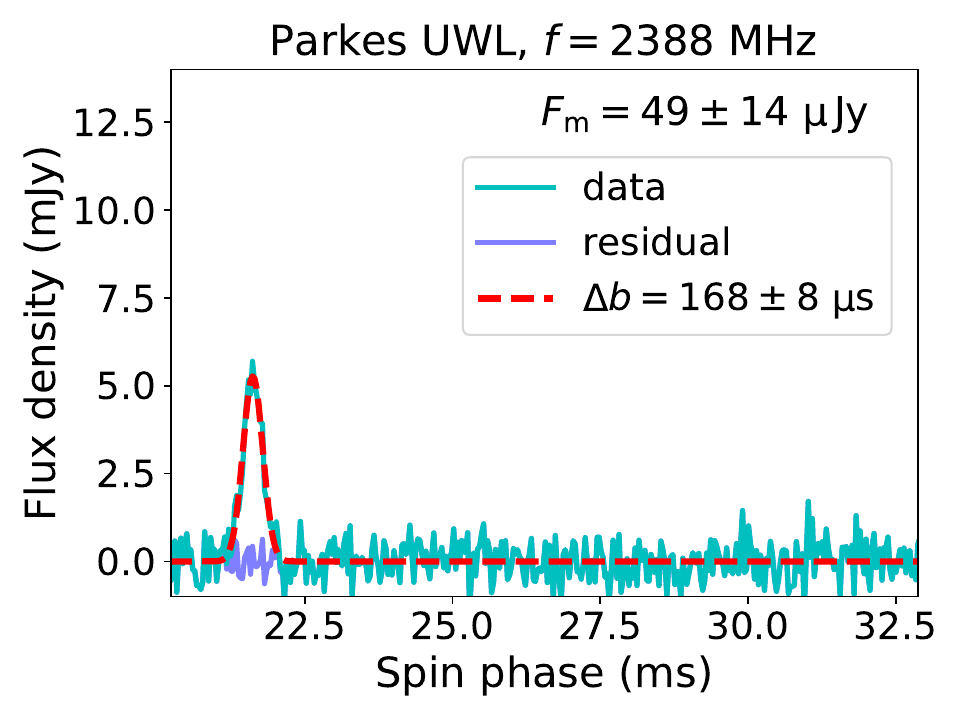}
    \includegraphics[width=1.9\columnwidth, clip]{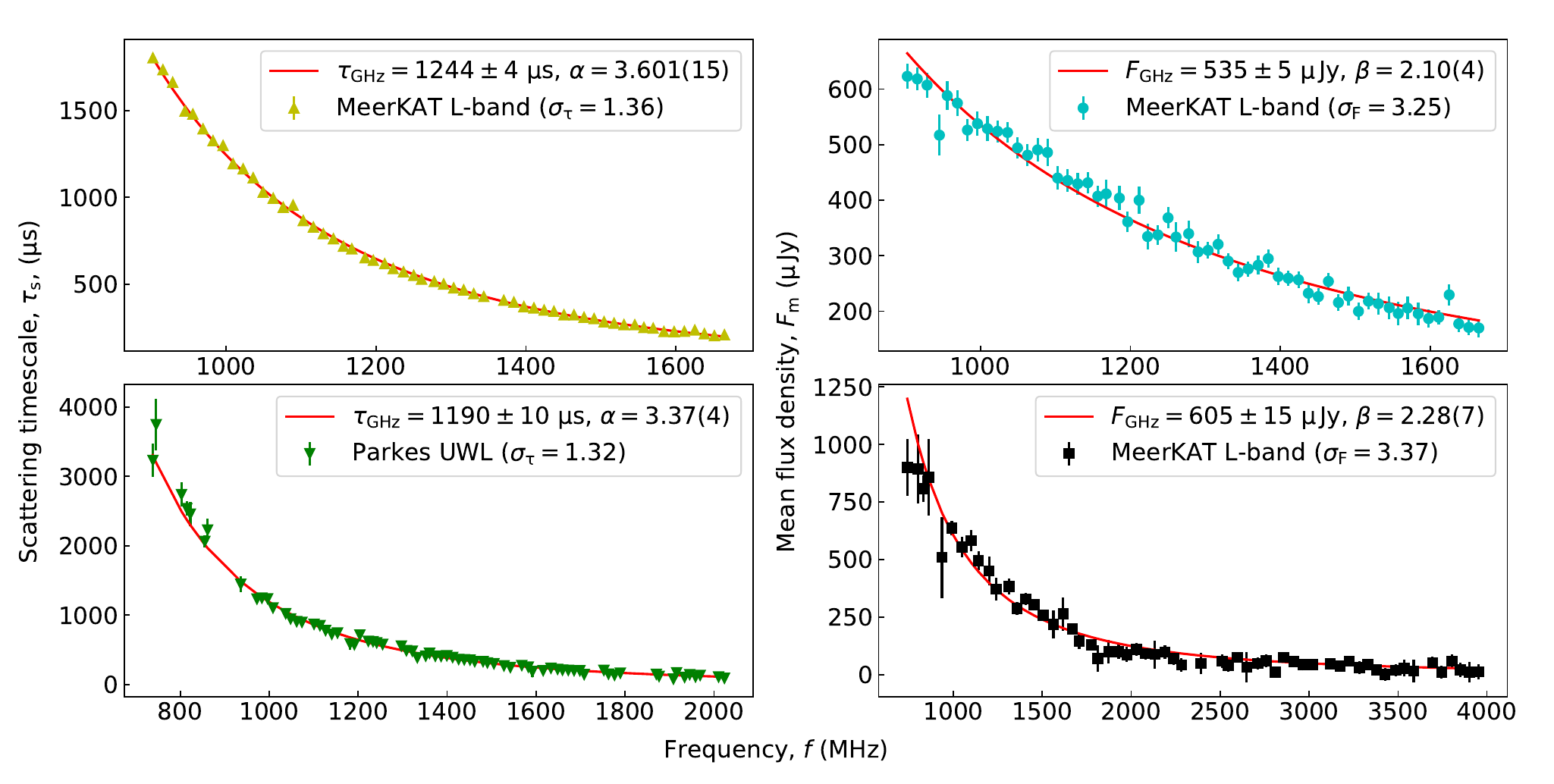}
	\caption{\textbf{Top:} pulse profiles of J1227$-$6208 at $f=902$ MHz and $f=1397$ MHz as seen by the MeerKAT L-band receiver (left and middle plots), and at $f=2388$ MHz as seen by the Parkes UWL receiver (right plot). The data (cyan lines) is shown against the best model (red dashes, fitted with equation~\ref{profile_model}) and the fit residuals (blue lines). The fit parameters $\tau_\textrm{s}$ and $\Delta b$ from equation~(\ref{profile_model}), and the mean flux density $F_\textrm{m}$ are quoted on each profile. \textbf{Bottom:} measurements of $\Delta\tau$ and $F_\textrm{m}$ at each frequency on the MeerKAT L-band and Parkes UWL profiles (dots with uncertainties), and the best fits of the power laws~(\ref{scattering_index}) and (\ref{intensity_evolution}) (red lines). The data points are displayed accounting for the error factors $\sigma_\textrm{\texttau}$ and $\sigma_\textrm{F}$, and the relevant $\tau_\textrm{GHz}$, $\alpha$, $F_\textrm{GHz}$ and $\beta$ values are quoted in the legend.}
	\label{profile_fits}
\end{figure*}

The MeerKAT data sets provide the most precise ToAs owing to the large sensitivity of the telescope. Most of the data were recorded with the L-band receivers (26 hours, 856$-$1712 MHz), but 19 hours were recorded with the new S-band receivers in the S1 configuration \citep[1968$-$2843 MHz,][]{barr2018sband} on 12$-$28 May 2023 to ensure a measurement of the Shapiro delay with a significant reduction of DM noise (see Table~\ref{data}). The data were recorded with the pulsar timing user-supplied equipment \citep[PTUSE,][]{bailes2020meerKAT} machines as part of the MeerTIME science program \citep{bailes2020meerKAT}, and included an orbital campaign on the dates of 4$-$11 October 2020 that accumulated 13 hours over nine L-band observations, with a dedicated five-hour-long observation at superior conjunction on 11 October 2020, aimed at constraining the Shapiro delay.

Both data sets were recorded with 1,024 frequency channels, four polarisation channels and coherent dedisperion at $\textrm{DM}\approx363$~cm$^{-3}$\,pc. The L-band data were processed, cleaned of RFI and calibrated with the \texttt{meerpipe} pipeline\footnote{\url{https://zenodo.org/records/7961071}}, which resulted in the trimming of the outer edges, leaving only 926 channels of useful data with a bandwidth of 775.5~MHz. Any remaining RFI was manually removed with the \texttt{PSRCHIVE/paz} command. Observations were then fully summed in polarization, and scrunched into eight frequency channels and 20-min-long subintegrations to extract frequency-resolved ToAs with the \texttt{PSRCHIVE/pat} command. Analogous to the Parkes data sets (Section~\ref{Parkes}), the frequency-resolved timing template was derived using the \texttt{PSRCHIVE/paas} on frequency-resolved standard profiles resulting from the time integration of the observations in the October 4$-$11 October 2020 orbital campaign. Unfortunately, calibration files were not available for the S-band data set, as it was taken jointly with commissioning data, and therefore it could not be calibrated nor band-pass corrected. This, however, is unlikely to affect our timing precision. S-band data were cleaned of RFI with the \texttt{clfd} software, and it was sub-banded to four frequency channels for frequency-resolved timing, with the profile being created from the integration of all of the observations, following the same steps as the L-band data.

\section{Emission analysis}\label{emission_analysis}

\subsection{Profile evolution}

J1227$-$6208 suffers from significant scattering and has a steep flux density spectrum. Fig.~\ref{freq_intensity_plots} shows time-integrated, frequency-resolved standard profiles from the four different receivers listed in Table~\ref{data}. The most obvious feature is scattering at the MeerKAT L-band and the bottom of the Parkes UWL band, resulting in an exponential, frequency-dependent tail of the pulse profile. In addition, the pulse brightness fades quickly at the top of the Parkes UWL band. It should be noted that the dark bands in the Parkes UWL profile are a result of the RFI excision processes, and are not intrinsic to the pulsar emission. To measure the scattering timescale $\tau_\textrm{s}$ at each frequency, we model the pulse at a given frequency $f$ as a single Gaussian function convolved with a scattering exponential tail
\begin{equation}\label{profile_model}S_b=\int A\times\exp\left(\frac{\left(b'-b_0\right)^2}{2\times\Delta b^2}\right)\times\exp\left(-\frac{b-b'}{\tau_\textrm{s}}\right)\textrm{d}b'
\textrm{,}\end{equation}
where $b$ is the spin phase in bins, and $b_0$ and $\Delta b$ stand for the Gaussian center and standard deviation. We measure the scattering index $\alpha$ by fitting the evolution of $\tau_\textrm{s}$ as a power law function of $f$,
\begin{equation}\label{scattering_index}\tau_\textrm{s}\left(f\right)=\tau_\textrm{GHz}\left(\frac{f}{\textrm{GHz}}\right)^{-\alpha}\textrm{,}\end{equation}
where $\tau_\textrm{GHz}$ is the reference value at $f=1$~GHz. For the flux density spectral index $\beta$, we measure the mean flux density  $F_\textrm{m}$ across the pulse phase at each sub-band, deriving the uncertainty from the off-pulse baseline noise, and fit another power law
\begin{equation}\label{intensity_evolution}F_\textrm{m}\left(f\right)=F_\textrm{GHz}\left(\frac{f}{\textrm{GHz}}\right)^{-\beta}\textrm{,}\end{equation}
where $F_\textrm{GHz}$ is the reference value at $f=1$~GHz.

We perform least-$\chi^2$ fits of equations~(\ref{profile_model}),(\ref{scattering_index}) and (\ref{intensity_evolution}) on the frequency-resolved, time-integrated MeerKAT L-band and Parkes UWL profiles (second and third plots in Fig.~\ref{freq_intensity_plots}) with the \texttt{python} module \texttt{scipy} \citep{virtanen2020scipy}. As shown in the upper part of Fig~\ref{profile_fits}, the scattered Gaussian modelling of the profile adjusts well to the data, with only some minor residual structure after subtraction of the model. The intrinsic pulse width stays consistent at $150<\Delta b<170$~\textmu s across the whole band, and the parameter $\tau_\textrm{s}$ becomes redundant in the modelling of the pulse at $f>2$~GHz in the Parkes UWL dataset, where a simple Gaussian function provides an adequate description. Thus, we restrict the measurement of $\tau_\textrm{GHz}$ and $\alpha$ at $f<2$~GHz. To ensure a proper uncertainty estimation on the scattering and spectral parameters, we multiply the individual measurement uncertainties of $\tau_\textrm{s}$ and $F_\textrm{m}$ by the error factors $\sigma_\textrm{\texttau}$ and $\sigma_\textrm{F}$ ($\delta^\prime=\sigma\times\delta$, where $\delta$ is the set of measurement uncertainties). Subsequently, we tune the values of $\sigma_\textrm{\texttau}$ and $\sigma_\textrm{F}$ to achieve a reduced $\chi^2=1$ in the fit of equations~(\ref{scattering_index})~and~(\ref{intensity_evolution}).

Our fits confirm the steep spectrum of J1227$-$6208 and measure a scattering index $\alpha<4$. The lower part of Fig.~\ref{profile_fits} shows the fit parameters for $\tau_\textrm{GHz}$, $\alpha$, $F_\textrm{m}$ and $\beta$ from equations~(\ref{scattering_index}) and (\ref{intensity_evolution}) on the measurements of $\tau_\textrm{s}$ and $F_\textrm{m}$ in the MeerKAT L-band and Parkes UWL datasets. The power law describes $\tau_\textrm{s}$ accurately for the MeerKAT L-band data set, with $\tau_\textrm{GHz}=1244\pm4$~\textmu s and $\alpha=3.601(15)$. However, this is different from the measurement performed in the Parkes UWL profile, which results in $\tau_\textrm{GHz}=1190\pm10$~\textmu s and $\alpha=3.37(15)$. The difference likely comes from the different frequency coverage and the low quality of measurements at $f<900$~MHz. Nonetheless, both fits agree on $\alpha<4$, far from the $\alpha=4.4$ predicted from a Kolmogorov medium and the $\alpha=4$ from Gaussian inhomogeneities \citep{rickett1977scattering,romani1986}, as already seen in some pulsars behind complex ISM environments \citep[e.g.][]{lohmer2004frequency,krishnakumar2019scatter,oswald2021scattering}. On the other hand, our $F_\textrm{m}$ measurements are not as precise, as indicated by the $\sigma>3$ values. The most likely explanation is the lower $S/N$ of the Parkes standard profiles due to its lower gain and the large degree of zapping of frequency channels during RFI excision. Nonetheless, both the MeerKAT L-band and Parkes UWL agree on a very steep spectrum, with $\beta=2.10(4)$ and $\beta=2.28(7)$, respectively, on the steeper side the typically observed values of $\beta=1.4\pm1.0$ \citep{bates2013spectral}. The derived reference flux densities are $F_\textrm{GHz}=535\pm5$~\textmu Jy and $F_\textrm{GHz}=605\pm15$~\textmu Jy. Regarding the peak flux density, the combined effect of scattering and spectral behaviour results in the brightest observed peak flux density of 13$-$14~mJy -and highest $S/N$- at $f\approx1200-1300$~MHz.

\subsection{Polarised emission}

\begin{figure}
	\centering
	\includegraphics[width=\linewidth, trim={180 60 220 85}, clip]{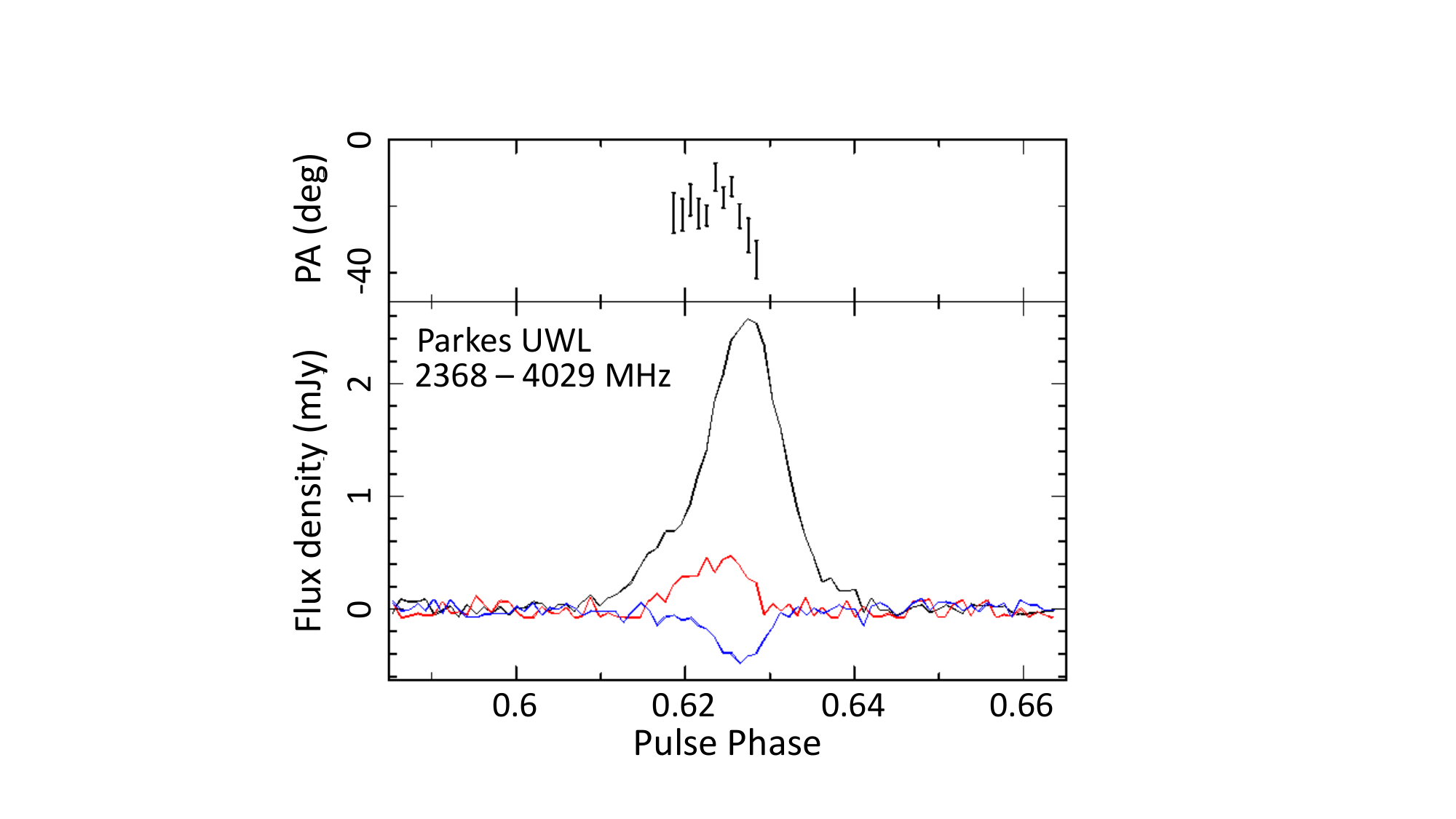}
	\includegraphics[width=\linewidth, trim={180 40 220 85}, clip]{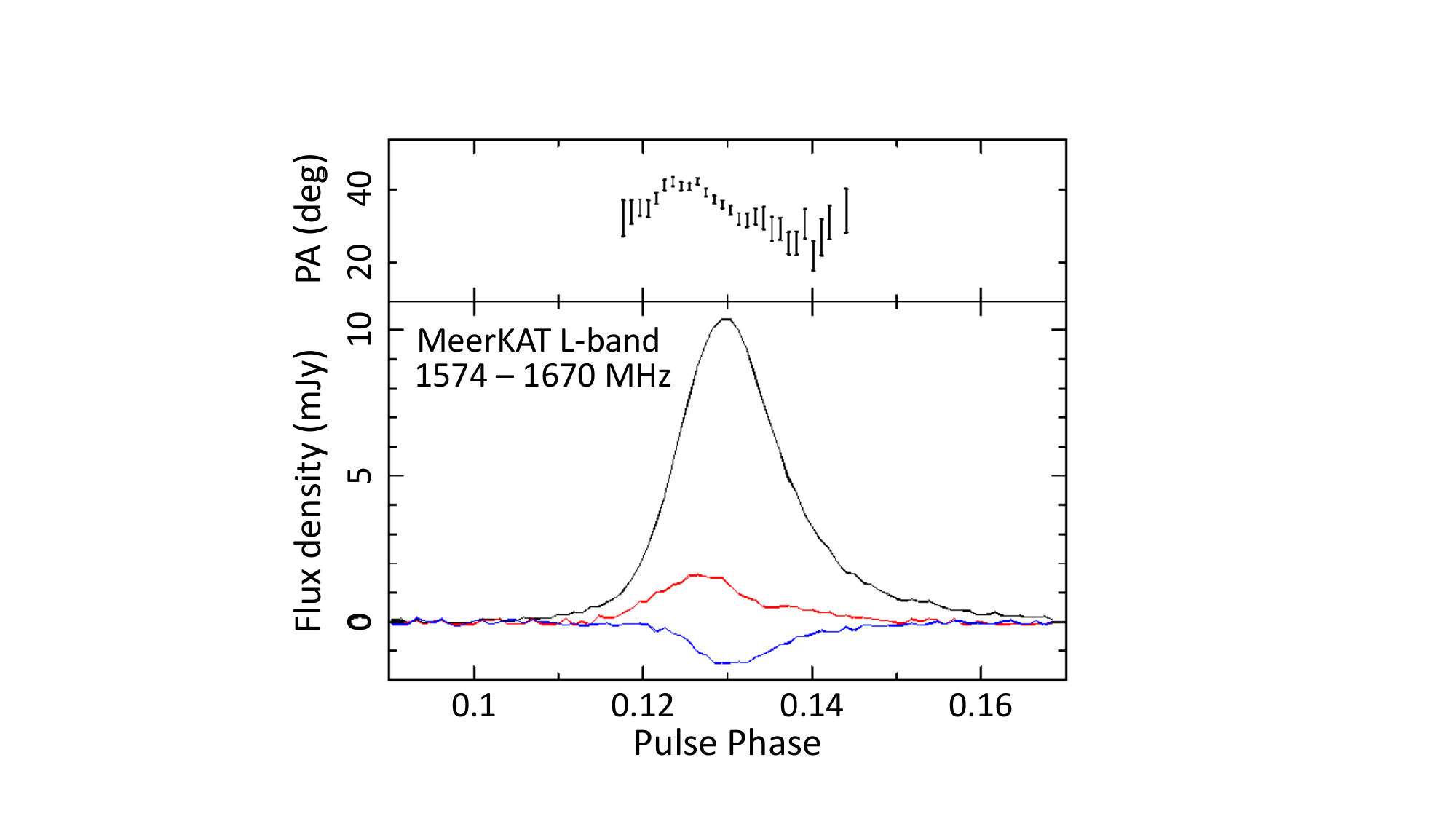}
	\caption{Parkes UWL (top) and MeerKAT L-band (bottom) polarisation profiles of J1227$-$6208, RM-corrected at 47.3~rad\,cm$^{-1}$. Total flux density (black lines), linearly polarised intensity (red lines), circularly polarised intensity (blue lines) and the PA angle (black error bars) and are plotted as a function of the pulse phase.}
	\label{polarisation}
\end{figure}

\begin{table*}[h!]
\caption[]{\label{fits} Resulting noise models as derived by \texttt{TEMPO2/TempoNEST}.}
\centering
    \begin{tabular}{lcccccccccccccccc}
    \hline
    \hline \\[-1.5ex]
    Model & $\nu_\textrm{cut,spin}^{-1}$ & $\nu_\textrm{cut,DM}^{-1}$ & $\log{E}$ & $\chi^2$ & $\log{A_\textrm{spin}}$ & $\gamma_\textrm{spin}$ & $\log{A_\textrm{DM}}$ & $\gamma_\textrm{DM}$ \\[0.3ex]
     & (days) & (days) &  &  &  &  &  & \\
    \hline \\[-1.5ex]
    \textbf{L}   & ... & 100 & 16310.61(4) & 1.013 & ...    & ...   & -10.02 & 1.148 \\
    \textbf{Ls}  & 500 & 100 & 16387.3(1)  & 0.973 & -11.71 & 1.485 &  -9.99 & 1.320 \\
    \textbf{C}   & ... &  50 & 16326.21(8) & 0.983 & ...    & ...   & -10.09 & 1.113 \\
    \textbf{Cs}  & 100 &  50 & 16416.80(2) & 0.966 & -11.75 & 1.499 & -10.01 & 1.573 \\
    \textbf{VC}  & ... &  30 & 16329.1(2)  & 0.944 & ...    & ...   & -10.09 & 1.432 \\
    \textbf{VCs} &  30 &  30 & 16430.9(5)  & 0.932 & -11.81 & 1.097 & -10.04 & 1.307 \\ 

    \hline
    \hline
    \end{tabular}
\tablefoot{
The best-fit reduced $\chi^2$ is derived from a subsequent \texttt{TEMPO2}
fit.}
\end{table*}

We perform a rotation measure (RM) fit on the MeerKAT L-band and Parkes UWL profiles with \texttt{PSRCHIVE/rmfit}, finding the maximum of integrated linearly polarised flux density at $\textrm{RM}=47.7$~and~46.2~rad\,cm$^{-1}$, with half-widths full-maximums of 50 and 30~rad\,cm$^{-1}$, respectively. These two values are consistent with each other, and they imply an average magnetic field parallel to the line of sight (LOS) of 0.16~\textmu G. Fig.~\ref{polarisation} shows the polarised emission of J1227$-$6208 at $f=1574-1670$~MHz from the MeerKAT L-band profile, and at $f=2368-4029$~MHz from the Parkes UWL profile. Approximately 15\% of the total intensity is linearly polarised, and a similar fraction of circular polarisation is detected. Linear polarisation is more prominent in the first half of the pulse, peaking before the total intensity, while circular polarisation peaks along with the total intensity. The position angle (PA) of the linear polarisation shows an increase before the linearly polarised emission peaks, and a decrease afterwards, instead of following the rotating vector model from \cite{radhakrishnan1969magnetic}. In the MeerKAT L-band, scattering drags some of the polarised emission into the scattering tail, resulting in some spurious PA angle measurements towards the end of the pulse. In the Parkes data set, an extra feature of the profile is revealed at the beginning of the pulse, making it slightly asymmetric. This feature is coincident with the off-centre linearly polarised emission.

\section{Timing analysis}\label{timing_anlysis}

\subsection{Noise modelling}\label{noise_section}

We fit the entire ToA time series with the DDH model, a modified version of the Damour-Deurelle timing model \cite[DD,][]{damour1986general} which includes the orthometric parametrization of the Shapiro delay \citep{freire2010orthometric}. The fit itself was performed with the pulsar timing software \texttt{TEMPO2}\footnote{\url{https://bitbucket.org/psrsoft/tempo2}}\citep{edwards2006tempo2,hobbs2006tempo2} and its plug-in \texttt{TEMPO2/TempoNEST}
\footnote{\url{https://github.com/LindleyLentati/TempoNest}}, which implements multi-nested Bayesian sampling of the highly multidimensional space of the timing model, including the pulsar spin and astrometric parameters, the DM evolution, the five Keplerian parameters and five independent PK parameters. In addition, \texttt{TEMPO2/TempoNEST} also fits for Gaussian white noise and correlated red noise. 

The ToA time series is affected by Gaussian instrumental noise and red spin and DM noise, which can significantly contaminate the measurement of other parameters in the timing model. In particular, there is a significant DM variability as measured in different observations, an effect that becomes very prominent in the MeerKAT L-band and Parkes UWL data sets. Following the method first described in \cite{vanHaasteren2009grav}, we model the presence of red noise as a power spectral density in the Fourier domain described with a power law 
\begin{equation}\label{powerlaw}S\left(\nu\right)=A^2\left(\frac{\nu}{\textrm{yr}^{-1}}\right)^{-\gamma}\textrm{,}\end{equation}
where $A=\{A_\textrm{spin},A_\textrm{DM}\}$ is the dimensionless amplitude of the correlation matrix, $\nu$ is the spectral frequency and $\gamma=\{\gamma_\textrm{spin},\gamma_\textrm{DM}\}$ is the power law index. We account for two possible sources of red noise: spin noise caused by the rotational variations originating within the pulsar, and DM noise originated by variations within the interstellar medium (ISM) along the LOS. The spin noise is frequency-independent (achromatic), while the DM noise is defined by its frequency dependence (chromatic), with
\begin{equation}\label{DMtoRes}A_{\textrm{DM},f} =A_\textrm{DM}\times\left(\frac{f}{1,400\,\textrm{MHz}}\right)^{-2}\textrm{ ,}\end{equation}
where $A_\textrm{DM}$ corresponds to the amplitude of red DM noise at $f=1,400$~MHz. To account for uncorrelated Gaussian (white) instrumental and pulsar noise for each telescope backend $k$ (Section~\ref{backends}), we add the EQUAD$_k$ and EFAC$_k$ parameters that re-scale the ToA uncertainties $\sigma_\textrm{k}$ into
\begin{equation}\hat\sigma_k^2=(\textrm{EFAC}_k\,\sigma_k)^2+\textrm{EQUAD}_k^2\textrm{,}\end{equation}
where $k$ is specific to each telescope backend. As a starting point, \texttt{TEMPO2/TempoNEST} only requires the extra assumption of a cut-off frequency for the correlated noises as a user-given input. We performed six different \texttt{TEMPO2/TempoNEST} runs: half of them include white noise and red DM noise, while the other half also include red spin noise. Each of these runs has different noise cut-offs, chosen from $\nu^{-1}_\textrm{cut}=30, 50, 100$~and~$500$~days. The resulting noise models, listed in Table~\ref{fits}, are named according to the cut-offs and inclusion or exclusion of red spin noise: loose (\textbf{L}), loose with spin (\textbf{Ls}), constrained (\textbf{C}), constrained with spin (\textbf{Cs}), very constrained (\textbf{VC}) and very constrained with spin (\textbf{VCs}).

All noise models provide an adequate description of the data, including those that do not include red spin noise, and red DM noise is always dominant over red spin noise when both are included. As shown in Table~\ref{fits}, models that include both noises are favoured by the nested importance sampling global log-evidence ($\log{E}$), and $A_\textrm{DM}$ is two orders of magnitude larger than $A_\textrm{spin}$. That is not surprising, as J1227$-$6028 is a recycled pulsar with an expected high rotational stability, and it shows that the main source of timing noise is indeed the ISM. Furthermore, Table~\ref{fits} also shows that models with shorter $\nu_\textrm{cut}^{-1}$ values are favoured by the $\log{E}$. Nonetheless, the resulting $\chi^2<0.95$ values for the \textbf{VC} and \textbf{VCs} models hint towards the possibility of over-fitting when accounting for large frequencies in the red noises.

\subsection{PK parameters constraints}\label{PKparameters_section}

\begin{table*}

\caption[]{\label{physics} Constraints on the PK parameters and the component masses from the global fits.}
\centering
    \begin{tabular}{lcccccccccccccccccc}
    \hline
    \hline \\[-1.5ex]
    Noise & $\dot\omega$ & $h_3$ & $\varsigma$ & $\dot P_\textrm{b}$ & $\dot x$ & $M_\textrm{p}$ & $M_\textrm{c}$ & $i$ \\[0.3ex]
    model & (deg\,yr$^{-1}$) & (\textmu s) &  & $\times10^{-13}$~s\,s$^{-1}$ & $\times10^{-15}$~ls\,s$^{-1}$ & ($M_\odot$) & ($M_\odot$) & (deg) \\
    \hline \\[-1.5ex]
    \textbf{L}   & 0.0171(9) & 3.8(3) & 0.84(4) & $0.02\pm2.4$ & $-1.6\pm6.6$ & 1.58(11) & 1.41(5) & 79.2(6) \\[0.5ex]
    \textbf{Ls}  & 0.0173(8) & 3.7(3) & 0.84(3) & $2.9\pm2.3$ & $-3.5\pm6.5$ & 1.59(10) & 1.42(5) & 78.6(7) \\[0.5ex]
    \textbf{C}   & 0.0170(9) & 3.7(3) & 0.85(3) & $1.0\pm2.6$ & $-5.0\pm6.9$ & 1.53(11) & 1.38(5) & 79.3(7) \\[0.5ex]
    \textbf{Cs}  & 0.0171(8) & 3.5(2) & 0.86(3) & $3.7\pm2.5$ & $-9.1\pm7.2$ & 1.53(10) & 1.39(5) & 78.4(7) \\[0.5ex]
    \textbf{VC}  & 0.0171(9) & 3.6(3) & 0.86(3) & $1.0\pm2.6$ & $-3.7\pm7.0$ & 1.54(11) & 1.39(5) & 79.2(7) \\[0.5ex]
    \textbf{VCs} & 0.0170(8) & 3.4(2) & 0.85(3) & $4.6\pm2.7$ & $-6.5\pm8.0$ & 1.51(11) & 1.38(5) & 78.2(7) \\[0.5ex]
    \textbf{extended} & 0.0171(11) & 3.6(5) & 0.85(5) & ... & ...            & 1.54(15) & 1.40(7) & $78.7\pm1.2$ \\[0.5ex]
    \hline
    \hline
    \end{tabular}
\tablefoot{
The PK parameters were measured in a global fit with \texttt{TEMPO2} under the assumption of the noise models derived by \texttt{TEMPO2/TempoNEST} (Table~\ref{fits}). The component masses and the inclination angle are derived from the $\chi^2$ mapping of DDGR solutions (Section~\ref{massesSection}). The \textbf{extended} row shows comprehensive uncertainties accounting for all measurements across the different noise models.
}
\end{table*}

We fit five PK parameters to constrain the component masses and the effects of proper motion. These are the rate of periastron advance $\dot\omega$, the orthometric shape and range parameters of the Shapiro delay $h_3$ and $\varsigma$ as defined in \cite{freire2010orthometric}, the orbital period derivative $\dot P_\textrm{b}$ and the projected semi-major axis derivative $\dot x$. From the preliminary timing solution, we expect $\dot\omega$ to be dominated by the relativistic precession of the Keplerian orbit as predicted by GR,
\begin{equation}\label{omdot}\dot\omega=3\left(\frac{P_\textrm{b}}{2\pi}\right)^{-5/3}\frac{\left(T_\odot M_\textrm{t}\right)^{2/3}}{1-e^2}\textrm{,}\end{equation}
where the total system mass $M_\textrm{t}$ is expressed in units of $M_\odot$, and $T_\odot=GM_\odot/c^3=4.92549094764$~\textmu s as defined in \cite{prvsa2016values}.
The Shapiro delay describes the periodic time delay of the pulses as they propagate through the gravitational field of the companion \citep{shapiro1964test}. However, as its periodicity is equal to $P_\textrm{b}$, part of its signal is degenerate with the R{\o}mer delay for systems with low inclination angles ($i$). Therefore, we use the orthometric parameters $h_3$ and $\varsigma$ to model the residual unabsorbed component of the Shapiro delay instead \citep{freire2010orthometric}. The orthometric amplitude
\begin{equation}\label{h3}h_3=T_\odot M_\textrm{c}\left(\frac{1-\cos{i}}{1+\cos{i}}\right)^{3/2}\end{equation}
describes the amplitude of the unabsorbed component, which depends on the companion mass $M_\textrm{c}$ and the inclination angle $i$, while the orthometric shape
\begin{equation}\label{stig}\varsigma=\left(\frac{1-\cos{i}}{1+\cos{i}}\right)^{1/2}\end{equation}
describes the shape of the delay in the orbital phase, depending only on $i$.

The remaining PK parameters, $\dot P_\textrm{b}$ and $\dot x$, have the estimated GR contributions of $\dot P_\textrm{b}\sim10^{-15}$~s\,s$^{-1}$ and $\dot x\sim10^{-20}$~ls\,s$^{-1}$, but likely much more dominant contributions are those of the Galactic acceleration field and the proper motion. For $\dot P_\textrm{b}$, we expect the Shklovskii effect and the Galactic acceleration field to be the dominant contributors \citep{shklovskii1970possible,damour1992orbital}, which introduce a derivative based on the Doppler factor ($D$) derivatives:
\begin{equation}\label{shk_acc}\frac{\dot P_\textrm{b}}{P_\textrm{b}}=-\left( \frac{\dot D}{D}\right)_\textrm{Shkl}-\left(\frac{\dot D}{D}\right)_\textrm{Gal}=\frac{1}{c} \left[ |\vec{\mu}|d+\vec{K_0}(\vec{a}_\textrm{PSR}-\vec{a}_\textrm{SSB}) \right] \textrm{ ,}\end{equation}
where $\vec{\mu}=(\mu_\textrm{RA},\mu_\textrm{DEC})$ is the sky proper motion of the pulsar, $\vec{K_0}$ is the Solar System barycenter (SSB) to pulsar system unit vector and $\vec{a_\textrm{PSR}}$ and  $\vec{a_\textrm{SSB}}$ are the Galactic acceleration field measured at the pulsar system and the SSB, respectively. On the other hand, the contribution to $\dot x$ is expected to be dominated by the geometric effect introduced by the proper motion of the system on the sky \citep{kopeikin1996proper}:
\begin{equation}\label{xdotPM}\frac{\dot x}{x}=1.54\times10^{-16}\cot{i}\left(-\frac{\mu_\textrm{RA}}{\textrm{mas\,yr}^{-1}}\sin{\Omega_\textrm{a}}+\frac{\mu_\textrm{DEC}}{\textrm{mas\,yr}^{-1}}\cos{\Omega_\textrm{a}}\right)\textrm{\,,}\end{equation}
where $\Omega_\textrm{a}$ is the longitude of ascending node. Finally, we do not include the amplitude of the Einstein delay $\gamma_\textrm{E}$, as for circular systems like J1227$-$6208 it is expected to be highly degenerate with $\dot x$, and its inclusion would lead to the non-detection of $\dot x$ instead (see \citealt{ridolfi2019possible} for a detailed discussion of this phenomenon).

\begin{figure*}
\centering
  \includegraphics[width=2\columnwidth]{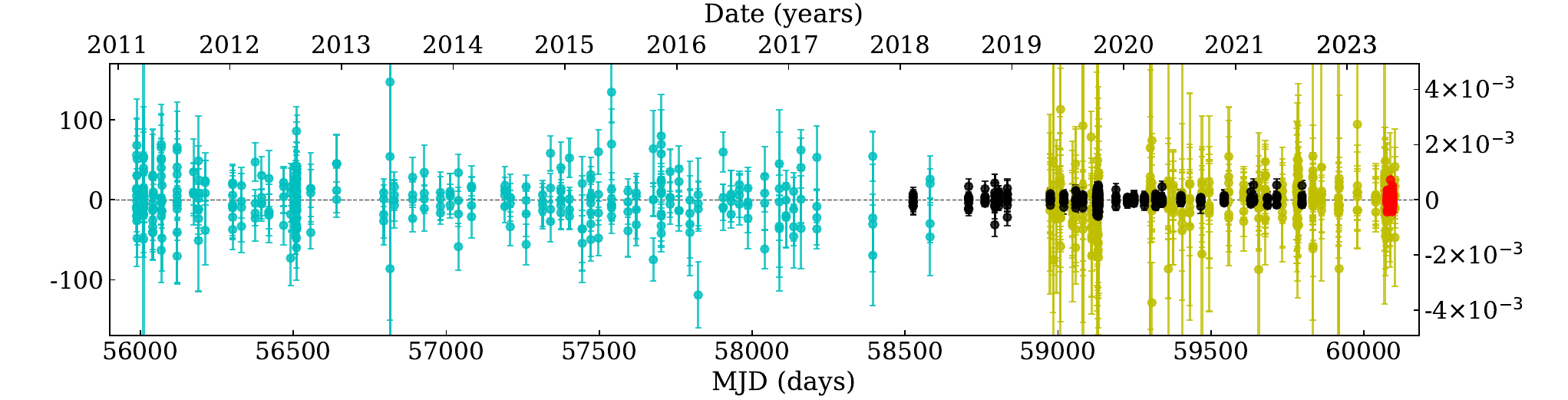}
 \includegraphics[width=2\columnwidth]{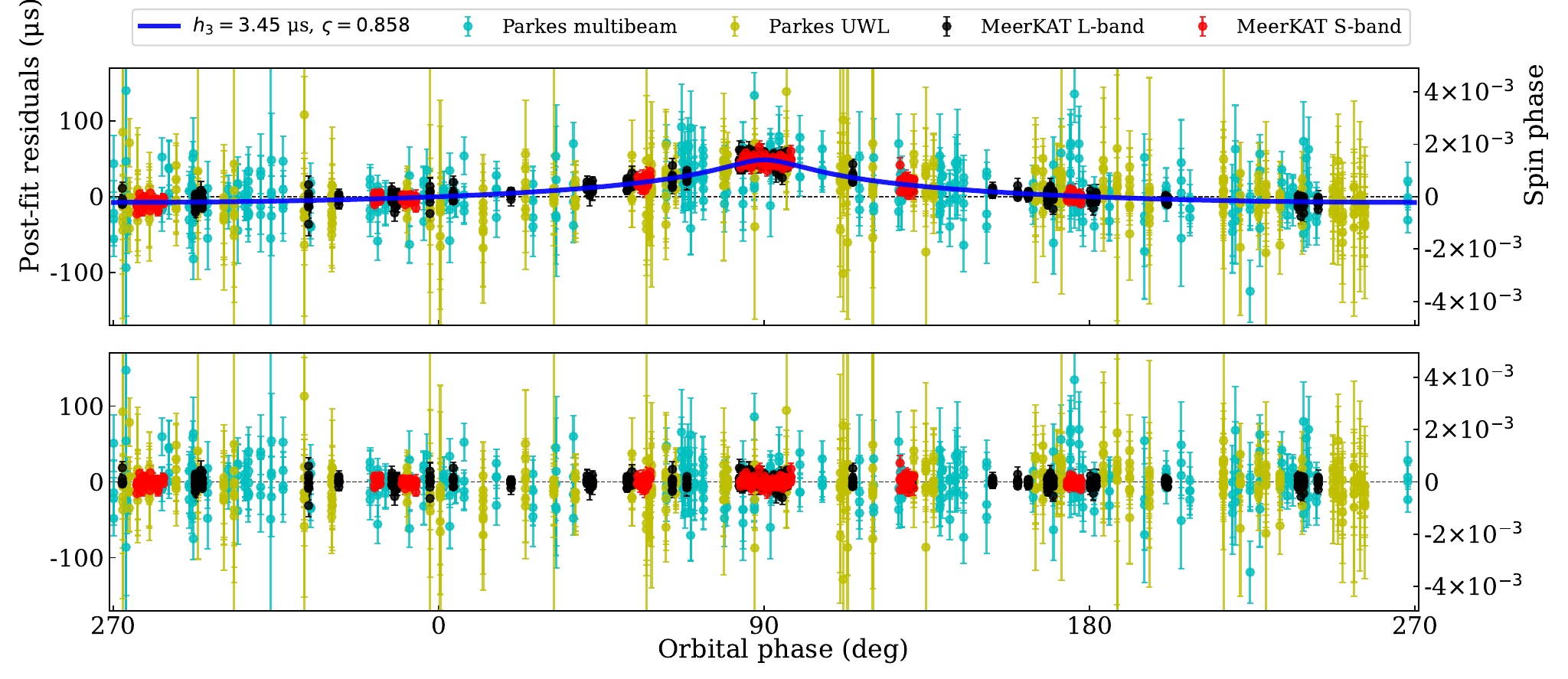}
 \caption{\textbf{Top:} ToA residuals as a function of MJD under the assumption of the \textbf{Cs} noise model, showing flat, Gaussian residuals. \textbf{Middle:} ToA residuals as a function of the orbital phase (orbital position from the ascending node, true anomaly $+$ periastron) excluding the $h_3$ and $\varsigma$ parameters from the timing model, showing the full amplitude of the Shapiro delay signal. The continuous blue line depicts the predicted Shapiro delay described by $h_3=3.45$~\textmu s and $\varsigma=0.858$. \textbf{Bottom:} the orthometric Shapiro delay parameters are now included in the model, resulting in flat, Gaussian residuals.}
 \label{shapiro_residuals}
\end{figure*}

To measure the PK parameters, we assume the noise models found by \texttt{TEMPO2/TempoNEST} and perform a global re-fit of all model parameters with \texttt{TEMPO2}. We quote the fit values and 1\textsigma~uncertainties as reported by the least-$\chi^2$ \texttt{TEMPO2} fit, which are consistent with the ones reported by \texttt{TEMPO2/TempoNEST} but with slightly larger uncertainties. We choose this because the \texttt{TEMPO2} fits are more consistent across noise models, and because it provides a more conservative uncertainty estimate. Additionally, it is consistent with the use of \texttt{TEMPO2} in the mass measurements presented in the following sections.

The measured PK parameters are consistent across all noise models. Table~\ref{physics} presents the measured value of each PK parameter for each assumed noise model in the global fit. The rate of periastron advance presents a very consistent value of $\dot\omega=0.0171(11)$~deg\,yr$^{-1}$ if we consider extended uncertainties from all of the measurements listed in Table~\ref{physics}, which results in a $\sim$15\textsigma~detection and indicates a total system mass of $M_\textrm{t}=2.9\pm0.3$~$M_\odot$. The Shapiro delay parameters $h_3=3.6\pm0.5$~\textmu s and $\varsigma=0.85\pm0.05$ are measured with high significance for the first time, presenting $\sim$7\textsigma~and $\sim$17\textsigma~detections if we consider extended uncertainties from across all the noise models.

The red DM noise is a major source of uncertainty in our measurement. The parameters most affected by the choice of the noise model is $h_3$, ranging from $3.8(3)$~\textmu s in the \textbf{L} model to $3.4(2)$~\textmu s in the \textbf{VCs}. In general, we measure lower values in models with high-frequency cut-offs and with both spin and DM noise. A possible explanation is that unmodelled red noise can bias the Shapiro delay measurement to higher amplitude values, especially achromatic spin noise. A similar phenomenon can occur in the opposite direction, with noise models removing power from the Shapiro delay signal when the orbital period is close to $\nu_\textrm{cut}$. That may be a potential explanation for the $\chi^2<0.95$ value in the \textbf{VC} and \textbf{VCs} models (Table~\ref{fits}). Nonetheless, the measured PK values of all parameters are consistent across all noise models within the 1\textsigma~uncertainty ranges.

The MeerKAT data set dominates the measurement of the orthometric Shapiro delay parameters. Fig.~\ref{shapiro_residuals} shows the timing residuals as a function of the orbital phase under the assumption of the \textbf{Cs} noise model. Despite some small gaps in orbital coverage, only the MeerKAT-derived ToAs have enough precision to detect the Shapiro delay with high significance, and therefore to constrain the component masses in this system. To corroborate this, we have attempted to fit the $h_3$ and $\varsigma$ parameters on each ToA dataset. The MeerKAT L-band ToAs reproduce the measurement from the global fit, while the S-band ToAs yield only a slight loss of precision owing to the sparser orbital coverage. On the other hand, the Parkes multibeam data set provides only very loose constraints as discussed in Section~\ref{discrepancies} and the Parkes UWL dataset is unable to converge on significant values.

The two remaining PK parameters, $\dot P_\textrm{b}$ and $\dot x$, do not yield a significant detection with the current timing baseline but still offer useful constraints. Table~\ref{measurements_orbital} presents the values for all the timing parameters of the global fit using the \textbf{Cs} noise model, as well as thee expected contributions to $\dot P_\textrm{b}$ and $\dot x$. The DM-derived distance from the NE2001 \citep{cordes2002ne2001} and YMW16 \citep{yao2017new} electron density models are $d=8.3$~kpc and $d=8.5$~kpc, respectively. With the detected proper motion vector of $\vec{\mu}=(-6.1\pm0.3,0.41\pm0.36)$~mas\,yr$^{-1}$ and the Galactic gravitational potential from \cite{mcmillan2017distribution} in eq.~(\ref{shk_acc}), we predict an order of magnitude of $\dot P_\textrm{b}\sim10^{-13}$~s\,s$^{-1}$. Likewise, assuming an inclination angle of $i\approx79$~deg (Section~\ref{massesSection}) in equation~(\ref{xdotPM}), we predict a maximum value of $|\dot x_\textrm{max}|\approx4.2\times10^{-15}$~ls\,s$^{-1}$ in case of a favourable $\Omega_\textrm{a}$ value. These estimates are of the same order of magnitude as the constraints on $\dot P_\textrm{b}$ and $\dot x$ presented in Table~\ref{physics}, which also present a consistent sign across all noise models, suggesting that the constraints can provide some physical information. Taking this into account, in Section~\ref{geometrySection} we translate the constraint on $\dot x$ into constraints on the orbital geometry of the system, and in Section~\ref{spin&period} we discuss the possible implications of the $\dot P_\textrm{b}$ constraints for the measurement of the spin period derivative $\dot P_\textrm{s}$.

\subsection{Mass constraints}\label{massesSection}

\begin{figure*}[h!]
\centering
 \includegraphics[width=2\columnwidth]{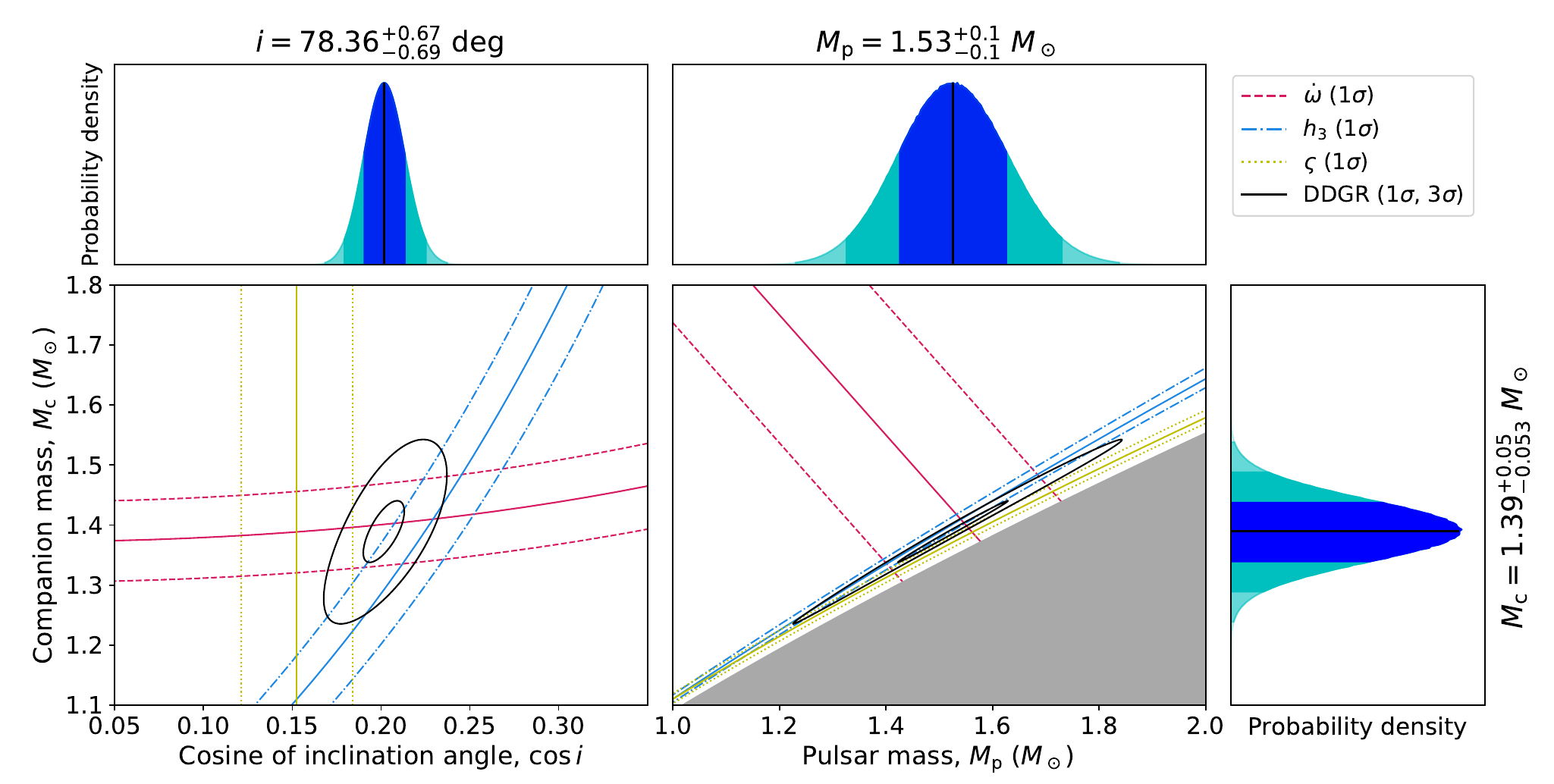}
 \caption{\textbf{Central plots}: mass and inclination angle constraints from the DDH PK measurements and the $\chi^2$ mapping with the DDGR model, both from the global fit and under the assumption of the \textbf{Cs} noise model (Table~\ref{fits}). The coloured solid lines represent the nominal values of the PK parameters, the coloured dashed lines their 1\textsigma~limits, and the solid black lines the 1\textsigma~and 3\textsigma~limits from DDGR. The shaded grey area in the right plot is the region excluded by the mass function ($i>90$~deg). \textbf{Corner plots}: marginalised one-dimensional probability densities for $M_\textrm{p}$, $M_\textrm{c}$ and $\cos{i}$ from DDGR $\chi^2$ mapping, showcasing the median value (black solid line) and the 31.4\%, 47.4\% and 49.9\% percentiles on both sides (shaded areas under the curve).}
 \label{mass_diagram}
\end{figure*}

In a common Bayesian approach \citep[e.g.][]{splaver2002masses}, we enforce the consistency of the PK parameters with GR and derive constraints on the pulsar mass $M_\textrm{p}$, the companion mass $M_\textrm{c}$ and the inclination angle $i$. We use the DDGR model, a modified version of the DD model that implements the PK effects directly from $M_\textrm{t}$ and $M_\textrm{c}$ as predicted by GR \citep{taylor1989further}. This allows us to explore a two-dimensional space of parameters to constrain the mass components and orbital inclination. For each noise model derived by \texttt{TEMPO2/TempoNEST}, we perform \texttt{TEMPO2} fits with the DDGR model in a uniform grid in the $M_\textrm{t}-\cos{i}$ space, where the uniform spacing on $\cos{i}$ is chosen to achieve a uniform sampling of the possible orbital geometries of the system. At each point in the grid, we register the resulting $\chi^2$ and transform it into a probability value, deriving a likelihood distribution. Subsequently, the distribution is integrated into marginal one-dimensional probability distributions for $M_\textrm{p}$, $M_\textrm{c}$ and $i$. Given that $\dot P_\textrm{b}$ and $\dot x$ are influenced by the proper motion of J1227$-$6208 and the Galactic acceleration field, we include them as independent excess parameters that do not need to be consistent with GR.

The values of $M_\textrm{p}$, $M_\textrm{c}$ and $i$, listed in Table~\ref{physics}, are consistent across all noise models, but there is some tension between the DDGR constraints and the $\varsigma$ parameter. Fig.~\ref{mass_diagram} depicts the constraints on the $M_\textrm{c}-\cos{i}$ and $M_\textrm{c}-M_\textrm{p}$ spaces as derived both with the DDGR $\chi^2$ mapping method and the constraints imposed by the independently measured $\dot\omega$, $h_3$ and $\varsigma$ parameters, in both cases under the assumption of the \textbf{Cs} noise model. The resulting DDGR constraints are 1\textsigma~consistent with the measurement of $\dot\omega$ and $h_3$, and 2\textsigma~consistent with $\varsigma$. This is also observed with the noise models \textbf{Ls}, \textbf{C}, \textbf{VC} and \textbf{VCs}. Only the assumption of the \textbf{L} noise model results in self-consistent constraints. The median values and the 68.2\% percentiles of the marginal one-dimensional distributions for $M_\textrm{p}$, $M_\textrm{c}$ and $i$ are quoted as the measurements and their 1\textsigma~uncertainties in Table~\ref{physics}, where it is seen that all noise models present 1\textsigma~consistent constraints. We note that models with high-frequency cut-offs result in slightly lower mass ranges, going from $M_\textrm{p}=1.59(10)~M_\odot$ and $M_\textrm{c}=1.42(5)~M_\odot$ in \textbf{Ls} (Fig.~\ref{DM100_R500_global}) to $M_\textrm{p}=1.51(11)~M_\odot$ and $M_\textrm{c}=1.38(5)~M_\odot$ in \textbf{VCs} (Fig.~\ref{DM30_R30_global}). That is consistent with the variation of the $h_3$ values across noise models as exposed in Section~\ref{PKparameters_section}, and shows how the DM noise is the main limiting factor in the precision of our mass measurements. On the other hand, the constraints on $i$ are more sensitive to the inclusion or exclusion of red spin noise, with $i=79.2(7)$~deg for the \textbf{L}, \textbf{C} and \textbf{VC} models and ranging from $i=78.6(7)$~deg to $i=78.2(7)$~deg from \textbf{Ls} to \textbf{VCs} models. Using extended uncertainty ranges from all the values listed in Table~\ref{physics}, we quote $M_\textrm{t}=2.9\pm0.3$~$M_\odot$, $M_\textrm{p}=1.54(15)~M_\odot$, $M_\textrm{c}=1.40(7)~M_\odot$ and $i=78.7\pm1.2$~deg, which constitute 10\textsigma, 10\textsigma, 20\textsigma~and 65\textsigma~constraints.

\subsection{Discrepancies between PK parameters and GR}\label{discrepancies}

\begin{table*}
\caption[]{\label{measurements_spinometric} Data reduction, model fit, spin, DM and astrometric parameters from the DDH \texttt{TEMPO2/TempoNEST} fits.}
\centering
\begin{tabular}{lccc}
\hline
\hline \\[-1.5ex]
Data set  &  Global & multibeam & MeerKAT+UWL \\
\hline \\[-1.5ex]
Data reduction parameters & & & \\
\hline \\[-1.5ex]
Number of ToAs & 1545 & 365 & 1180 \\
First ToA (MJD) & 55987.54 & 55987.54 & 58526.27  \\
Last ToA (MJD) & 60101.30 & 58581.61 & 60101.30 \\
Solar System ephemeris & DE430 & DE430 & DE430 \\
Timescale & TCB & TCB & TCB \\
Shortest correlated spin noise timescale, $\nu_\textrm{spin}^{-1}$ (days) & 100 & 100 & 100 \\
Shortest correlated DM noise timescale, $\nu_\textrm{DM}^{-1}$ (days) & 50 & 50 & 50 \\[0.5ex]
\hline \\[-1.5ex]
Model fit paramaters & & & \\
\hline \\[-1.5ex]
Log correlated spin noise amplitude, $\log{A_\textrm{spin}}$ (\textmu s) & -11.7524 & -12.0384 & -11.8007 \\
Log correlated DM noise amplitude, $\log{A_\textrm{DM}}$ (\textmu s) & -10.0075 & -11.7218 & -9.993 \\
Spin noise spectral index, $\gamma_\textrm{spin}$ (\textmu s) & 1.49911 & 0.281109 & 1.0046 \\
DM noise spectral index, $\gamma_\textrm{DM}$ (\textmu s) & 1.57333 & -1.80875 & 1.45055 \\
Weighted root mean square of the timing residuals (\textmu s) & 6.995 & 23.571 & 6.094 \\
Reduced $\chi^2$ & 0.9650 & 1.0298 & 0.9643 \\[0.5ex]
\hline \\[-1.5ex]
Spin and astrometric parameters &  &  \\
\hline \\[-1.5ex]
Reference epoch for spin, position and DM (MJD) & 55991.22 & 55991.22 & 58526 \\
Right ascension, RA (J2000, hh:mm:ss) & 12:27:00.4414(4) & 12:27:00.4412(4) & 12:27:00.441(2) \\
Declination, DEC (J2000, dd:mm:ss) & -62:08:43.791(3) & -62:08:43.788(2) & -62:08:43.80(1)\\
Proper motion in RA ($\mu_\textrm{RA}$, mas\,yr$^{-1}$) & $-6.1(3)$ & $-5.2(7)$ & $-5.9\pm1.1$ \\
Proper motion in DEC ($\mu_\textrm{DEC}$, mas\,yr$^{-1}$) & $0.41\pm0.36$ & $-0.4(7)$ & $1.4\pm1.2$ \\
Spin frequency, $F_0$ (Hz) & 28.962140551274(9) & 28.96214055129(2) & 28.96214051691(1) \\
Spin frequency derivative, $F_1$ ($10^{-16}$~Hz\,s$^{-1}$) & $-1.5677(4)$ & $-1.570(2)$  & $-1.5636(14)$ \\
Dispersion measure, DM$_0$ (pc~cm$^{-3}$) & 362.816(12) & 362.808(11) & 362.895(7) \\
First derivative of DM, DM$_1$ ($10^{-2}$~pc~cm$^{-3}$~yr$^{-1}$) & 1.72(5) & 2.8(9) & $-1.86(9)$ \\
Second derivative of DM, DM$_2$ ($10^{-3}$~pc~cm$^{-3}$~yr$^{-2}$) & -2.3(9) & $-3.1\pm2.1$ & $3.1\pm2.1$ \\
\hline \\[-1.5ex]
Derived parameters & & \\
\hline \\[-1.5ex]
Galactic longitude, $l$ (deg) & 300.082 & ... & ... \\
Galactic latitude, $b$ (deg) & 0.591 & ... & ... \\
Spin period, $P_\textrm{s}$ (ms) & 34.527834647774(10) & 34.52783464775(2) & 34.527834688742(19) \\
Spin period derivative, $\dot P_\textrm{s}$ ($10^{-19}$~s\,s$^{-1}$) & 1.8690(5) & 1.872(3) & 1.8640(17) \\
NE2001 DM-derived distance, $d$ (kpc) & 8.3 & ... & ... \\
YMW16 DM-derived distance, $d$ (kpc)  & 8.5 & ... & ... \\
Shklovskii contribution to $\dot P_\textrm{s}$ ($10^{-19}$~s\,s$^{-1}$) & $\sim0.26$ & ... & ... \\
Galactic potential contribution to $\dot P_\textrm{s}$ ($10^{-19}$~s\,s$^{-1}$) & $\sim(-0.26)$ & ... & ... \\
Characteristic age, $\tau_\textrm{c}$ (Gyr) & 2.9 & ... & ... \\
Spin-down luminosity, $|\dot E|$ (erg\,s$^{-1}$) & $6.3\times10^{32}$ & ... & ... \\
Surface magnetic field strength, $B_\textrm{surf}$ (G) & $2.6\times10^9$ & ... & ... \\

\hline
\hline
\end{tabular}

\tablefoot{
The derived parameters are derived only for the global fit, the most con straining solution, and left black for the multibeam and MeerKAT+UWL fit. The value of $\dot P_\textrm{s}$ has not been corrected for the possible Shklovskii and Galactic potential contributions.
}

\end{table*}

\begin{table*}
\caption[]{\label{measurements_orbital} Orbital Keplerian and PK parameters from the DDH \texttt{TEMPO2/TempoNEST} fits, and mass constraints from the DDGR $\chi^2$ mapping fits.}
\centering
\begin{tabular}{lccc}
\hline
\hline \\[-1.5ex]
Data set & Global & multibeam & MeerKAT+UWL \\
\hline \\[-1.5ex]
Keplerian orbital parameters &  &  \\
\hline \\[-1.5ex]
Orbital period, $P_\textrm{b}$ (days) & 6.7210192(3) & 6.72102(2) & 6.7210189(4) \\
Orbital eccentricity, $e$ & 0.00114918(4) & 0.0011492(4) & 0.00114918(5) \\
Longitude of periastron, $\omega$ (deg) & 27.090(9) & 27.07(3) & 27.211(12) \\
Projected semi-major axis of the pulsar orbit, $x$ (ls) & 23.200666(3) & 23.20066(3) & 23.200665(3) \\
Epoch of periastron, $T_0$ (MJD) & 55991.6995(2) & 55991.6991(6) & 58525.52384(9) \\
\hline \\[-1.5ex]
Post-Keplerian (PK) orbital parameters &  &  \\
\hline \\[-1.5ex]
Rate of periastron advance $\dot\omega$ (deg\,yr$^{-1}$) & 0.01710(83) & 0.0252(48) & 0.0163(11) \\
Orthometric amplitude of Shapiro delay, $h_3$ (\textmu s) & 3.45(23) & $3.5\pm1.8$ & 3.44(23) \\[.3ex]
Orthometric ratio of Shapiro delay, $\varsigma$ & 0.858(27) & 0.76(27) & 0.853(30) \\[.3ex]
Derivative of orbital period, $\dot P_\textrm{b}$ ($10^{-13}$~s\,s$^{-1}$) & $3.7\pm2.5$ & $-10\pm30$ & $8\pm40$ \\[.3ex]
Estimated Shklovskii contribution to $\dot P_\textrm{b}$ ($10^{-13}$s\,s$^{-1}$) & $\sim4.5$ & ... & ... \\
Estimated Galactic potential contribution to $\dot P_\textrm{b}$ ($10^{-13}$s\,s$^{-1}$) & $\sim(-4.5)$ & ... & ... \\
Derivative of projected semi-major axis, $\dot x$ ($10^{-15}$~ls\,s$^{-1}$) & $-9.1\pm7.2$ & $-88\pm42$ & $-17\pm11$ \\
Estimated maximum proper motion contribution to $\dot x$, $|\dot x_\textrm{max}|$ ($10^{-15}$~ls\,s$^{-1}$) & $\sim4.2$ & ... & ... \\
\hline \\[-1.5ex]
Mass and orbital geometry constraints &  &  \\
\hline \\[-1.5ex]
Mass function, $f_M$ ($M_\odot$) & 0.297 & 0.297 & 0.297 \\
Total mass, $M_\textrm{t}$ ($M_\odot$) & $3.1^{+0.6}_{-0.8}$& $5.5\pm1.7$  & $3.0^{+0.7}_{-1.0}$ \\[0.4ex]
Companion mass, $M_\textrm{c}$ ($M_\odot$) & 1.39(5) & ... & 1.31(7) \\[0.4ex]
Pulsar mass, $M_\textrm{p}$ ($M_\odot$) & 1.53(10) & ... & 1.37(14) \\[0.4ex]
Inclination angle, $i$ (deg) & 78.4(7) & ... & 78.9(8) \\[0.4ex]
Longitude of ascending note, $\Omega_\textrm{a}$ (deg) & 86/266$^{+76}_{-78}$ & ... & ... \\ [0.5ex]
\hline
\hline
\end{tabular}

\tablefoot{
The non-GR contributions to $\dot P_\textrm{b}$ and $\dot x$ are derived only from the global fit constraints, where the proper motion measurement is the most significant (Table~\ref{measurements_spinometric}). Mass and orbital geometry constraints are derived from the $\chi^2$ mapping of DDGR solutions, as exposed in Sections~\ref{massesSection}, \ref{discrepancies} and \ref{geometrySection}.
}

\end{table*}

\begin{table*}

\caption[]{\label{reduced_physics} Constraints on the PK parameters and the component masses from the MeerKAT+UWL fits.}
\centering
    \begin{tabular}{lcccccccccccccccc}
    \hline
    \hline \\[-1.5ex]
    Model & $\dot\omega$ & $h_3$ & $\varsigma$ & $M_\textrm{p}$ & $M_\textrm{c}$ & i \\[0.3ex]
     & (deg\,yr$^{-1}$) & (\textmu s) &  & ($M_\odot$) & ($M_\odot$) & (deg) \\
    \hline \\[-1.5ex]
    \textbf{L}    & 0.0163(12) & 3.9(3) & 0.82(4) & 1.42(14) & 1.33(7) & 79.6(8) \\[0.5ex]
    \textbf{Ls}   & 0.0166(11) & 3.7(3) & 0.84(3) & 1.43(14) & 1.34(7) & 79.2(8) \\[0.5ex]
    \textbf{C}    & 0.0163(12) & 3.7(3) & 0.84(4) & 1.39(15) & 1.32(7) & 79.4(8) \\[0.5ex]
    \textbf{Cs}   & 0.0164(11) & 3.4(3) & 0.86(3) & 1.37(14) & 1.31(7) & 78.9(8) \\[0.5ex]
    \textbf{VC}   & 0.0162(12) & 3.7(3) & 0.84(4) & 1.38(15) & 1.32(8) & 79.4(9) \\[0.5ex]
    \textbf{VCs} & 0.0160(11) & 3.4(2) & 0.85(3) & 1.31(13) & 1.28(7) & 78.9(8) \\[0.5ex]
    \textbf{extended} & 0.0163(14) & 3.7(5) & 0.84(6) & 1.36(20) & 1.31(10) & 79.2$\pm$1.1 \\[0.5ex]
    \hline
    \hline
    \end{tabular}
\tablefoot{
The PK parameters were measured in the MeerKAT+UWL fit with \texttt{TEMPO2} under the assumption of the noise models as derived by \texttt{TEMPO2/TempoNEST} (Table~\ref{fits}). The component masses were derived from the $\chi^2$ mapping of DDGR solutions (Section~\ref{discrepancies}). The \textbf{extended} row shows comprehensive uncertainties accounting for all measurements across the different noise models.
}
\end{table*}

\begin{figure*}
\centering
 \includegraphics[width=2\columnwidth]{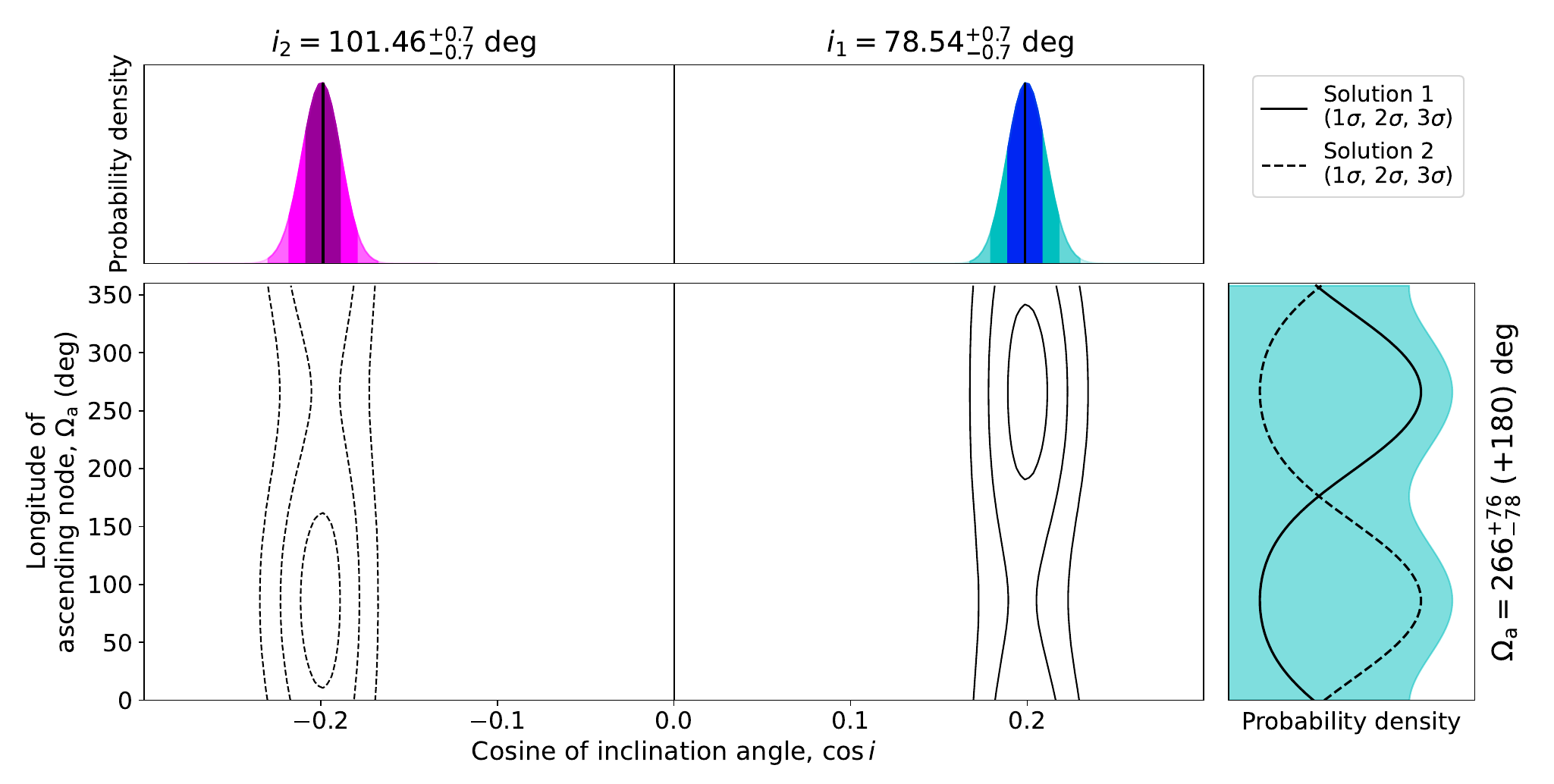}
 \caption{\textbf{Central plots}: constraints on the $\Omega_\textrm{a}-\cos{i}$ space derived from the $\chi^2$ mapping of DDGR solutions enforcing consistency with eq.~(\ref{xdotPM}) in the global fit and assuming the \textbf{Cs} noise model (trable~\ref{fits}). The solid contours represent the explored space ($\cos{i}>0$), while the dashed contours are derived from symmetries in eq.~(\ref{xdotPM}). \textbf{Corner plots}: marginalised one-dimensional probability densities for $\cos{i}$ and $\Omega_\textrm{a}$, showcasing the median value (black solid line) and the 31.4\%, 47.4\% and 49.9\% percentiles on both sides (shaded areas under the curve).}
 \label{geometry_diagram}
\end{figure*}

As shown in Fig.~\ref{mass_diagram}, $\varsigma$ is only 2\textsigma~consistent with the DDGR constraints, and the $h_3$ and $\dot\omega$ parameters. To explore the cause of this slight discrepancy, we analyse the individual contributions to the PK parameters from segregated data sets. Keeping the noise models derived by \texttt{TEMPO2/TempoNEST}, we create two new isolated ToA time series: one with the ToAs derived from Parkes multibeam observations (multibeam), and another one with the ToAs derived from MeerKAT and the Parkes UWL observations (MeerKAT+UWL). These two sets differ significantly both in timing baseline and in ToA quality, with the Parkes multibeam-derived ToAs providing the longest baseline, but with the MeerKAT+UWL ToAs being much more precise and providing observing frequency information (Figs.~\ref{freq_intensity_plots}~and~\ref{shapiro_residuals}). Subsequently, we repeat the \texttt{TEMPO2} fits on each side. 

The Shapiro delay measurement is not significantly affected by the split. In the multibeam fits, the orthometric Shapiro delay parameters are measured with larger uncertainties across all noise models ($h_3=4.1-4.7\pm2.1$~\textmu s, $\varsigma=0.69-0.76\pm0.29$), but they are still consistent with the measurements from the global fit. On the other hand, the MeerKAT+UWL $h_3$ and $\varsigma$ values listed in Table~\ref{reduced_physics} are almost identical to the ones from the global fit quoted in Table~\ref{physics}. That is consistent with the Shapiro delay being primarily constrained by the MeerKAT ToAs owing to their higher precision (see Section~\ref{PKparameters_section}).

The tension between the PK parameters is reduced in the MeerKAT+UWL fits. In the multibeam fits, the periastron advance $\dot \omega$ increases to $\dot \omega=0.026(5)$~deg\,yr$^{-1}$, which is more than 1\textsigma~away from the global fit. Not only is this an unrealistic value, as it implies total system mass of $M_\textrm{t}=5.5\pm1.7~M_\odot$, but it also suggests that the multibeam data set is somehow biased towards a larger $\dot\omega$ value. However, the MeerKAT+UWL fits give values much more consistent with the global fits, with $\dot\omega=0.0163(14)$~deg\,yr$^{-1}$ from the extended uncertainties. The median value of $\dot\omega$ is slightly reduced within the 1\textsigma~uncertainties consistently across all noise models (Table~\ref{reduced_physics}) with only a small increase in uncertainty compared to the global fit ($\dot \omega=0.0171(11)$~deg\,yr$^{-1}$, Table~\ref{physics}). With this subtle change, $\dot \omega$, $h_3$ and $\varsigma$ become 1\textsigma~consistent with each other across all noise models.

In an attempt to understand the origin of the apparent biases introduced by the the multibeam data set, we have performed two extra \texttt{TEMPO2/TempoNEST} runs with $\nu_\textrm{spin}^{-1}=100$~days and $\nu_\textrm{DM}^{-1}=50$~days on the multibeam and MeerKAT+UWL data sets, and compared the resulting noise and timing parameters with those of the \textbf{Cs} model derived in the global fit (Table~\ref{fits}). Table~\ref{measurements_spinometric} and \ref{measurements_orbital} show the global fit, the multibeam fit and the MeerKAT+UWL fit results side to side. In Table~\ref{measurements_spinometric}, it becomes readily evident that \texttt{TEMPO2/TempoNEST} converges on similar timing noise models for the global and MeerKAT+UWL data sets. Table~\ref{measurements_orbital} also shows that the MeerKAT+UWL PK parameter measurements are also virtually identical to those derived with the assumption of the \textbf{Cs} model, presented in Table~\ref{reduced_physics}. However, the noise model diverges in the isolated multibeam data set and it is unable to converge on realistic values, showing that the Parkes multibeam data does not have enough timing precision and bandwidth to constrain the spin and DM timing noise. This could explain why the multibeam gives a discrepant $\dot\omega$ measurement, and why it could be biasing the global fit measurements as well.

In light of this, we repeat the $\chi^2$ mapping of DDGR solutions with the MeerKAT+UWL data set only and with the assumption of the noise models derived in the global \texttt{TEMPO2/TempoNEST} fits (Table~\ref{fits}), deriving the $M_\textrm{p}$, $M_\textrm{c}$ and $i$ constraints quoted in Table~\ref{reduced_physics}. Similar to the global fits, $\dot\omega$ and $h_3$ dominate the constraints, but this time, the DDGR limits are 1\textsigma~consistent with all of the PK parameters. Like in the global fits, the more constraining noise models result in lower mass ranges, ranging from $M_\textrm{p}=1.31(31)~M_\odot$ and $M_\textrm{c}=1.28(7)~M_\odot$ with the \textbf{VCs} noise model (Fig.~\ref{DM30_R30_mkt+uwl}) to $M_\textrm{p}=1.43(14)~M_\odot$ and $M_\textrm{c}=1.34(7)~M_\odot$ with the \textbf{Ls} noise model (Fig.~\ref{DM100_R500_mkt+uwl}). For reference, Fig.~\ref{DM50_R100_mkt+uwl} in the Appendix~(\ref{appendix}) shows the counterpart for Fig.~\ref{mass_diagram}, with the assumption of the \textbf{Cs} noise model on the MeerKAT+UWL data set. Nonetheless, the uncertainty ranges are still 1\textsigma~consistent with the global fits presented in Table~\ref{physics}. Accounting for extended uncertainties from all measurements in Table~\ref{reduced_physics}, we quote $M_\textrm{t}=2.7\pm0.4$~$M_\odot$, $M_\textrm{p}=1.36(21)~M_\odot$, $M_\textrm{c}=1.30(9)~M_\odot$ and $i=79.3\pm1.1$~deg from the MeerKAT+UWL global fits, which are 7\textsigma, 6\textsigma, 14\textsigma~and 77\textsigma~measurements and are in 1\textsigma~consistency with the ones quoted in Section~\ref{massesSection}.

Finally, we attest that $\dot P_\textrm{b}$ and $\dot x$ can only be constrained in the global fit. Both the multibeam and the MeerKAT+UWL fits are unable to constrain to $\dot P_\textrm{b}$ and $\dot x$. As shown in Table~\ref{measurements_orbital}, the uncertainties on both parameters increase by an order of magnitude. Therefore, it is evident that only the accumulated baseline of the global fit can constrain them.

\subsection{Orbital geometry constraints}\label{geometrySection}

In Section~\ref{PKparameters_section} it was noted that the uncertainty in the constraints of $\dot x$ across have a similar size as the $|\dot x_\textrm{max}|\approx4.2\times10^{-15}$ expected from the proper motion, as quoted in Table~\ref{measurements_orbital}. Therefore, we investigate how the DDGR solutions are affected upon enforcement of consistency between $\dot x$, $\vec{\mu}$ and the orbital geometry of the system. Assuming the noise models derived with \texttt{TEMPO2/TempoNEST}, we implement the same likelihood approach from Section~\ref{massesSection} with two major modifications. First, the mapping is now done in a uniform three-dimensional grid on the $M_\textrm{t}-\cos{i}-\Omega_\textrm{a}$ space. Second, equation~(\ref{xdotPM}) is implemented by forcing the excess $\dot x$ value to be consistent with the $i$, $\Omega_\textrm{a}$ and $\vec{\mu}=(\mu_\textrm{RA},\mu_\textrm{DEC})$ values. To compensate for the extra dimensionality, the grid resolution is reduced to avoid a manifold increase of the computational running time of this experiment, with only 64 points along the $M_\textrm{t}$ axis across 10\textsigma, 60 points along the $\cos{i}$ across 12\textsigma~and 180 points across the $0\leq\Omega_\textrm{a}<360$ range. In addition, equation~(\ref{xdotPM}) is symmetric with respect to the transformation $i\rightarrow180\textrm{~deg}-i$ and $\Omega_\textrm{a}\rightarrow\Omega_\textrm{a}+180$~deg. Therefore, the $\cos{i}<0$ side of the explored space is derived by implementing this transformation. This last point is true only because the distance to J1227$-$6208 is large enough so that the orbital motion of Earth does not introduce year-long periodic contributions to $\dot x$. Finally, we integrate the likelihood distribution along the $M_\textrm{t}$ axis to derive two-dimensional likelihood distribution on the $\Omega_\textrm{a}-\cos{i}$ space, from which the marginal probability distributions for $\Omega_\textrm{a}$ and $\cos{i}$ are derived.

The constraints on $\cos{i}$ and $\Omega_\textrm{a}$ are drawn on Fig.~\ref{geometry_diagram}. As expected, the constraints on $i$ are consistent with those derived in Section~\ref{massesSection}. The value of $\Omega_\textrm{a}$ is not constrained owing to the large overlap of the $\cos{i}>0$ and $\cos{i}<0$ probability distributions, but the probability density contours show that two regions in the $\cos{i}-\Omega_\textrm{a}$ are preferred. For $\cos{i}>0$, the longitude of ascending node is constrained at $\Omega_\textrm{a}=266\pm78$~deg, while for $\cos{i}<0$ it is constrained at $\Omega_\textrm{a}=86\pm78$~deg. From equation~\ref{xdotPM}, using the values of $x$, $i$, $\mu_\textrm{RA}$ and $\mu_\textrm{DEC}$ presented in Table~\ref{measurements_spinometric} and $\Omega_\textrm{a}=266$~deg, we predict a value of $\dot x\approx-4.2\times10^{-15}$~ls\,s$^{-1}$, in full consistency with the values presented in Table~\ref{physics}. Therefore, it is clear that the measured excess $\dot x$ originates from the proper motion of the system in the sky.

\subsection{Spin and orbital period evolution constraints}\label{spin&period}

As shown in Table~\ref{measurements_spinometric}, in the global fit we measure the spin evolution parameters $P_\textrm{s}$ and $\dot P_\textrm{s}$ with high significance, showing that J1227$-$6208 is indeed a mildly recycled pulsar. However, the Shklovskii effect and Galactic potential field affect $\dot P_\textrm{s}$ in the same way as they affect $\dot P_\textrm{b}$, following equation~(\ref{shk_acc}). The estimated values of these contributions are listed at the end of Table~\ref{measurements_spinometric}, making it evident that albeit they are smaller than the measurement ($\sim$$10^{-20}$s\,s$^{-1}$ against $\dot P_\textrm{s}=1.8690(5)\times10^{-19}$~s\,s$^{-1}$), they can nonetheless be significant. Thus, we should cautious of pulsar parameters derived from $P_\textrm{s}$ and $\dot P_\textrm{s}$. Since a parallax distance measurement is out of the question, the best prospect for constraining the true combined contribution of the Shklovskii effect and Galactic potential field is to improve the $\dot P_\textrm{b}$ measurement, on which these effects are orders of magnitude larger than the GR prediction. Therefore, a measurement of $\dot P_\textrm{b}$ will be direct measurement of the Doppler factor derivative. Nonetheless, with the current observations, the consistently positive sign of $\dot P_\textrm{b}$ across all noise models (Table~\ref{physics}) already suggests a dominance of the Shklovskii effect over the Galactic potential contribution.

\section{Astrophysical Implications}\label{astrophysics}

\begin{figure*}
\centering
 \includegraphics[width=\columnwidth]{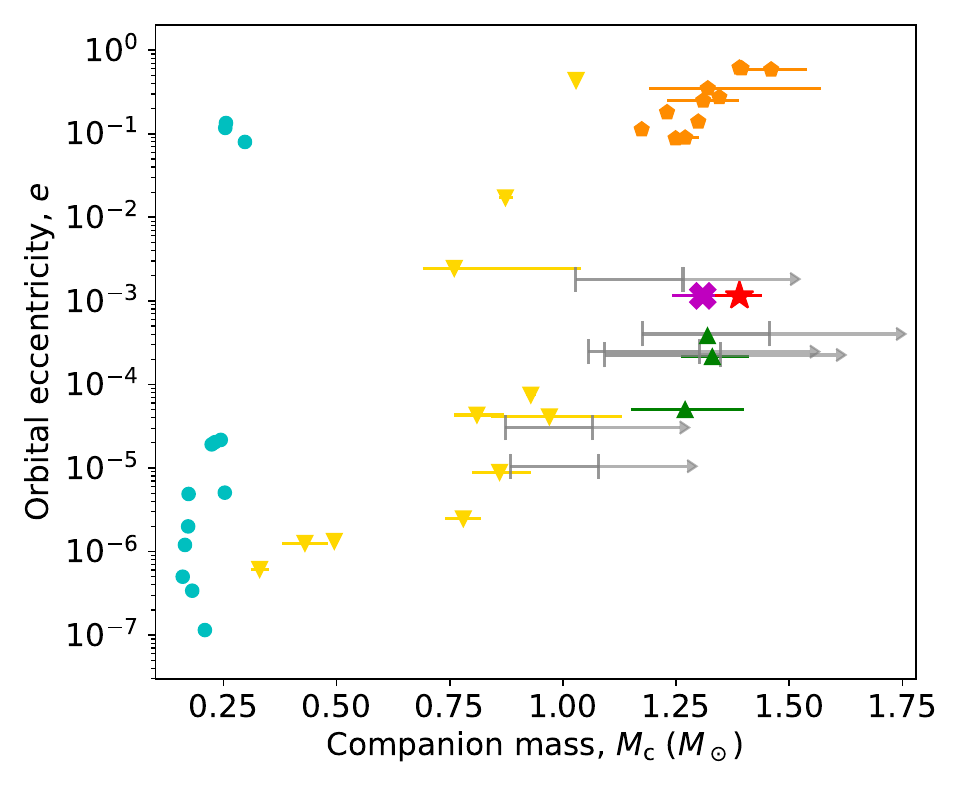}
 \includegraphics[width=\columnwidth]{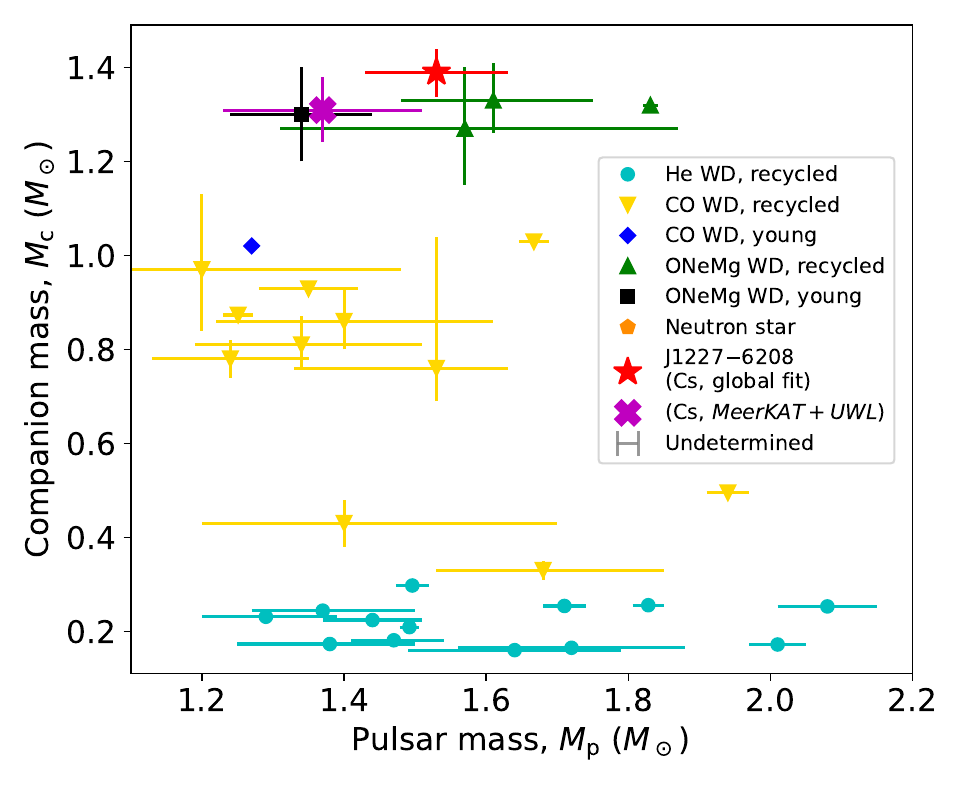}
 \caption{\textbf{Left:} orbital eccentricity against companion mass from recycled binary pulsars in the Galactic field with companion mass constraints, distinguished by the nature of their companions. For systems without timing mass measurements ("Undetermined" in the legend), the lower constraints on the minimum and median $M_\textrm{c}$ values derived from the mass function ($i=90$, 60~deg) are plotted. \textbf{Right:} companion mass against pulsar mass for all PSR$-$WD systems with mass measurements from timing in the Galactic field. The constraints on J1227$-$6208 constraints plotted twice, from the global fit (red star) and the MeerKAT+UWL fit (magenta star). In both cases, we use the \textbf{Cs} noise model results. Constraints on systems besides J1227$-$6208 are taken from \url{https://www3.mpifr-bonn.mpg.de/staff/pfreire/NS_masses.html} and references therein, and from \cite{camilo2001discovery} (PSR J1435$-$6100), \cite{cruces2021fast} (PSR J2338+4818), \cite{edwards2001recycled} (PSR J1157$-$5112), \cite{gautam2022relativistic} (PSR J1952+2630), Jang et al., in prep. (PSR J1439$-$5501), \cite{martinez2019discovery} (PSR J0709+0458), \cite{parent2019palfa} (PSR J1932+1756) and \cite{tan2020timing} (PSR J1658+3630).}
 
 \label{WDvsNS}
\end{figure*}

\subsection{J1227$-$6208 as a DNS system}

The mass ranges derived in Sections~\ref{massesSection} and \ref{discrepancies} and presented in Tables~\ref{physics} and \ref{reduced_physics} are consistent with the mass distribution of the double-NS (DNS) population, with $M_\textrm{c}$ being significantly above the current lower mass limit of a NS \citep[1.17~$M_\odot$,][]{martinez2015dns}, making a NS nature a hypothesis in need of consideration.

Recycled pulsars with massive WD and NS companions follow a similar evolutionary path. With original companions more massive than 5~$M_\odot$ \citep{lazarus2014massive}, these systems go through a intermediate-mass or high-mass X-ray binary (IMXB or HMXB) stage, resulting in a dynamically unstable mass transfer and the formation of a common envelope \citep[CE, e.g.][]{van2018high}. After the hydrogen-rich CE is expelled, the system becomes a circular PSR$-$He naked star. Stable mass accretion on to the NS occurs when the companion leaves the He main sequence and the system enters a short-lasted (less than 100~kyr) Case BB RLO stage, leading to the partial recycling of the pulsar \citep{lazarus2014massive}. Afterwards, and if the stripped He star has retained enough mass ($\gtrsim$1.45~$M_\odot$), the companion will undergo an electron-capture supernova, forming a second NS \citep{tauris2015sn,tauris2017dns}. This process entails the loss $0.2-0.4$~$M_\odot$ and a supernova kick of $<50$~km\,s$^{-1}$ that either disrupts the system or introduces a significant orbital eccentricity \citep[$e\gtrsim0.1$,][]{tauris2017dns}. On the other hand, if the stripped companion is unable to trigger a supernova, it becomes a massive WD instead \citep{lazarus2014massive,tauris2012coII}.

We argue that J1227$-$6208 is very unlikely to be a DNS system based on its orbital eccentricity. The left plot of Fig.~\ref{WDvsNS} depicts the orbital eccentricity $e$ of recycled pulsars against $M_ \textrm{c}$ and it shows that, while the $M_\textrm{p}$ in J1227$-$6208 is consistent with the DNS population, its $e$ value is two orders of magnitude below those of known DNS systems. Instead, it lays among other massive recycled PSR$-$CO/ONeMg WD systems. While \cite{tauris2017dns} show that very low eccentricities ($e<0.01$) in DNS are technically possible, they also argue that this scenario is extremely unlikely, requiring an extraordinary fine-tuning of the supernova kick magnitude and direction. That is strong evidence against the companion of J1227$-$5936 having undergone a supernova, making it a massive WD instead.

\subsection{J1227$-$6208 as a massive PSR$-$WD system}

J1227$-$6208 thus belongs to an emerging class of massive recycled PSR$-$ONeMg WD systems. The right side of Fig.~\ref{WDvsNS} maps measured PSR$-$WD component masses in a $M_\textrm{c}-M_\textrm{p}$ diagram, showing that J1227$-$6208 lies among the well-studied massive systems PSR J2222$-$0137, PSR J1528$-$3146 and J1439$-$5501, represented by the green triangles. J1227$-$6208 also shares with these three systems similar spin properties ($P_\textrm{0}=34.52$~ms against 32.82, 60.82 and 28.64~ms), binary period ($P_\textrm{b}=6.72$~days against 2.45, 3.18, and 2.12 days) and orbital eccentricity ($e=1.15\times10^{-3}$ against 4.65$\times10^{-4}$, 2.13$\times10^{-4}$ and 4.99$\times10^{-5}$), suggesting a similar nature. Therefore, it is very plausible that J1227$-$6208 has followed the evolutionary path of recycled PSR$-$ONeMg WD systems ($M_\textrm{c}\gtrsim1.1~M_\odot$). It should be noted that the young system PSR B2303+46, shown as a black square in Figs.~\ref{WDvsNS}, is the result of an exotic evolution where the WD formed before the pulsar and thus no recycling has occurred in it \citep{vanKerkwijk1999massive,tauris2000young}.

As we argue in favour of a WD nature for the companion, the possibility of $M_\textrm{c}$ laying beyond the Chandrasekhar limit (Section~\ref{massesSection}) is brought into question. Theoretical models predict that fast-rotating WDs may exist in the $1.38<M_\textrm{WD}/M_\odot<1.48$ mass range without collapsing in a supernova \citep[e.g.][]{yoon2005rapidly}. However, these conditions can only be the result of accretion and the spin-up in the WD, or be the product of a WD merger. Both of these scenarios are difficult to reconcile with the evolutionary model proposed above and the low orbital eccentricity, and therefore the mass of the companion of J1227$-$6208 is likely lower than 1.38~$M_\odot$. Continued timing in the future will be essential to increase the precision of the mass constraints so that a more definitive statement can be made on this aspect.

J1227$-$6208 and its companion are the fourth partially recycled pulsar with a $M_\textrm{c}>1.1$~$M_\odot$ ONeMg WD companion with mass measurements from timing. As seen on the right plot of Fig.~\ref{WDvsNS}, it is tempting to see this population as an emerging class distinct from partially recycled pulsars with the massive CO WD companions, in addition to the already-known separation between massive CO WD systems ($M_\textrm{c}\gtrsim0.7$~$M_\odot$) and the fully recycled millisecond pulsars with light He and CO WD companions ($M_\textrm{c}\lesssim0.5$~$M_\odot$) \cite{mckee2020precise,shamohammadi2023manyS}. However, a gap between CO WD and ONeMg WD masses is not predicted by theory. Furthermore, the left plot of Fig.~\ref{WDvsNS} shows that a handful of known massive systems without mass measurements can potentially fill this gap: PSR J1435$-$6100 \citep{camilo2001discovery}, PSR J2338+4818 \citep{cruces2021fast}, PSR J1157$-$5112 \citep{edwards2001recycled}, PSR J1952+2630 (Jang et al., in prep.), PSR J0709+0458 \citep{martinez2019discovery}, PSR J1932+1756 \citep{parent2019palfa} and PSR J1658+3630 \citep{tan2020timing}. Thus, this gap could be spurious, arising from the reduced size of the sample of mass measurements in massive systems. New mass measurements and the discovery and follow-up of further massive partially recycled PSR$-$ONeMg WD systems will be key in confirming whether the white dwarf mass distribution is trimodal or if this notion is just a statistical fluke.

\subsection{The pulsar mass of J1227$-$6208}

Theoretical models for recycled pulsars with the most massive WD companions suggest that these systems went through a short phase of accretion during post-CE Case BB RLO from their naked He-star companion, resulting in observed spin-up periods of $P_\textrm{s}\gtrsim20$~ms. According to \cite{tauris2012coII}, the amount of mass accreted by a recycled pulsar given a $P_\textrm{s}$ can be estimated with the following equation:
\begin{equation}\label{spin-up_equation}\Delta M_\textrm{p}\approx0.22~M_\odot\frac{\left(M_\textrm{p}/M_\odot\right)^{1/3}}{\left(P_\mathrm{s}/\mathrm{ms}\right)^{4/3}}\mathrm{.}\end{equation}
Assuming its current spin period of $P_\textrm{s}\approx34.5$~ms results in the accretion of 0.0022 $M_\odot$ from its companion during recycling. On the other hand, assuming a possible initial spin-up period of $P_\textrm{s}\approx20$~ms results in the accretion of at most 0.0045 $M_\odot$ from its companion. In both cases, the implication is that the current $M_\textrm{p}$ measurement is close its birth mass. Therefore a more precise measurement of $M_\textrm{p}$ will contribute to the statistics of NS birth masses, like in the similar systems PSR J2222$-$0137, PSR J1534$-$3146 and PSR J1439$-$5501. In fact, the measurement of $M_\textrm{p}=1.76(6)$~$M_\odot$ PSR J2222$-$0137 in \cite{cognard2017massive} ($M_\textrm{p}=1.831(10)$~$M_\odot$ in \citealt{guo2021improved}) has already expanded such distribution on its upper end.

If the amount of accreted matter estimated from theory is correct, the mass-transfer rate was likely Eddington or super-Eddington. For a NS of 1.5$~M_\odot$, the Eddington luminosity limit is $L\approx1.5\times10^{38}$~erg\,s$^{-1}$ and the accretion-to-luminosity efficiency is $L\approx0.1\times\dot Mc^2$ \citep{shakura1973blackholes,poutaten2007supercritically}, resulting in a maximum accretion rate of $\dot M\approx0.03~M_\odot$\,Myr$^{-1}$. Simulations find that the duration of the accretion stage during the Case BB RLO is $\lesssim$100,000~yr for massive PSR$-$WD systems \citep[e.g.]{lazarus2014massive,tauris2015sn}, and \cite{cognard2017massive} even found that the accretion time was $\lesssim$20,000~yr for the massive PSR$-$ONeMg WD system in PSR J2222$-$0137. The Eddington limit would allow for the accretion of at most 0.003~$M_\odot$ in the span of 100,000~yr, and $5\times10^{-4}$~$M_\odot$ in 20,000~yr, implying that if the recycling process occurred on a similar time span for J1227$-$6208, then it has accreted either close to the Eddington limit or significantly above it like in PSR J2222$-$0137. This is a plausible scenario, as super-Eddington accretion onto NSs has been directly observed in extra-galactic ultra-luminous X-ray sources such as NGC 5907 X-1 \citep{israel2017ulx}, NGC 300 ULX1 \citep{carpano2018ulx} and M82 X-2 \citep{bachetti2014ulx}. However, an alternative explanation is that the spin-up efficiency is larger than what is predicted by \cite{tauris2012coII}, requiring less mass accretion to reach a similar $P_\textrm{s}$.

\section{Future prospects}\label{prospects}

\subsection{Prospects for timing}

\begin{figure}
\centering
 \includegraphics[width=\columnwidth]{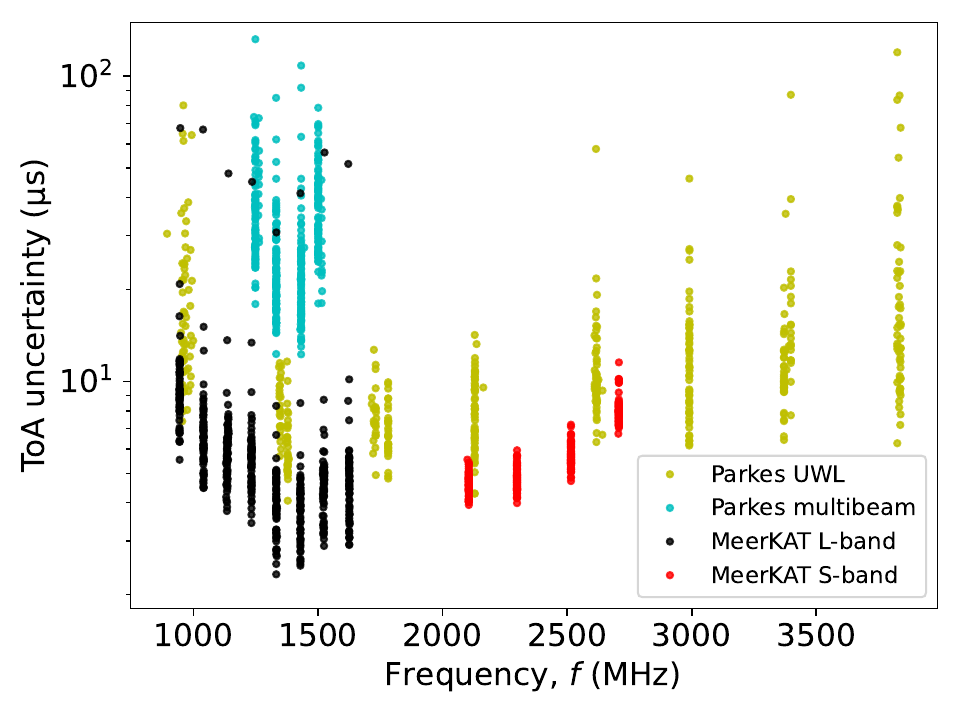}
 \caption{ToA uncertainty against observing frequency for each telescope receiver. The highest and lowest frequencies of the Parkes multibeam and MeerKAT S-band data sets are biased upwards because the signal drops off at the edge of the bands (see Fig.~\ref{freq_intensity_plots}).}
 \label{precision}
\end{figure}

Future timing observations of J1227$-$6208 should seek to strike a balance between its steep spectrum and the ISM-induced effects. Fig.~\ref{precision} portrays how ToA uncertainty evolves with $f$ across different receivers, and it shows that scattering affects the ToA quality negatively at $f<1,300$~MHz, leading to an increase of timing uncertainty. The best ToA quality is found at the $f=1.3 - 2.2$~GHz range, while at $f>2,200$~MHz the steep spectrum starts hampering the $S/N$, leading to another increase in timing uncertainty. Another factor to take into account is that, as reported in Section~\ref{noise_section}, red DM noise is very prominent, making it preferable to observe at high frequencies where its amplitude is lower. However, doing so comes at the price of achieving a sub-optimal $S/N$ and ToA quality due to the steep spectral index reported in Section~\ref{emission_analysis}. Therefore, MeerKAT observations with the S-band receivers at the S0 configuration (1750$-$2625~MHz) will offer the best compromise between a high ToA precision and a reduced ISM influence, with $A_{\textrm{DM,}1700}=0.68\times A_\textrm{DM}$ and $A_{\textrm{DM,}2600}=0.28\times A_\textrm{DM}$, thus enabling a precision increase on the measurement of PK parameters. In complement to this, UWL observations will be essential to track the DM evolution to properly model the still-present DN noise in the MeerKAT S-band ToAs.

The significance of the PK parameters will improve with an increasing baseline. The most significant improvement in the mass constraints will come from an improved measurement of $\dot\omega$. Assuming that the uncertainty on $\dot\omega$ scales with the timing baseline according to the power law \citep{damour1992tests}
\begin{equation}\Delta\dot\omega^\prime\approx\Delta\dot\omega\times\left(\frac{T}{4.31~\textrm{yr}}\right)^{-3/2}\textrm{,}\end{equation}
where 4.31~yr is the timing baseline of the MeerKAT+UWL data set, we expect $\Delta\dot\omega^\prime\approx\Delta\dot\omega/2$ with just two extra years of timing. This improvement will reduce the total mass measurement uncertainty by 65\% if we assume the current MeerKAT ToA uncertainty, aiding a more robust settlement or confirmation of any discrepancies between the PK parameters in the global fit.

Tighter constraints on $\dot P_\textrm{b}$ and $\dot x$ will also lead to an independent estimate of the distance of J1227$-$6208 and a better constraint on its orbital geometry. However, since these measurements are dependent on the global 11.26 year baseline, reducing uncertainty will require more time. Assuming once again a power law evolution of the uncertainties
$\dot P_\textrm{b}$ \citep{damour1992tests}
\begin{equation}\Delta\dot P_\textrm{b}^\prime\approx\Delta\dot P_\textrm{b}\times\left(\frac{T}{11.26~\textrm{yr}}\right)^{-5/2}\textrm{,}\end{equation}
then another decade of timing will be required to achieve a 3\textsigma~detection of $\dot P_\textrm{b}$. Constraints on the orbital geometry will also require an improvement of the measurement of the proper motion. Following the power law evolution of the uncertainty on $\dot x$ \citep{damour1992tests}
\begin{equation}\Delta\dot x^\prime\approx\Delta\dot x\times\left(\frac{T}{11.26~\textrm{yr}}\right)^{-3/2}\textrm{ ,}\end{equation}
two more decades of timing will be necessary to achieve a 3\textsigma~detection on $\dot x$. However, as seen in Section~\ref{geometrySection}, constraints on the $\Omega_\textrm{a}-\cos{i}$ can already be achieved even with a low-significance detection. Furthermore, the proper motion vector measurement will continue to improve alongside it, leading to more significant constraints in the $\Omega_\textrm{a}-\cos{i}$ space before that time. 

\subsection{Prospects for an optical detection}

There would be obvious benefits to having an optical detection of the companion of J1227$-$6208. Besides the confirmation of its WD nature, a measurement of its colour index and magnitude could be used to constrain its age with the help of WD evolutionary models. This would provide an independent measurement of the mass and age of the companion of J1227$-$6208 like other massive PSR$-$WD systems, such as PSR J1528$-$3146 \citep{jacoby2006detection} and PSR J1439$-$5501 \citep{pallanca2013optical}. That would allow, for example, for a true estimate the initial spin-up period.

As of now, we are unable to report an optical detection. \textit{Gaia} DR3 is sensitive to sources with magnitude $G\approx20.7$ \citep{gaia2016mission,gaia2023dr3}. Assuming the YMW16-derived DM distance of $d=8.5~kpc$, detection would require an absolute magnitude of at most $M_\textrm{G}\approx6$ to be detected, too bright for any WD. The closest reported \textit{Gaia} DR3 source\footnote{Designation: \textit{Gaia} DR3 6054663745667894016} is 0.29~arcsec away from the timing position, with a color index $\textrm{B}-\textrm{R}=1.59$ and magnitude $\textrm{G}=17.4$. Its parallax distance is $d=5.3\pm2.0$, giving it an absolute magnitude of $M_\textrm{G}=3-5$, and putting it either on the main sequence or the giant branch. Therefore, this source is most likely not associated with J1227$-$6208.

A potential detection will be possible only with the most powerful optical and infrared telescopes available. Assuming $d=8.5$~kp, we use the cooling tracks provided by \cite{bergeron2011cooling}\footnote{Tables available here: \url{https://www.astro.umontreal.ca/~bergeron/CoolingModels/}} for a $M_\textrm{WD}=1.3~M_\odot$ with a He atmosphere to predict visible and infrared magnitudes given an age. If the age of the system is consistent with the characteristic age of $\tau_\textrm{c}=2.93$~Gyr, then the companion will have a optical, red and infrared magnitudes of  $\textrm{V, R, I}>30$ and it will be undetectable. But if, like in PSR J1439$-$5501, the true age is significantly younger than $\tau_\textrm{c}$ for instance only 100~Myr old \citep{pallanca2013optical}, the magnitudes could reach $\textrm{V, R, I}<27$, making is potentially detectable in long exposure by sensitive optical and infrared telescopes such as the Very Large Telescope, the \textit{Hubble} Space Telescope or the \textit{James Webb} Space Telescope.

\section{Conclusions}\label{conclusion}

This study adds J1227$-$6208 to the rare class of confirmed mildly recycled pulsars with massive ONeMg WD companions ($M_\textrm{WD}>1.1~M_\odot$), along with PSR J2222$-$0137, PSR J1528$-$3146 and PSR J1439$-$5501. In addition, we have provided a detailed study of the pulsar's emission and the origin of the PK parameters measured in pulsar timing. The study includes 33 hours of Parkes multibeam data since 2012, 72 hours of Parkes ULW data since 2020, 25 hours of MeerKAT data since 2019 and 19 hours of MeerKAT S-band data from 2023, producing a timing baseline of more than 11 years.

Taking advantage of the large frequency coverage offered by these receivers to study the profile evolution of J1227$-$6208, we measure a steep flux density spectral index of $2.06<\alpha<2.35$ and a flat scattering index $3.33<\beta<3.62$, with a scattering timescale of $\tau_\textrm{s}\approx1.2$~ms and a mean flux density of $F_\textrm{m}=530-620$~\textmu Jy at the reference frequency of 1 GHz. Due to the combined effect of scattering and spectral properties, the largest peak flux density is observed at the frequency of $1.2-1.3$~GHz, with $F_\textrm{max}=13-14$~mJy. Around 15\% of the emission of J1227$-$6208 is linearly polarised, with a behaviour that does not resemble the rotating vector model, and a further $\sim$15\% of the emission is circularly polarized.

The timing of J1227$-$6208 suffers from red spin noise and, more prominently, red DM noise. We have implemented a Bayesian nested sampling algorithm to measure the parameters of the DDH timing model, including the spectral modelling of red DM, red spin noise and white noise, and the timing model parameters. We derived six different noise models: three with correlated spin noise and three without it, using different frequency cut-offs in all cases. We find that the red DM noise amplitude at 1400 MHz is almost two orders of magnitude larger than the red spin noise amplitude, and that models with both correlated spin and correlated DM noise with high-frequency cut-offs are favoured by the sampling evidence. Nonetheless, all of the resulting PK parameters are consistent across models within 1\textsigma~uncertainty. The Shapiro delay orthometric parameters, $h_3$ and $\varsigma$, and the periastron advance $\dot \omega$ are dominated by the MeerKAT-derived ToAs, while the constraints of the orbital period derivative $\dot P_\textrm{b}$ and the projected semi-major axis derivative $\dot x$ are possible only with the accumulated timing baseline.

We argue that the constraints on $\dot P_\textrm{b}$ and $\dot x$ are physical despite their low significance. $\dot P_\textrm{b}$ presents a consistent sign across all noise models and is about the same order of magnitude as the combined prediction of the Shklovskii effect given by the DM distance and the proper motion vector and the Galactic acceleration fields ($\dot P_\textrm{b}\sim10^{-13}$~s\,s$^{-1}$), therefore implying that a more significant detection will occur in the following years. This detection will also clarify the magnitude of the Shklovskii effect and the Galactic acceleration on the spin period derivative $\dot P_\textrm{s}$. The situation is similar for $\dot x$, as we predict $|\dot x_\textrm{max}|\approx4.2\times10^{-15}$~ls\,s$^{-1}$ for from the proper motion vector, $i=79$~deg and a favourable longitude of ascending node angle $\Omega_\textrm{a}$, which is of the same order of magnitude as the timing constraints. The mapping of solutions that implement consistency with the excess $\dot x$, the proper motion, $i$ and $\Omega_\textrm{a}$ indicate that the resulting $\chi^2$ is indeed sensitive to the location in the $\Omega_\textrm{a}-\cos{i}$ grid, with a preference for either $\Omega_\textrm{a}=266\pm78$~deg and $i=79$~deg, or $\Omega_\textrm{a}=86\pm78$~deg and $i=101$~deg. This parameter is also likely to get a more significant detection in the following years.

With the noise models from the DDH fits, we measure the pulsar mass $M_\textrm{p}$, companion mass $M_\textrm{c}$ and inclination angle $i$ by performing a mapping of solutions under the assumption of GR across all noise models, and attest that they are consistent with the PK parameters. In the global fit, the $h_\textrm{3}=3.6\pm0.5$~\textmu s and $\dot \omega=0.0171(11)$~deg\,yr$^{-1}$ measurements are 1\textsigma~consistent with the assumption of GR, while $\varsigma=0.85\pm0.05$ is 2\textsigma~consistent with GR, resulting in $M_\textrm{p}=1.54(15)~M_\odot$, $M_\textrm{c}=1.40(7)~M_\odot$ and $i=78.7\pm1.2$~deg in a conservative uncertainty range that includes all noise models. However, if we exclude the Parkes multibeam observations from the analysis, the median of the periastron advance is reduced to $\dot\omega=0.0163(14)$~deg\,yr$^{-1}$, resulting in 1\textsigma~consistency between the three PK parameters and GR, and the mass constraints of $M_\textrm{p}=1.36(20)~M_\odot$, $M_\textrm{c}=1.31(10)~M_\odot$ and $i=79.2\pm1.1$~deg. All of these measurements are within 1\textsigma~uncertainty overlap, and they result in the total ranges of possible values of $2.3<M_\textrm{t}/M_\odot<3.2$, $1.16<M_\textrm{p}/M_\odot<1.69$, $1.21<M_\textrm{c}/M_\odot<1.47$~$M_\odot$ and $75.5<i/\textrm{deg}<80.3$.
This wide range of possible values demonstrates that DM variability and its modelling as red DM noise is the main limiting factor of the precision of our timing analysis.

Despite the companion mass uncertainty range extending into the regime of known NS masses, the probability of it being a NS is extremely low due to the very small orbital eccentricity ($e=1.15\times10^{-3}$), which is much more consistent with the massive recycled PSR$-$CO/ONeMg WD population. The WD mass measured in this work is significant as it overlaps with the companion masses measured for three other similar systems. This measurement explores the potential multi-modality of the WD mass distribution noted in previous works, which suggests that the WD mass distribution is not continuous, but characterised by discrete clumps. Whether WDs with $M_\textrm{WD}\gtrsim1.1$~$M_\odot$ are segregated from their $0.7\lesssim M/{M_\odot}\lesssim1.0$ counterparts or not will be decided by further measurements of systems in these mass ranges. Regarding the pulsar mass, given the relatively large spin period of 34~ms, J1227$-$6208 has likely accreted no more than 0.0045~$M_\odot$ during the recycling process, that a precise measurement of $M_\textrm{p}$ will contribute to the NS birth amass distribution. However, if accretion occurred in a time period significantly shorter than 100,000 years like in the case of PSR J2222$-$0137 \citep{cognard2017massive}, this would imply either super-Eddington accretion or a larger spin-up efficiency than predicted in \cite{tauris2012coII}.

Future timing of this system will be key for constraining the PK parameters and the mass constraints even further. The most precise ToAs are produced in the $1.3-2.2$~GHz frequency range due to the balance between the steep spectral index and scattering. However, the presence of red DM noise requires that future observations are taken at higher frequencies, making the $1.7-2.6$~GHz MeerKAT S-band receiver on the S0 configuration the best observing band in the future. Optical detection of the WD companion may be possible in long exposures with the most powerful optical/infrared telescopes if the system is younger than a few hundred Myr, leading to an independent confirmation of the nature of the system, or otherwise to a lower limits of its true age.

\begin{acknowledgements}

The MeerKAT telescope is operated by the South African Radio Astronomy Observatory (SARAO), which is a facility of the National Research Foundation, an agency of the Department of Science and Innovation. SARAO acknowledges the ongoing advice and calibration of GPS systems by the National Metrology Institute of South Africa (NMISA) and the time space reference systems department of the Paris Observatory. The Parkes radio telescope is part of the Australia Telescope National Facility (ATNF), which is funded by the Australian Government for operation as a National Facility managed by the Commonwealth Scientific and Industrial Research Organisation (CSIRO). We acknowledge the Wiradjuri people as the Traditional Owners of the Observatory site. This work used the OzSTAR national facility at Swinburne University of Technology. OzSTAR is funded by Swinburne University of Technology and the National Collaborative Research Infrastructure Strategy (NCRIS). All authors affiliated with the Max Planck Institute for radioastronomy (MPIfR) acknowledge the continuing valuable support from the Max-Planck Society (MPG). Vivek Venkatraman Krishnan acknowledges financial support from the European Research Council (ERC) starting grant 'COMPACT' (grant agreement number: 101078094). Alessandro Ridolfi is supported by the Italian National Institute for Astrophysics (INAF) through an ``IAF - Astrophysics Fellowship in Italy'' fellowship (Codice Unico di Progetto: C59J21034720001; Project ``MINERS''). We also give our sincere thanks to Huanchen Hu for her comments on this article.

\end{acknowledgements}

\bibliography{J1227.bib}

\begin{thebibliography}{92}
\expandafter\ifx\csname natexlab\endcsname\relax\def\natexlab#1{#1}\fi

\bibitem[{{Bachetti} {et~al.}(2014){Bachetti}, {Harrison}, {Walton},
  {Grefenstette}, {Chakrabarty}, {F{\"u}rst}, {Barret}, {Beloborodov}, {Boggs},
  {Christensen}, {Craig}, {Fabian}, {Hailey}, {Hornschemeier}, {Kaspi},
  {Kulkarni}, {Maccarone}, {Miller}, {Rana}, {Stern}, {Tendulkar}, {Tomsick},
  {Webb}, \& {Zhang}}]{bachetti2014ulx}
{Bachetti}, M., {Harrison}, F.~A., {Walton}, D.~J., {et~al.} 2014, \nat, 514,
  202

\bibitem[{{Bailes} {et~al.}(2020){Bailes}, {Jameson}, {Abbate}, {Barr}, {Bhat},
  {Bondonneau}, {Burgay}, {Buchner}, {Camilo}, {Champion}, {Cognard},
  {Demorest}, {Freire}, {Gautam}, {Geyer}, {Griessmeier}, {Guillemot}, {Hu},
  {Jankowski}, {Johnston}, {Karastergiou}, {Karuppusamy}, {Kaur}, {Keith},
  {Kramer}, {van Leeuwen}, {Lower}, {Maan}, {McLaughlin}, {Meyers},
  {Os{\l}owski}, {Oswald}, {Parthasarathy}, {Pennucci}, {Posselt}, {Possenti},
  {Ransom}, {Reardon}, {Ridolfi}, {Schollar}, {Serylak}, {Shaifullah},
  {Shamohammadi}, {Shannon}, {Sobey}, {Song}, {Spiewak}, {Stairs}, {Stappers},
  {van Straten}, {Szary}, {Theureau}, {Venkatraman Krishnan}, {Weltevrede},
  {Wex}, {Abbott}, {Adams}, {Burger}, {Gamatham}, {Gouws}, {Horn}, {Hugo},
  {Joubert}, {Manley}, {McAlpine}, {Passmoor}, {Peens-Hough}, {Ramudzuli},
  {Rust}, {Salie}, {Schwardt}, {Siebrits}, {Van Tonder}, {Van Tonder}, \&
  {Welz}}]{bailes2020meerKAT}
{Bailes}, M., {Jameson}, A., {Abbate}, F., {et~al.} 2020, \pasa, 37, e028

\bibitem[{{Barr}(2018)}]{barr2018sband}
{Barr}, E.~D. 2018, in Pulsar Astrophysics the Next Fifty Years, ed.
  P.~{Weltevrede}, B.~B.~P. {Perera}, L.~L. {Preston}, \& S.~{Sanidas}, Vol.
  337, 175--178

\bibitem[{{Bates} {et~al.}(2013){Bates}, {Lorimer}, \&
  {Verbiest}}]{bates2013spectral}
{Bates}, S.~D., {Lorimer}, D.~R., \& {Verbiest}, J.~P.~W. 2013, \mnras, 431,
  1352

\bibitem[{{Bates} {et~al.}(2015){Bates}, {Thornton}, {Bailes}, {Barr}, {Bassa},
  {Bhat}, {Burgay}, {Burke-Spolaor}, {Champion}, {Flynn}, {Jameson},
  {Johnston}, {Keith}, {Kramer}, {Levin}, {Lyne}, {Milia}, {Ng}, {Petroff},
  {Possenti}, {Stappers}, {van Straten}, \& {Tiburzi}}]{bates2015discovery}
{Bates}, S.~D., {Thornton}, D., {Bailes}, M., {et~al.} 2015, \mnras, 446, 4019

\bibitem[{{Bergeron} {et~al.}(2011){Bergeron}, {Wesemael}, {Dufour},
  {Beauchamp}, {Hunter}, {Saffer}, {Gianninas}, {Ruiz}, {Limoges}, {Dufour},
  {Fontaine}, \& {Liebert}}]{bergeron2011cooling}
{Bergeron}, P., {Wesemael}, F., {Dufour}, P., {et~al.} 2011, \apj, 737, 28

\bibitem[{{Berthereau} {et~al.}(2023){Berthereau}, {Guillemot}, {Freire},
  {Kramer}, {Venkatraman Krishnan}, {Cognard}, {Theureau}, {Bailes}, {i
  Bernadich}, \& {Lower}}]{berthereau2023j1528}
{Berthereau}, A., {Guillemot}, L., {Freire}, P.~C.~C., {et~al.} 2023, \aap,
  674, A71

\bibitem[{{Boyles} {et~al.}(2013){Boyles}, {Lynch}, {Ransom}, {Stairs},
  {Lorimer}, {McLaughlin}, {Hessels}, {Kaspi}, {Kondratiev}, {Archibald},
  {Berndsen}, {Cardoso}, {Cherry}, {Epstein}, {Karako-Argaman}, {McPhee},
  {Pennucci}, {Roberts}, {Stovall}, \& {van Leeuwen}}]{boyles2013gbt}
{Boyles}, J., {Lynch}, R.~S., {Ransom}, S.~M., {et~al.} 2013, \apj, 763, 80

\bibitem[{{Caiazzo} {et~al.}(2021){Caiazzo}, {Burdge}, {Fuller}, {Heyl},
  {Kulkarni}, {Prince}, {Richer}, {Schwab}, {Andreoni}, {Bellm}, {Drake},
  {Duev}, {Graham}, {Helou}, {Mahabal}, {Masci}, {Smith}, \&
  {Soumagnac}}]{caiazzo2021wd}
{Caiazzo}, I., {Burdge}, K.~B., {Fuller}, J., {et~al.} 2021, \nat, 595, 39

\bibitem[{{Camilo} {et~al.}(2001){Camilo}, {Lyne}, {Manchester}, {Bell},
  {Stairs}, {D'Amico}, {Kaspi}, {Possenti}, {Crawford}, \&
  {McKay}}]{camilo2001discovery}
{Camilo}, F., {Lyne}, A.~G., {Manchester}, R.~N., {et~al.} 2001, \apjl, 548,
  L187

\bibitem[{{Carpano} {et~al.}(2018){Carpano}, {Haberl}, {Maitra}, \&
  {Vasilopoulos}}]{carpano2018ulx}
{Carpano}, S., {Haberl}, F., {Maitra}, C., \& {Vasilopoulos}, G. 2018, \mnras,
  476, L45

\bibitem[{{Cognard} {et~al.}(2017){Cognard}, {Freire}, {Guillemot}, {Theureau},
  {Tauris}, {Wex}, {Graikou}, {Kramer}, {Stappers}, {Lyne}, {Bassa},
  {Desvignes}, \& {Lazarus}}]{cognard2017massive}
{Cognard}, I., {Freire}, P. C.~C., {Guillemot}, L., {et~al.} 2017, \apj, 844,
  128

\bibitem[{{Cordes}(2004)}]{cordes2002ne2001}
{Cordes}, J.~M. 2004, ASP Conference Series, 317, 211

\bibitem[{{Cruces} {et~al.}(2021){Cruces}, {Champion}, {Li}, {Kramer}, {Zhu},
  {Wang}, {Cameron}, {Chen}, {Hobbs}, {Freire}, {Graikou}, {Krco}, {Liu},
  {Miao}, {Niu}, {Pan}, {Qian}, {Xue}, {Xie}, {You}, {Yu}, {Yuan}, {Yue},
  {Zhu}, {Zhu}, {Lackeos}, {Porayko}, {Wongphecauxon}, {Main}, \& {Crafts
  Collaboration}}]{cruces2021fast}
{Cruces}, M., {Champion}, D.~J., {Li}, D., {et~al.} 2021, \mnras, 508, 300

\bibitem[{{Damour} \& {Deruelle}(1986)}]{damour1986general}
{Damour}, T. \& {Deruelle}, N. 1986, Ann. Inst. Henri Poincar{\'e} Phys.
  Th{\'e}or, 44, 263

\bibitem[{{Damour} \& {Taylor}(1991)}]{damour1992orbital}
{Damour}, T. \& {Taylor}, J.~H. 1991, \apj, 366, 501

\bibitem[{{Damour} \& {Taylor}(1992)}]{damour1992tests}
{Damour}, T. \& {Taylor}, J.~H. 1992, \prd, 45, 1840

\bibitem[{{Deller} {et~al.}(2013){Deller}, {Boyles}, {Lorimer}, {Kaspi},
  {McLaughlin}, {Ransom}, {Stairs}, \& {Stovall}}]{deller2013vlbi}
{Deller}, A.~T., {Boyles}, J., {Lorimer}, D.~R., {et~al.} 2013, \apj, 770, 145

\bibitem[{{Edwards} \& {Bailes}(2001)}]{edwards2001recycled}
{Edwards}, R.~T. \& {Bailes}, M. 2001, \apj, 553, 801

\bibitem[{{Edwards} {et~al.}(2006){Edwards}, {Hobbs}, \&
  {Manchester}}]{edwards2006tempo2}
{Edwards}, R.~T., {Hobbs}, G.~B., \& {Manchester}, R.~N. 2006, \mnras, 372,
  1549

\bibitem[{{Faulkner} {et~al.}(2004){Faulkner}, {Stairs}, {Kramer}, {Lyne},
  {Hobbs}, {Possenti}, {Lorimer}, {Manchester}, {McLaughlin}, {D'Amico},
  {Camilo}, \& {Burgay}}]{faulkner2004finding}
{Faulkner}, A.~J., {Stairs}, I.~H., {Kramer}, M., {et~al.} 2004, \mnras, 355,
  147

\bibitem[{{Fonseca} {et~al.}(2021){Fonseca}, {Cromartie}, {Pennucci}, {Ray},
  {Kirichenko}, {Ransom}, {Demorest}, {Stairs}, {Arzoumanian}, {Guillemot},
  {Parthasarathy}, {Kerr}, {Cognard}, {Baker}, {Blumer}, {Brook}, {DeCesar},
  {Dolch}, {Dong}, {Ferrara}, {Fiore}, {Garver-Daniels}, {Good}, {Jennings},
  {Jones}, {Kaspi}, {Lam}, {Lorimer}, {Luo}, {McEwen}, {McKee}, {McLaughlin},
  {McMann}, {Meyers}, {Naidu}, {Ng}, {Nice}, {Pol}, {Radovan},
  {Shapiro-Albert}, {Tan}, {Tendulkar}, {Swiggum}, {Wahl}, \&
  {Zhu}}]{Fonseca_2021}
{Fonseca}, E., {Cromartie}, H.~T., {Pennucci}, T.~T., {et~al.} 2021, \apjl,
  915, L12

\bibitem[{{Freire} \& {Wex}(2010)}]{freire2010orthometric}
{Freire}, P. C.~C. \& {Wex}, N. 2010, \mnras, 409, 199

\bibitem[{{Gaia Collaboration} {et~al.}(2016){Gaia Collaboration}, {Prusti},
  {de Bruijne}, {Brown}, {Vallenari}, {Babusiaux}, {Bailer-Jones}, {Bastian},
  {Biermann}, {Evans}, {Eyer}, {Jansen}, {Jordi}, {Klioner}, {Lammers},
  {Lindegren}, {Luri}, {Mignard}, {Milligan}, {Panem}, {Poinsignon},
  {Pourbaix}, {Randich}, {Sarri}, {Sartoretti}, {Siddiqui}, {Soubiran},
  {Valette}, {van Leeuwen}, {Walton}, {Aerts}, {Arenou}, {Cropper}, {Drimmel},
  {H{\o}g}, {Katz}, {Lattanzi}, {O'Mullane}, {Grebel}, {Holland}, {Huc},
  {Passot}, {Bramante}, {Cacciari}, {Casta{\~n}eda}, {Chaoul}, {Cheek}, {De
  Angeli}, {Fabricius}, {Guerra}, {Hern{\'a}ndez}, {Jean-Antoine-Piccolo},
  {Masana}, {Messineo}, {Mowlavi}, {Nienartowicz}, {Ord{\'o}{\~n}ez-Blanco},
  {Panuzzo}, {Portell}, {Richards}, {Riello}, {Seabroke}, {Tanga},
  {Th{\'e}venin}, {Torra}, {Els}, {Gracia-Abril}, {Comoretto},
  {Garcia-Reinaldos}, {Lock}, {Mercier}, {Altmann}, {Andrae}, {Astraatmadja},
  {Bellas-Velidis}, {Benson}, {Berthier}, {Blomme}, {Busso}, {Carry},
  {Cellino}, {Clementini}, {Cowell}, {Creevey}, {Cuypers}, {Davidson}, {De
  Ridder}, {de Torres}, {Delchambre}, {Dell'Oro}, {Ducourant}, {Fr{\'e}mat},
  {Garc{\'\i}a-Torres}, {Gosset}, {Halbwachs}, {Hambly}, {Harrison}, {Hauser},
  {Hestroffer}, {Hodgkin}, {Huckle}, {Hutton}, {Jasniewicz}, {Jordan},
  {Kontizas}, {Korn}, {Lanzafame}, {Manteiga}, {Moitinho}, {Muinonen},
  {Osinde}, {Pancino}, {Pauwels}, {Petit}, {Recio-Blanco}, {Robin}, {Sarro},
  {Siopis}, {Smith}, {Smith}, {Sozzetti}, {Thuillot}, {van Reeven}, {Viala},
  {Abbas}, {Abreu Aramburu}, {Accart}, {Aguado}, {Allan}, {Allasia},
  {Altavilla}, {{\'A}lvarez}, {Alves}, {Anderson}, {Andrei}, {Anglada Varela},
  {Antiche}, {Antoja}, {Ant{\'o}n}, {Arcay}, {Atzei}, {Ayache}, {Bach},
  {Baker}, {Balaguer-N{\'u}{\~n}ez}, {Barache}, {Barata}, {Barbier}, {Barblan},
  {Baroni}, {Barrado y Navascu{\'e}s}, {Barros}, {Barstow}, {Becciani},
  {Bellazzini}, {Bellei}, {Bello Garc{\'\i}a}, {Belokurov}, {Bendjoya},
  {Berihuete}, {Bianchi}, {Bienaym{\'e}}, {Billebaud}, {Blagorodnova},
  {Blanco-Cuaresma}, {Boch}, {Bombrun}, {Borrachero}, {Bouquillon}, {Bourda},
  {Bouy}, {Bragaglia}, {Breddels}, {Brouillet}, {Br{\"u}semeister},
  {Bucciarelli}, {Budnik}, {Burgess}, {Burgon}, {Burlacu}, {Busonero}, {Buzzi},
  {Caffau}, {Cambras}, {Campbell}, {Cancelliere}, {Cantat-Gaudin}, {Carlucci},
  {Carrasco}, {Castellani}, {Charlot}, {Charnas}, {Charvet}, {Chassat},
  {Chiavassa}, {Clotet}, {Cocozza}, {Collins}, {Collins}, {Costigan}, {Crifo},
  {Cross}, {Crosta}, {Crowley}, {Dafonte}, {Damerdji}, {Dapergolas}, {David},
  {David}, {De Cat}, {de Felice}, {de Laverny}, {De Luise}, {De March}, {de
  Martino}, {de Souza}, {Debosscher}, {del Pozo}, {Delbo}, {Delgado},
  {Delgado}, {di Marco}, {Di Matteo}, {Diakite}, {Distefano}, {Dolding}, {Dos
  Anjos}, {Drazinos}, {Dur{\'a}n}, {Dzigan}, {Ecale}, {Edvardsson}, {Enke},
  {Erdmann}, {Escolar}, {Espina}, {Evans}, {Eynard Bontemps}, {Fabre},
  {Fabrizio}, {Faigler}, {Falc{\~a}o}, {Farr{\`a}s Casas}, {Faye}, {Federici},
  {Fedorets}, {Fern{\'a}ndez-Hern{\'a}ndez}, {Fernique}, {Fienga}, {Figueras},
  {Filippi}, {Findeisen}, {Fonti}, {Fouesneau}, {Fraile}, {Fraser}, {Fuchs},
  {Furnell}, {Gai}, {Galleti}, {Galluccio}, {Garabato}, {Garc{\'\i}a-Sedano},
  {Gar{\'e}}, {Garofalo}, {Garralda}, {Gavras}, {Gerssen}, {Geyer}, {Gilmore},
  {Girona}, {Giuffrida}, {Gomes}, {Gonz{\'a}lez-Marcos},
  {Gonz{\'a}lez-N{\'u}{\~n}ez}, {Gonz{\'a}lez-Vidal}, {Granvik}, {Guerrier},
  {Guillout}, {Guiraud}, {G{\'u}rpide}, {Guti{\'e}rrez-S{\'a}nchez}, {Guy},
  {Haigron}, {Hatzidimitriou}, {Haywood}, {Heiter}, {Helmi}, {Hobbs},
  {Hofmann}, {Holl}, {Holland}, {Hunt}, {Hypki}, {Icardi}, {Irwin}, {Jevardat
  de Fombelle}, {Jofr{\'e}}, {Jonker}, {Jorissen}, {Julbe}, {Karampelas},
  {Kochoska}, {Kohley}, {Kolenberg}, {Kontizas}, {Koposov}, {Kordopatis},
  {Koubsky}, {Kowalczyk}, {Krone-Martins}, {Kudryashova}, {Kull}, {Bachchan},
  {Lacoste-Seris}, {Lanza}, {Lavigne}, {Le Poncin-Lafitte}, {Lebreton},
  {Lebzelter}, {Leccia}, {Leclerc}, {Lecoeur-Taibi}, {Lemaitre}, {Lenhardt},
  {Leroux}, {Liao}, {Licata}, {Lindstr{\o}m}, {Lister}, {Livanou}, {Lobel},
  {L{\"o}ffler}, {L{\'o}pez}, {Lopez-Lozano}, {Lorenz}, {Loureiro},
  {MacDonald}, {Magalh{\~a}es Fernandes}, {Managau}, {Mann}, {Mantelet},
  {Marchal}, {Marchant}, {Marconi}, {Marie}, {Marinoni}, {Marrese},
  {Marschalk{\'o}}, {Marshall}, {Mart{\'\i}n-Fleitas}, {Martino}, {Mary},
  {Matijevi{\v{c}}}, {Mazeh}, {McMillan}, {Messina}, {Mestre}, {Michalik},
  {Millar}, {Miranda}, {Molina}, {Molinaro}, {Molinaro}, {Moln{\'a}r},
  {Moniez}, {Montegriffo}, {Monteiro}, {Mor}, {Mora}, {Morbidelli}, {Morel},
  {Morgenthaler}, {Morley}, {Morris}, {Mulone}, {Muraveva}, {Musella},
  {Narbonne}, {Nelemans}, {Nicastro}, {Noval}, {Ord{\'e}novic},
  {Ordieres-Mer{\'e}}, {Osborne}, {Pagani}, {Pagano}, {Pailler}, {Palacin},
  {Palaversa}, {Parsons}, {Paulsen}, {Pecoraro}, {Pedrosa}, {Pentik{\"a}inen},
  {Pereira}, {Pichon}, {Piersimoni}, {Pineau}, {Plachy}, {Plum}, {Poujoulet},
  {Pr{\v{s}}a}, {Pulone}, {Ragaini}, {Rago}, {Rambaux}, {Ramos-Lerate},
  {Ranalli}, {Rauw}, {Read}, {Regibo}, {Renk}, {Reyl{\'e}}, {Ribeiro},
  {Rimoldini}, {Ripepi}, {Riva}, {Rixon}, {Roelens}, {Romero-G{\'o}mez},
  {Rowell}, {Royer}, {Rudolph}, {Ruiz-Dern}, {Sadowski}, {Sagrist{\`a}
  Sell{\'e}s}, {Sahlmann}, {Salgado}, {Salguero}, {Sarasso}, {Savietto},
  {Schnorhk}, {Schultheis}, {Sciacca}, {Segol}, {Segovia}, {Segransan},
  {Serpell}, {Shih}, {Smareglia}, {Smart}, {Smith}, {Solano}, {Solitro},
  {Sordo}, {Soria Nieto}, {Souchay}, {Spagna}, {Spoto}, {Stampa}, {Steele},
  {Steidelm{\"u}ller}, {Stephenson}, {Stoev}, {Suess}, {S{\"u}veges}, {Surdej},
  {Szabados}, {Szegedi-Elek}, {Tapiador}, {Taris}, {Tauran}, {Taylor},
  {Teixeira}, {Terrett}, {Tingley}, {Trager}, {Turon}, {Ulla}, {Utrilla},
  {Valentini}, {van Elteren}, {Van Hemelryck}, {van Leeuwen}, {Varadi},
  {Vecchiato}, {Veljanoski}, {Via}, {Vicente}, {Vogt}, {Voss}, {Votruba},
  {Voutsinas}, {Walmsley}, {Weiler}, {Weingrill}, {Werner}, {Wevers},
  {Whitehead}, {Wyrzykowski}, {Yoldas}, {{\v{Z}}erjal}, {Zucker}, {Zurbach},
  {Zwitter}, {Alecu}, {Allen}, {Allende Prieto}, {Amorim},
  {Anglada-Escud{\'e}}, {Arsenijevic}, {Azaz}, {Balm}, {Beck}, {Bernstein},
  {Bigot}, {Bijaoui}, {Blasco}, {Bonfigli}, {Bono}, {Boudreault}, {Bressan},
  {Brown}, {Brunet}, {Bunclark}, {Buonanno}, {Butkevich}, {Carret}, {Carrion},
  {Chemin}, {Ch{\'e}reau}, {Corcione}, {Darmigny}, {de Boer}, {de Teodoro}, {de
  Zeeuw}, {Delle Luche}, {Domingues}, {Dubath}, {Fodor}, {Fr{\'e}zouls},
  {Fries}, {Fustes}, {Fyfe}, {Gallardo}, {Gallegos}, {Gardiol}, {Gebran},
  {Gomboc}, {G{\'o}mez}, {Grux}, {Gueguen}, {Heyrovsky}, {Hoar}, {Iannicola},
  {Isasi Parache}, {Janotto}, {Joliet}, {Jonckheere}, {Keil}, {Kim},
  {Klagyivik}, {Klar}, {Knude}, {Kochukhov}, {Kolka}, {Kos}, {Kutka}, {Lainey},
  {LeBouquin}, {Liu}, {Loreggia}, {Makarov}, {Marseille}, {Martayan},
  {Martinez-Rubi}, {Massart}, {Meynadier}, {Mignot}, {Munari}, {Nguyen},
  {Nordlander}, {Ocvirk}, {O'Flaherty}, {Olias Sanz}, {Ortiz}, {Osorio},
  {Oszkiewicz}, {Ouzounis}, {Palmer}, {Park}, {Pasquato}, {Peltzer}, {Peralta},
  {P{\'e}turaud}, {Pieniluoma}, {Pigozzi}, {Poels}, {Prat}, {Prod'homme},
  {Raison}, {Rebordao}, {Risquez}, {Rocca-Volmerange}, {Rosen}, {Ruiz-Fuertes},
  {Russo}, {Sembay}, {Serraller Vizcaino}, {Short}, {Siebert}, {Silva},
  {Sinachopoulos}, {Slezak}, {Soffel}, {Sosnowska}, {Strai{\v{z}}ys}, {ter
  Linden}, {Terrell}, {Theil}, {Tiede}, {Troisi}, {Tsalmantza}, {Tur},
  {Vaccari}, {Vachier}, {Valles}, {Van Hamme}, {Veltz}, {Virtanen}, {Wallut},
  {Wichmann}, {Wilkinson}, {Ziaeepour}, \& {Zschocke}}]{gaia2016mission}
{Gaia Collaboration}, {Prusti}, T., {de Bruijne}, J.~H.~J., {et~al.} 2016,
  \aap, 595, A1

\bibitem[{{Gaia Collaboration} {et~al.}(2023){Gaia Collaboration}, {Vallenari},
  {Brown}, {Prusti}, {de Bruijne}, {Arenou}, {Babusiaux}, {Biermann},
  {Creevey}, {Ducourant}, {Evans}, {Eyer}, {Guerra}, {Hutton}, {Jordi},
  {Klioner}, {Lammers}, {Lindegren}, {Luri}, {Mignard}, {Panem}, {Pourbaix},
  {Randich}, {Sartoretti}, {Soubiran}, {Tanga}, {Walton}, {Bailer-Jones},
  {Bastian}, {Drimmel}, {Jansen}, {Katz}, {Lattanzi}, {van Leeuwen}, {Bakker},
  {Cacciari}, {Casta{\~n}eda}, {De Angeli}, {Fabricius}, {Fouesneau},
  {Fr{\'e}mat}, {Galluccio}, {Guerrier}, {Heiter}, {Masana}, {Messineo},
  {Mowlavi}, {Nicolas}, {Nienartowicz}, {Pailler}, {Panuzzo}, {Riclet}, {Roux},
  {Seabroke}, {Sordo}, {Th{\'e}venin}, {Gracia-Abril}, {Portell}, {Teyssier},
  {Altmann}, {Andrae}, {Audard}, {Bellas-Velidis}, {Benson}, {Berthier},
  {Blomme}, {Burgess}, {Busonero}, {Busso}, {C{\'a}novas}, {Carry}, {Cellino},
  {Cheek}, {Clementini}, {Damerdji}, {Davidson}, {de Teodoro}, {Nu{\~n}ez
  Campos}, {Delchambre}, {Dell'Oro}, {Esquej}, {Fern{\'a}ndez-Hern{\'a}ndez},
  {Fraile}, {Garabato}, {Garc{\'\i}a-Lario}, {Gosset}, {Haigron}, {Halbwachs},
  {Hambly}, {Harrison}, {Hern{\'a}ndez}, {Hestroffer}, {Hodgkin}, {Holl},
  {Jan{\ss}en}, {Jevardat de Fombelle}, {Jordan}, {Krone-Martins}, {Lanzafame},
  {L{\"o}ffler}, {Marchal}, {Marrese}, {Moitinho}, {Muinonen}, {Osborne},
  {Pancino}, {Pauwels}, {Recio-Blanco}, {Reyl{\'e}}, {Riello}, {Rimoldini},
  {Roegiers}, {Rybizki}, {Sarro}, {Siopis}, {Smith}, {Sozzetti}, {Utrilla},
  {van Leeuwen}, {Abbas}, {{\'A}brah{\'a}m}, {Abreu Aramburu}, {Aerts},
  {Aguado}, {Ajaj}, {Aldea-Montero}, {Altavilla}, {{\'A}lvarez}, {Alves},
  {Anders}, {Anderson}, {Anglada Varela}, {Antoja}, {Baines}, {Baker},
  {Balaguer-N{\'u}{\~n}ez}, {Balbinot}, {Balog}, {Barache}, {Barbato},
  {Barros}, {Barstow}, {Bartolom{\'e}}, {Bassilana}, {Bauchet}, {Becciani},
  {Bellazzini}, {Berihuete}, {Bernet}, {Bertone}, {Bianchi}, {Binnenfeld},
  {Blanco-Cuaresma}, {Blazere}, {Boch}, {Bombrun}, {Bossini}, {Bouquillon},
  {Bragaglia}, {Bramante}, {Breedt}, {Bressan}, {Brouillet}, {Brugaletta},
  {Bucciarelli}, {Burlacu}, {Butkevich}, {Buzzi}, {Caffau}, {Cancelliere},
  {Cantat-Gaudin}, {Carballo}, {Carlucci}, {Carnerero}, {Carrasco},
  {Casamiquela}, {Castellani}, {Castro-Ginard}, {Chaoul}, {Charlot}, {Chemin},
  {Chiaramida}, {Chiavassa}, {Chornay}, {Comoretto}, {Contursi}, {Cooper},
  {Cornez}, {Cowell}, {Crifo}, {Cropper}, {Crosta}, {Crowley}, {Dafonte},
  {Dapergolas}, {David}, {David}, {de Laverny}, {De Luise}, {De March}, {De
  Ridder}, {de Souza}, {de Torres}, {del Peloso}, {del Pozo}, {Delbo},
  {Delgado}, {Delisle}, {Demouchy}, {Dharmawardena}, {Di Matteo}, {Diakite},
  {Diener}, {Distefano}, {Dolding}, {Edvardsson}, {Enke}, {Fabre}, {Fabrizio},
  {Faigler}, {Fedorets}, {Fernique}, {Fienga}, {Figueras}, {Fournier},
  {Fouron}, {Fragkoudi}, {Gai}, {Garcia-Gutierrez}, {Garcia-Reinaldos},
  {Garc{\'\i}a-Torres}, {Garofalo}, {Gavel}, {Gavras}, {Gerlach}, {Geyer},
  {Giacobbe}, {Gilmore}, {Girona}, {Giuffrida}, {Gomel}, {Gomez},
  {Gonz{\'a}lez-N{\'u}{\~n}ez}, {Gonz{\'a}lez-Santamar{\'\i}a},
  {Gonz{\'a}lez-Vidal}, {Granvik}, {Guillout}, {Guiraud},
  {Guti{\'e}rrez-S{\'a}nchez}, {Guy}, {Hatzidpamitriou}, {Hauser}, {Haywood},
  {Helmer}, {Helmi}, {Sarmiento}, {Hidalgo}, {Hilger}, {H{\l}adczuk}, {Hobbs},
  {Holland}, {Huckle}, {Jardine}, {Jasniewicz}, {Jean-Antoine Piccolo},
  {Jim{\'e}nez-Arranz}, {Jorissen}, {Juaristi Campillo}, {Julbe}, {Karbevska},
  {Kervella}, {Khanna}, {Kontizas}, {Kordopatis}, {Korn}, {K{\'o}sp{\'a}l},
  {Kostrzewa-Rutkowska}, {Kruszy{\'n}ska}, {Kun}, {Laizeau}, {Lambert},
  {Lanza}, {Lasne}, {Le Campion}, {Lebreton}, {Lebzelter}, {Leccia}, {Leclerc},
  {Lecoeur-Taibi}, {Liao}, {Licata}, {Lindstr{\o}m}, {Lister}, {Livanou},
  {Lobel}, {Lorca}, {Loup}, {Madrero Pardo}, {Magdaleno Romeo}, {Managau},
  {Mann}, {Manteiga}, {Marchant}, {Marconi}, {Marcos}, {Marcos Santos},
  {Mar{\'\i}n Pina}, {Marinoni}, {Marocco}, {Marshall}, {Martin Polo},
  {Mart{\'\i}n-Fleitas}, {Marton}, {Mary}, {Masip}, {Massari},
  {Mastrobuono-Battisti}, {Mazeh}, {McMillan}, {Messina}, {Michalik}, {Millar},
  {Mints}, {Molina}, {Molinaro}, {Moln{\'a}r}, {Monari}, {Mongui{\'o}},
  {Montegriffo}, {Montero}, {Mor}, {Mora}, {Morbidelli}, {Morel}, {Morris},
  {Muraveva}, {Murphy}, {Musella}, {Nagy}, {Noval}, {Oca{\~n}a}, {Ogden},
  {Ordenovic}, {Osinde}, {Pagani}, {Pagano}, {Palaversa}, {Palicio},
  {Pallas-Quintela}, {Panahi}, {Payne-Wardenaar}, {Pe{\~n}alosa Esteller},
  {Penttil{\"a}}, {Pichon}, {Piersimoni}, {Pineau}, {Plachy}, {Plum}, {Poggio},
  {Pr{\v{s}}a}, {Pulone}, {Racero}, {Ragaini}, {Rainer}, {Raiteri}, {Rambaux},
  {Ramos}, {Ramos-Lerate}, {Re Fiorentin}, {Regibo}, {Richards}, {Rios Diaz},
  {Ripepi}, {Riva}, {Rix}, {Rixon}, {Robichon}, {Robin}, {Robin}, {Roelens},
  {Rogues}, {Rohrbasser}, {Romero-G{\'o}mez}, {Rowell}, {Royer}, {Ruz Mieres},
  {Rybicki}, {Sadowski}, {S{\'a}ez N{\'u}{\~n}ez}, {Sagrist{\`a} Sell{\'e}s},
  {Sahlmann}, {Salguero}, {Samaras}, {Sanchez Gimenez}, {Sanna},
  {Santove{\~n}a}, {Sarasso}, {Schultheis}, {Sciacca}, {Segol}, {Segovia},
  {S{\'e}gransan}, {Semeux}, {Shahaf}, {Siddiqui}, {Siebert}, {Siltala},
  {Silvelo}, {Slezak}, {Slezak}, {Smart}, {Snaith}, {Solano}, {Solitro},
  {Souami}, {Souchay}, {Spagna}, {Spina}, {Spoto}, {Steele},
  {Steidelm{\"u}ller}, {Stephenson}, {S{\"u}veges}, {Surdej}, {Szabados},
  {Szegedi-Elek}, {Taris}, {Taylor}, {Teixeira}, {Tolomei}, {Tonello}, {Torra},
  {Torra}, {Torralba Elipe}, {Trabucchi}, {Tsounis}, {Turon}, {Ulla}, {Unger},
  {Vaillant}, {van Dillen}, {van Reeven}, {Vanel}, {Vecchiato}, {Viala},
  {Vicente}, {Voutsinas}, {Weiler}, {Wevers}, {Wyrzykowski}, {Yoldas}, {Yvard},
  {Zhao}, {Zorec}, {Zucker}, \& {Zwitter}}]{gaia2023dr3}
{Gaia Collaboration}, {Vallenari}, A., {Brown}, A.~G.~A., {et~al.} 2023, \aap,
  674, A1

\bibitem[{{Gautam} {et~al.}(2022){Gautam}, {Freire}, {Batrakov}, {Kramer},
  {Miao}, {Parent}, \& {Zhu}}]{gautam2022relativistic}
{Gautam}, T., {Freire}, P.~C.~C., {Batrakov}, A., {et~al.} 2022, \aap, 668,
  A187

\bibitem[{{Gregoris} \& {Ong}(2023)}]{gregoris2023chandrasekhar}
{Gregoris}, D. \& {Ong}, Y.~C. 2023, Annals of Physics, 452, 169287

\bibitem[{{Guo} {et~al.}(2021){Guo}, {Freire}, {Guillemot}, {Kramer}, {Zhu},
  {Wex}, {McKee}, {Deller}, {Ding}, {Kaplan}, {Stappers}, {Cognard}, {Miao},
  {Haase}, {Keith}, {Ransom}, \& {Theureau}}]{guo2021improved}
{Guo}, Y.~J., {Freire}, P.~C.~C., {Guillemot}, L., {et~al.} 2021, \aap, 654,
  A16

\bibitem[{{Hobbs} {et~al.}(2020){Hobbs}, {Manchester}, {Dunning}, {Jameson},
  {Roberts}, {George}, {Green}, {Tuthill}, {Toomey}, {Kaczmarek}, {Mader},
  {Marquarding}, {Ahmed}, {Amy}, {Bailes}, {Beresford}, {Bhat}, {Bock},
  {Bourne}, {Bowen}, {Brothers}, {Cameron}, {Carretti}, {Carter}, {Castillo},
  {Chekkala}, {Cheng}, {Chung}, {Craig}, {Dai}, {Dawson}, {Dempsey}, {Doherty},
  {Dong}, {Edwards}, {Ergesh}, {Gao}, {Han}, {Hayman}, {Indermuehle},
  {Jeganathan}, {Johnston}, {Kanoniuk}, {Kesteven}, {Kramer}, {Leach},
  {Mcintyre}, {Moss}, {Os{\l}owski}, {Phillips}, {Pope}, {Preisig}, {Price},
  {Reeves}, {Reilly}, {Reynolds}, {Robishaw}, {Roush}, {Ruckley}, {Sadler},
  {Sarkissian}, {Severs}, {Shannon}, {Smart}, {Smith}, {Smith}, {Sobey},
  {Staveley-Smith}, {Tzioumis}, {van Straten}, {Wang}, {Wen}, \&
  {Whiting}}]{hobbs2020uwl}
{Hobbs}, G., {Manchester}, R.~N., {Dunning}, A., {et~al.} 2020, \pasa, 37, e012

\bibitem[{{Hobbs} {et~al.}(2006){Hobbs}, {Edwards}, \&
  {Manchester}}]{hobbs2006tempo2}
{Hobbs}, G.~B., {Edwards}, R.~T., \& {Manchester}, R.~N. 2006, \mnras, 369, 655

\bibitem[{{Hollands} {et~al.}(2020){Hollands}, {Tremblay}, {G{\"a}nsicke},
  {Camisassa}, {Koester}, {Aungwerojwit}, {Chote}, {C{\'o}rsico}, {Dhillon},
  {Gentile-Fusillo}, {Hoskin}, {Izquierdo}, {Marsh}, \&
  {Steeghs}}]{hollands2020wd}
{Hollands}, M.~A., {Tremblay}, P.~E., {G{\"a}nsicke}, B.~T., {et~al.} 2020,
  Nature Astronomy, 4, 663

\bibitem[{{Hotan} {et~al.}(2004){Hotan}, {van Straten}, \&
  {Manchester}}]{hotan2004psrchive}
{Hotan}, A.~W., {van Straten}, W., \& {Manchester}, R.~N. 2004, \pasa, 21, 302

\bibitem[{{Hu} {et~al.}(2020){Hu}, {Kramer}, {Wex}, {Champion}, \&
  {Kehl}}]{hu2020constraining}
{Hu}, H., {Kramer}, M., {Wex}, N., {Champion}, D.~J., \& {Kehl}, M.~S. 2020,
  \mnras, 497, 3118

\bibitem[{{Israel} {et~al.}(2017){Israel}, {Belfiore}, {Stella}, {Esposito},
  {Casella}, {De Luca}, {Marelli}, {Papitto}, {Perri}, {Puccetti}, {Castillo},
  {Salvetti}, {Tiengo}, {Zampieri}, {D'Agostino}, {Greiner}, {Haberl},
  {Novara}, {Salvaterra}, {Turolla}, {Watson}, {Wilms}, \&
  {Wolter}}]{israel2017ulx}
{Israel}, G.~L., {Belfiore}, A., {Stella}, L., {et~al.} 2017, Science, 355, 817

\bibitem[{{Jacoby} {et~al.}(2007){Jacoby}, {Bailes}, {Ord}, {Knight}, \&
  {Hotan}}]{jacoby2007discovery}
{Jacoby}, B.~A., {Bailes}, M., {Ord}, S.~M., {Knight}, H.~S., \& {Hotan}, A.~W.
  2007, \apj, 656, 408

\bibitem[{{Jacoby} {et~al.}(2006){Jacoby}, {Chakrabarty}, {van Kerkwijk},
  {Kulkarni}, \& {Kaplan}}]{jacoby2006detection}
{Jacoby}, B.~A., {Chakrabarty}, D., {van Kerkwijk}, M.~H., {Kulkarni}, S.~R.,
  \& {Kaplan}, D.~L. 2006, \apjl, 640, L183

\bibitem[{{Jonas} \& {MeerKAT Team}(2016)}]{jonas2016meerkat}
{Jonas}, J. \& {MeerKAT Team}. 2016, in MeerKAT Science: On the Pathway to the
  SKA, 1

\bibitem[{{Kaplan} {et~al.}(2014){Kaplan}, {Boyles}, {Dunlap}, {Tendulkar},
  {Deller}, {Ransom}, {McLaughlin}, {Lorimer}, \&
  {Stairs}}]{kaplan2014companion}
{Kaplan}, D.~L., {Boyles}, J., {Dunlap}, B.~H., {et~al.} 2014, \apj, 789, 119

\bibitem[{{Knispel} {et~al.}(2013){Knispel}, {Eatough}, {Kim}, {Keane},
  {Allen}, {Anderson}, {Aulbert}, {Bock}, {Crawford}, {Eggenstein}, {Fehrmann},
  {Hammer}, {Kramer}, {Lyne}, {Machenschalk}, {Miller}, {Papa}, {Rastawicki},
  {Sarkissian}, {Siemens}, \& {Stappers}}]{knispel2013discovery}
{Knispel}, B., {Eatough}, R.~P., {Kim}, H., {et~al.} 2013, \apj, 774, 93

\bibitem[{{Kopeikin}(1996)}]{kopeikin1996proper}
{Kopeikin}, S.~M. 1996, \apjl, 467, L93

\bibitem[{{Kramer} {et~al.}(2021){Kramer}, {Stairs}, {Venkatraman Krishnan},
  {Freire}, {Abbate}, {Bailes}, {Burgay}, {Buchner}, {Champion}, {Cognard},
  {Gautam}, {Geyer}, {Guillemot}, {Hu}, {Janssen}, {Lower}, {Parthasarathy},
  {Possenti}, {Ransom}, {Reardon}, {Ridolfi}, {Serylak}, {Shannon}, {Spiewak},
  {Theureau}, {van Straten}, {Wex}, {Oswald}, {Posselt}, {Sobey}, {Barr},
  {Camilo}, {Hugo}, {Jameson}, {Johnston}, {Karastergiou}, {Keith}, \&
  {Os{\l}owski}}]{kramer2021relbin}
{Kramer}, M., {Stairs}, I.~H., {Venkatraman Krishnan}, V., {et~al.} 2021,
  \mnras, 504, 2094

\bibitem[{{Krishnakumar} {et~al.}(2019){Krishnakumar}, {Maan}, {Joshi}, \&
  {Manoharan}}]{krishnakumar2019scatter}
{Krishnakumar}, M.~A., {Maan}, Y., {Joshi}, B.~C., \& {Manoharan}, P.~K. 2019,
  \apj, 878, 130

\bibitem[{{K{\"u}lebi} {et~al.}(2010){K{\"u}lebi}, {Jordan}, {Nelan},
  {Bastian}, \& {Altmann}}]{kulebi2010wd}
{K{\"u}lebi}, B., {Jordan}, S., {Nelan}, E., {Bastian}, U., \& {Altmann}, M.
  2010, \aap, 524, A36

\bibitem[{{Kundu} \& {Mukhopadhyay}(2012)}]{kundu2012mass}
{Kundu}, A. \& {Mukhopadhyay}, B. 2012, Modern Physics Letters A, 27, 1250084

\bibitem[{{Lazarus} {et~al.}(2014){Lazarus}, {Tauris}, {Knispel}, {Freire},
  {Deneva}, {Kaspi}, {Allen}, {Bogdanov}, {Chatterjee}, {Stairs}, \&
  {Zhu}}]{lazarus2014massive}
{Lazarus}, P., {Tauris}, T.~M., {Knispel}, B., {et~al.} 2014, \mnras, 437, 1485

\bibitem[{{L{\"o}hmer} {et~al.}(2004){L{\"o}hmer}, {Mitra}, {Gupta}, {Kramer},
  \& {Ahuja}}]{lohmer2004frequency}
{L{\"o}hmer}, O., {Mitra}, D., {Gupta}, Y., {Kramer}, M., \& {Ahuja}, A. 2004,
  \aap, 425, 569

\bibitem[{{Lorimer} {et~al.}(2006){Lorimer}, {Faulkner}, {Lyne}, {Manchester},
  {Kramer}, {McLaughlin}, {Hobbs}, {Possenti}, {Stairs}, {Camilo}, {Burgay},
  {D'Amico}, {Corongiu}, \& {Crawford}}]{lorimer2006timing}
{Lorimer}, D.~R., {Faulkner}, A.~J., {Lyne}, A.~G., {et~al.} 2006, \mnras, 372,
  777

\bibitem[{Lorimer \& Kramer(2005)}]{lorimer2012handbook}
Lorimer, D.~R. \& Kramer, M. 2005, Handbook of pulsar astronomy (Cambridge
  University press)

\bibitem[{{Martinez} {et~al.}(2019){Martinez}, {Gentile}, {Freire}, {Stovall},
  {Deneva}, {Desvignes}, {Jenet}, {McLaughlin}, {Bagchi}, \&
  {Devine}}]{martinez2019discovery}
{Martinez}, J.~G., {Gentile}, P., {Freire}, P.~C.~C., {et~al.} 2019, \apj, 881,
  166

\bibitem[{{Martinez} {et~al.}(2015){Martinez}, {Stovall}, {Freire}, {Deneva},
  {Jenet}, {McLaughlin}, {Bagchi}, {Bates}, \& {Ridolfi}}]{martinez2015dns}
{Martinez}, J.~G., {Stovall}, K., {Freire}, P.~C.~C., {et~al.} 2015, \apj, 812,
  143

\bibitem[{{Mathew} \& {Nandy}(2021)}]{mathew2021chandrasekhar}
{Mathew}, A. \& {Nandy}, M.~K. 2021, Royal Society Open Science, 8, 210301

\bibitem[{{McKee} {et~al.}(2020){McKee}, {Freire}, {Berezina}, {Champion},
  {Cognard}, {Graikou}, {Guillemot}, {Keith}, {Kramer}, {Lyne}, {Stappers},
  {Tauris}, \& {Theureau}}]{mckee2020precise}
{McKee}, J.~W., {Freire}, P.~C.~C., {Berezina}, M., {et~al.} 2020, \mnras, 499,
  4082

\bibitem[{{McMillan}(2017)}]{mcmillan2017distribution}
{McMillan}, P.~J. 2017, \mnras, 465, 76

\bibitem[{{Mickaliger} {et~al.}(2012){Mickaliger}, {Lorimer}, {Boyles},
  {McLaughlin}, {Collins}, {Hough}, {Tehrani}, {Tenney}, {Liska}, \&
  {Swiggum}}]{mickaliger2012discovery}
{Mickaliger}, M.~B., {Lorimer}, D.~R., {Boyles}, J., {et~al.} 2012, \apj, 759,
  127

\bibitem[{{Miller} {et~al.}(2023){Miller}, {Caiazzo}, {Heyl}, {Richer},
  {El-Badry}, {Rodriguez}, {Vanderbosch}, \& {van Roestel}}]{miller2023hydaes}
{Miller}, D.~R., {Caiazzo}, I., {Heyl}, J., {et~al.} 2023, \apjl, 956, L41

\bibitem[{{Oswald} {et~al.}(2021){Oswald}, {Karastergiou}, {Posselt},
  {Johnston}, {Bailes}, {Buchner}, {Geyer}, {Keith}, {Kramer}, {Parthasarathy},
  {Reardon}, {Serylak}, {Shannon}, {Spiewak}, {van Straten}, \& {Venkatraman
  Krishnan}}]{oswald2021scattering}
{Oswald}, L.~S., {Karastergiou}, A., {Posselt}, B., {et~al.} 2021, \mnras, 504,
  1115

\bibitem[{{{\"O}zel} \& {Freire}(2016)}]{ozel2016masses}
{{\"O}zel}, F. \& {Freire}, P. 2016, \araa, 54, 401

\bibitem[{{Pallanca} {et~al.}(2013){Pallanca}, {Lanzoni}, {Dalessandro},
  {Ferraro}, {Possenti}, {Salaris}, \& {Burgay}}]{pallanca2013optical}
{Pallanca}, C., {Lanzoni}, B., {Dalessandro}, E., {et~al.} 2013, \apj, 773, 127

\bibitem[{{Parent} {et~al.}(2019){Parent}, {Kaspi}, {Ransom}, {Freire},
  {Brazier}, {Camilo}, {Chatterjee}, {Cordes}, {Crawford}, {Deneva}, {Ferdman},
  {Hessels}, {van Leeuwen}, {Lyne}, {Madsen}, {McLaughlin}, {Patel}, {Scholz},
  {Stairs}, {Stappers}, \& {Zhu}}]{parent2019palfa}
{Parent}, E., {Kaspi}, V.~M., {Ransom}, S.~M., {et~al.} 2019, \apj, 886, 148

\bibitem[{{Poutanen} {et~al.}(2007){Poutanen}, {Lipunova}, {Fabrika},
  {Butkevich}, \& {Abolmasov}}]{poutaten2007supercritically}
{Poutanen}, J., {Lipunova}, G., {Fabrika}, S., {Butkevich}, A.~G., \&
  {Abolmasov}, P. 2007, \mnras, 377, 1187

\bibitem[{{Pr{\v{s}}a} {et~al.}(2016){Pr{\v{s}}a}, {Harmanec}, {Torres},
  {Mamajek}, {Asplund}, {Capitaine}, {Christensen-Dalsgaard}, {Depagne},
  {Haberreiter}, {Hekker}, {Hilton}, {Kopp}, {Kostov}, {Kurtz}, {Laskar},
  {Mason}, {Milone}, {Montgomery}, {Richards}, {Schmutz}, {Schou}, \&
  {Stewart}}]{prvsa2016values}
{Pr{\v{s}}a}, A., {Harmanec}, P., {Torres}, G., {et~al.} 2016, \aj, 152, 41

\bibitem[{{Pshirkov} {et~al.}(2020){Pshirkov}, {Dodin}, {Belinski},
  {Zheltoukhov}, {Fedoteva}, {Voziakova}, {Potanin}, {Blinnikov}, \&
  {Postnov}}]{pshirkov2020wd}
{Pshirkov}, M.~S., {Dodin}, A.~V., {Belinski}, A.~A., {et~al.} 2020, \mnras,
  499, L21

\bibitem[{{Radhakrishnan} \& {Cooke}(1969)}]{radhakrishnan1969magnetic}
{Radhakrishnan}, V. \& {Cooke}, D.~J. 1969, \aplett, 3, 225

\bibitem[{{Rickett}(1977)}]{rickett1977scattering}
{Rickett}, B.~J. 1977, \araa, 15, 479

\bibitem[{{Ridolfi} {et~al.}(2019){Ridolfi}, {Freire}, {Gupta}, \&
  {Ransom}}]{ridolfi2019possible}
{Ridolfi}, A., {Freire}, P.~C.~C., {Gupta}, Y., \& {Ransom}, S.~M. 2019,
  \mnras, 490, 3860

\bibitem[{{Romani} {et~al.}(1986){Romani}, {Narayan}, \&
  {Blandford}}]{romani1986}
{Romani}, R.~W., {Narayan}, R., \& {Blandford}, R. 1986, \mnras, 220, 19

\bibitem[{{Sarkissian} {et~al.}(2011){Sarkissian}, {Carretti}, \& {van
  Straten}}]{sarkissian2011backends}
{Sarkissian}, J.~M., {Carretti}, E., \& {van Straten}, W. 2011, in American
  Institute of Physics Conference Series, Vol. 1357, Radio Pulsars: An
  Astrophysical Key to Unlock the Secrets of the Universe, ed. M.~{Burgay},
  N.~{D'Amico}, P.~{Esposito}, A.~{Pellizzoni}, \& A.~{Possenti}, 351--352

\bibitem[{{Shakura} \& {Sunyaev}(1973)}]{shakura1973blackholes}
{Shakura}, N.~I. \& {Sunyaev}, R.~A. 1973, \aap, 24, 337

\bibitem[{{Shamohammadi} {et~al.}(2023){Shamohammadi}, {Bailes}, {Freire},
  {Parthasarathy}, {Reardon}, {Shannon}, {Venkatraman Krishnan}, {Bernadich},
  {Cameron}, {Champion}, {Corongiu}, {Flynn}, {Geyer}, {Kramer}, {Miles},
  {Possenti}, \& {Spiewak}}]{shamohammadi2023manyS}
{Shamohammadi}, M., {Bailes}, M., {Freire}, P.~C.~C., {et~al.} 2023, \mnras,
  520, 1789

\bibitem[{{Shapiro}(1964)}]{shapiro1964test}
{Shapiro}, I.~I. 1964, \prl, 13, 789

\bibitem[{{Shklovskii}(1970)}]{shklovskii1970possible}
{Shklovskii}, I.~S. 1970, \sovast, 13, 562

\bibitem[{{Splaver} {et~al.}(2002){Splaver}, {Nice}, {Arzoumanian}, {Camilo},
  {Lyne}, \& {Stairs}}]{splaver2002masses}
{Splaver}, E.~M., {Nice}, D.~J., {Arzoumanian}, Z., {et~al.} 2002, \apj, 581,
  509

\bibitem[{{Staveley-Smith} {et~al.}(1996){Staveley-Smith}, {Wilson}, {Bird},
  {Disney}, {Ekers}, {Freeman}, {Haynes}, {Sinclair}, {Vaile}, {Webster}, \&
  {Wright}}]{staveley1996multibeam}
{Staveley-Smith}, L., {Wilson}, W.~E., {Bird}, T.~S., {et~al.} 1996, \pasa, 13,
  243

\bibitem[{{Tan} {et~al.}(2020){Tan}, {Bassa}, {Cooper}, {Hessels},
  {Kondratiev}, {Michilli}, {Sanidas}, {Stappers}, {van Leeuwen}, {Donner},
  {Grie{\ss}meier}, {Kramer}, {Tiburzi}, {Weltevrede}, {Ciardi}, {Hoeft},
  {Mann}, {Miskolczi}, {Schwarz}, {Vocks}, \& {Wucknitz}}]{tan2020timing}
{Tan}, C.~M., {Bassa}, C.~G., {Cooper}, S., {et~al.} 2020, \mnras, 492, 5878

\bibitem[{{Tauris} {et~al.}(2017){Tauris}, {Kramer}, {Freire}, {Wex}, {Janka},
  {Langer}, {Podsiadlowski}, {Bozzo}, {Chaty}, {Kruckow}, {van den Heuvel},
  {Antoniadis}, {Breton}, \& {Champion}}]{tauris2017dns}
{Tauris}, T.~M., {Kramer}, M., {Freire}, P.~C.~C., {et~al.} 2017, \apj, 846,
  170

\bibitem[{{Tauris} {et~al.}(2012){Tauris}, {Langer}, \&
  {Kramer}}]{tauris2012coII}
{Tauris}, T.~M., {Langer}, N., \& {Kramer}, M. 2012, \mnras, 425, 1601

\bibitem[{{Tauris} {et~al.}(2015){Tauris}, {Langer}, \&
  {Podsiadlowski}}]{tauris2015sn}
{Tauris}, T.~M., {Langer}, N., \& {Podsiadlowski}, P. 2015, \mnras, 451, 2123

\bibitem[{{Tauris} \& {Sennels}(2000)}]{tauris2000young}
{Tauris}, T.~M. \& {Sennels}, T. 2000, \aap, 355, 236

\bibitem[{{Tauris} \& {van den Heuvel}(2023)}]{tauris2023physics}
{Tauris}, T.~M. \& {van den Heuvel}, E. P.~J. 2023, {Physics of Binary Star
  Evolution. From Stars to X-ray Binaries and Gravitational Wave Sources}
  (Princeton University Press)

\bibitem[{{Taylor} \& {Weisberg}(1989)}]{taylor1989further}
{Taylor}, J.~H. \& {Weisberg}, J.~M. 1989, \apj, 345, 434

\bibitem[{{Thorsett} {et~al.}(1993){Thorsett}, {Arzoumanian}, {McKinnon}, \&
  {Taylor}}]{thorsett1993massive}
{Thorsett}, S.~E., {Arzoumanian}, Z., {McKinnon}, M.~M., \& {Taylor}, J.~H.
  1993, \apjl, 405, L29

\bibitem[{{Thorsett} \& {Chakrabarty}(1999)}]{thorsett1996masses}
{Thorsett}, S.~E. \& {Chakrabarty}, D. 1999, \apj, 512, 288

\bibitem[{{Tomaschitz}(2018)}]{tomaschitz2018chandrasekhar}
{Tomaschitz}, R. 2018, Physica A Statistical Mechanics and its Applications,
  489, 128

\bibitem[{{van den Heuvel}(2019)}]{van2018high}
{van den Heuvel}, E. P.~J. 2019, IAU Symposium, 346, 1

\bibitem[{{van Haasteren} {et~al.}(2009){van Haasteren}, {Levin}, {McDonald},
  \& {Lu}}]{vanHaasteren2009grav}
{van Haasteren}, R., {Levin}, Y., {McDonald}, P., \& {Lu}, T. 2009, \mnras,
  395, 1005

\bibitem[{{van Kerkwijk} \& {Kulkarni}(1999)}]{vanKerkwijk1999massive}
{van Kerkwijk}, M.~H. \& {Kulkarni}, S.~R. 1999, \apjl, 516, L25

\bibitem[{Virtanen {et~al.}(2020)Virtanen, Gommers, Oliphant, Haberland, Reddy,
  Cournapeau, Burovski, Peterson, Weckesser, Bright, {van der Walt}, Brett,
  Wilson, Millman, Mayorov, Nelson, Jones, Kern, Larson, Carey, Polat, Feng,
  Moore, {VanderPlas}, Laxalde, Perktold, Cimrman, Henriksen, Quintero, Harris,
  Archibald, Ribeiro, Pedregosa, {van Mulbregt}, \& {SciPy 1.0
  Contributors}}]{virtanen2020scipy}
Virtanen, P., Gommers, R., Oliphant, T.~E., {et~al.} 2020, Nature Methods, 17,
  261

\bibitem[{{Voisin} {et~al.}(2020){Voisin}, {Cognard}, {Freire}, {Wex},
  {Guillemot}, {Desvignes}, {Kramer}, \& {Theureau}}]{Voisin_2020}
{Voisin}, G., {Cognard}, I., {Freire}, P.~C.~C., {et~al.} 2020, \aap, 638, A24

\bibitem[{{Yao} {et~al.}(2017){Yao}, {Manchester}, \& {Wang}}]{yao2017new}
{Yao}, J.~M., {Manchester}, R.~N., \& {Wang}, N. 2017, \apj, 835, 29

\bibitem[{{Yoon} \& {Langer}(2005)}]{yoon2005rapidly}
{Yoon}, S.~C. \& {Langer}, N. 2005, \aap, 435, 967

\bibitem[{{Zhao} {et~al.}(2022){Zhao}, {Freire}, {Kramer}, {Shao}, \&
  {Wex}}]{Zhao_2022}
{Zhao}, J., {Freire}, P. C.~C., {Kramer}, M., {Shao}, L., \& {Wex}, N. 2022,
  Classical and Quantum Gravity, 39, 11LT01

\end{thebibliography}

\begin{appendix}\label{appendix}
\onecolumn

\section{Additional mass constraints plots}

\begin{figure}[!h]
\centering
 \includegraphics[width=1\columnwidth]{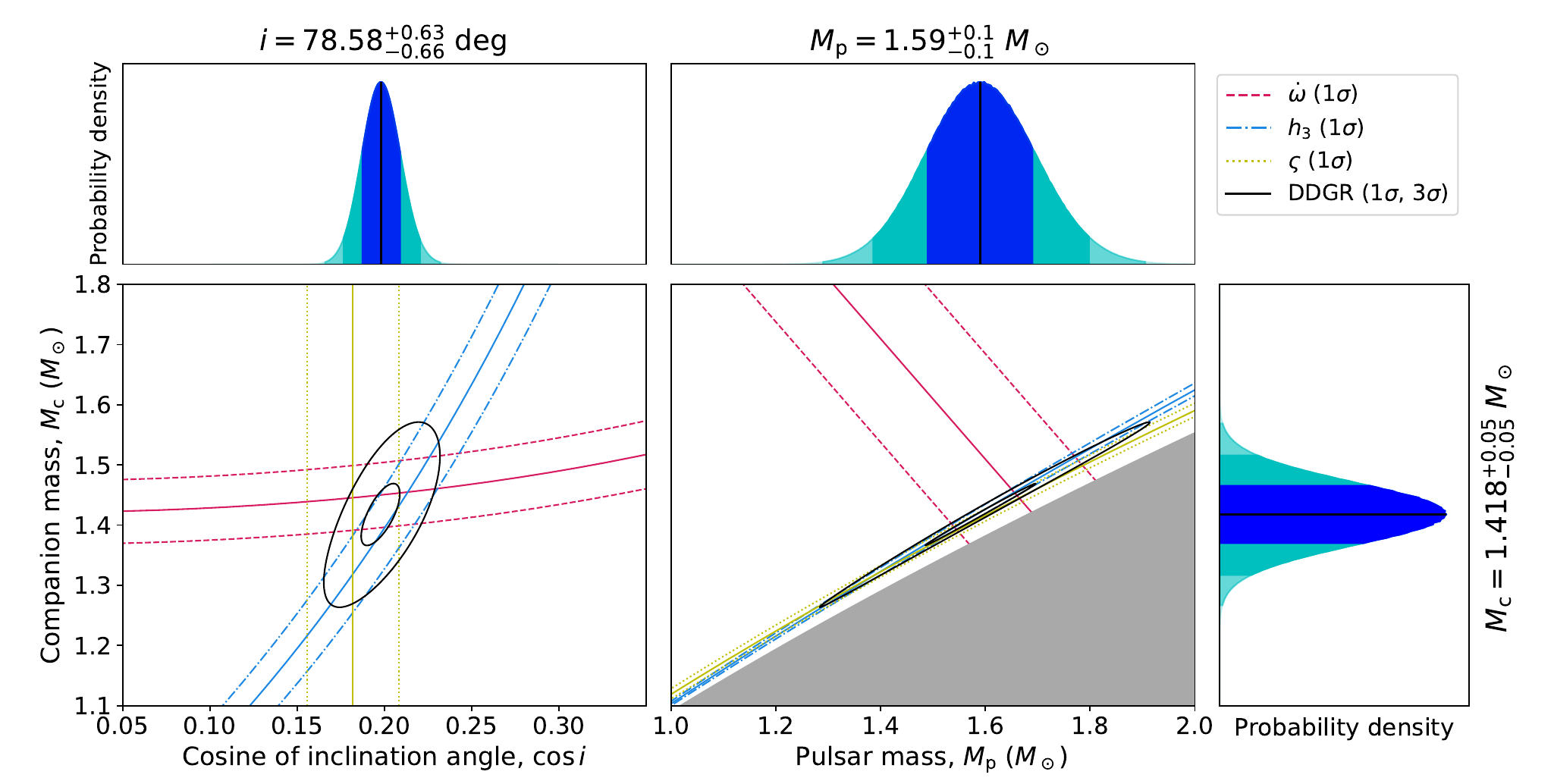}
 \caption{Mass and inclination angle constraints from the DDH PK measurements and the $\chi^2$ mapping with the DDGR model, both from the global fit and under the assumption of the \textbf{Ls} noise model (Table~\ref{fits}). Contours have been drawn following the same principles as in Fig.~\ref{mass_diagram}.}
 \label{DM100_R500_global}
\end{figure}

\begin{figure}[h]
\centering
 \includegraphics[width=1\columnwidth]{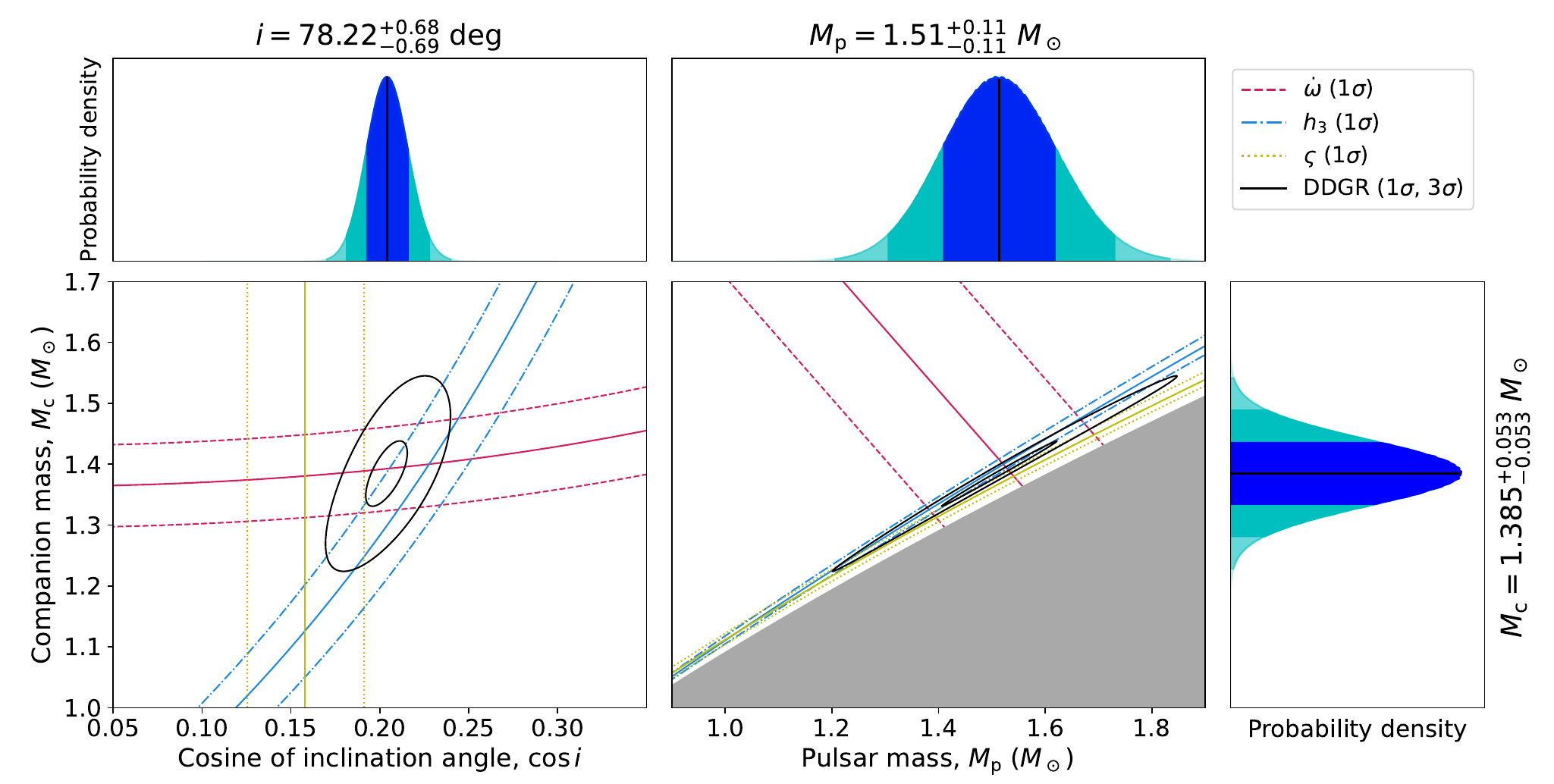}
 \caption{Mass and inclination angle constraints from the DDH PK measurements and the $\chi^2$ mapping with the DDGR model, both from the global fit and under the assumption of the \textbf{VCs} noise model (Table~\ref{fits}). Contours have been drawn following the same principles as in Fig.~\ref{mass_diagram}.}
  \label{DM30_R30_global}
\end{figure}

\begin{figure}[h]
\centering
 \includegraphics[width=1\columnwidth]{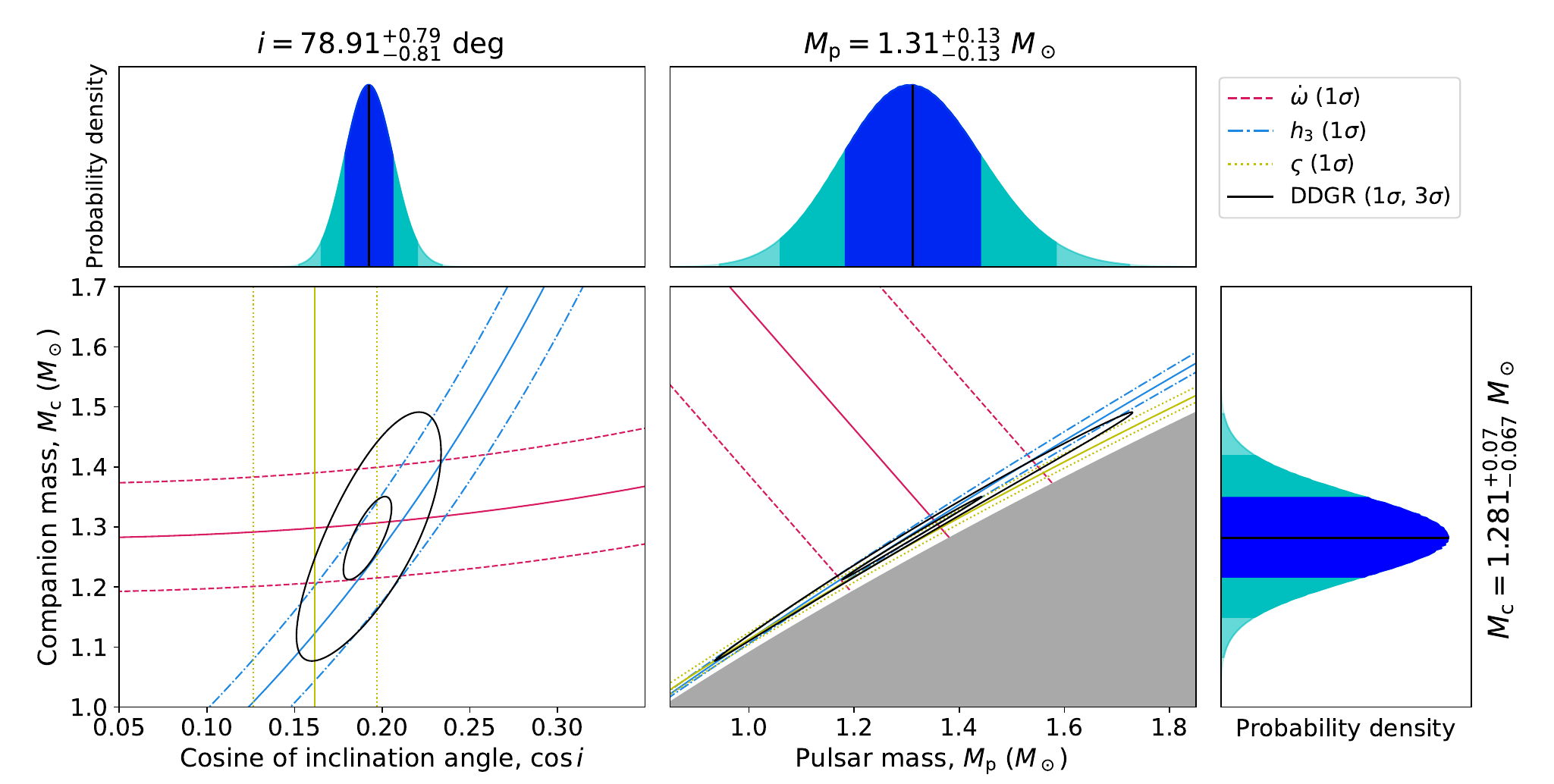}
 \caption{Mass and inclination angle constraints from the DDH PK measurements and the $\chi^2$ mapping with the DDGR model, both from the MeerKAT+UWL fit and under the assumption of the \textbf{VCs} noise model (Table~\ref{fits}). Contours have been drawn following the same principles as in Fig.~\ref{mass_diagram}.}
 \label{DM30_R30_mkt+uwl}
\end{figure}

\begin{figure}[h]
\centering
 \includegraphics[width=1\columnwidth]{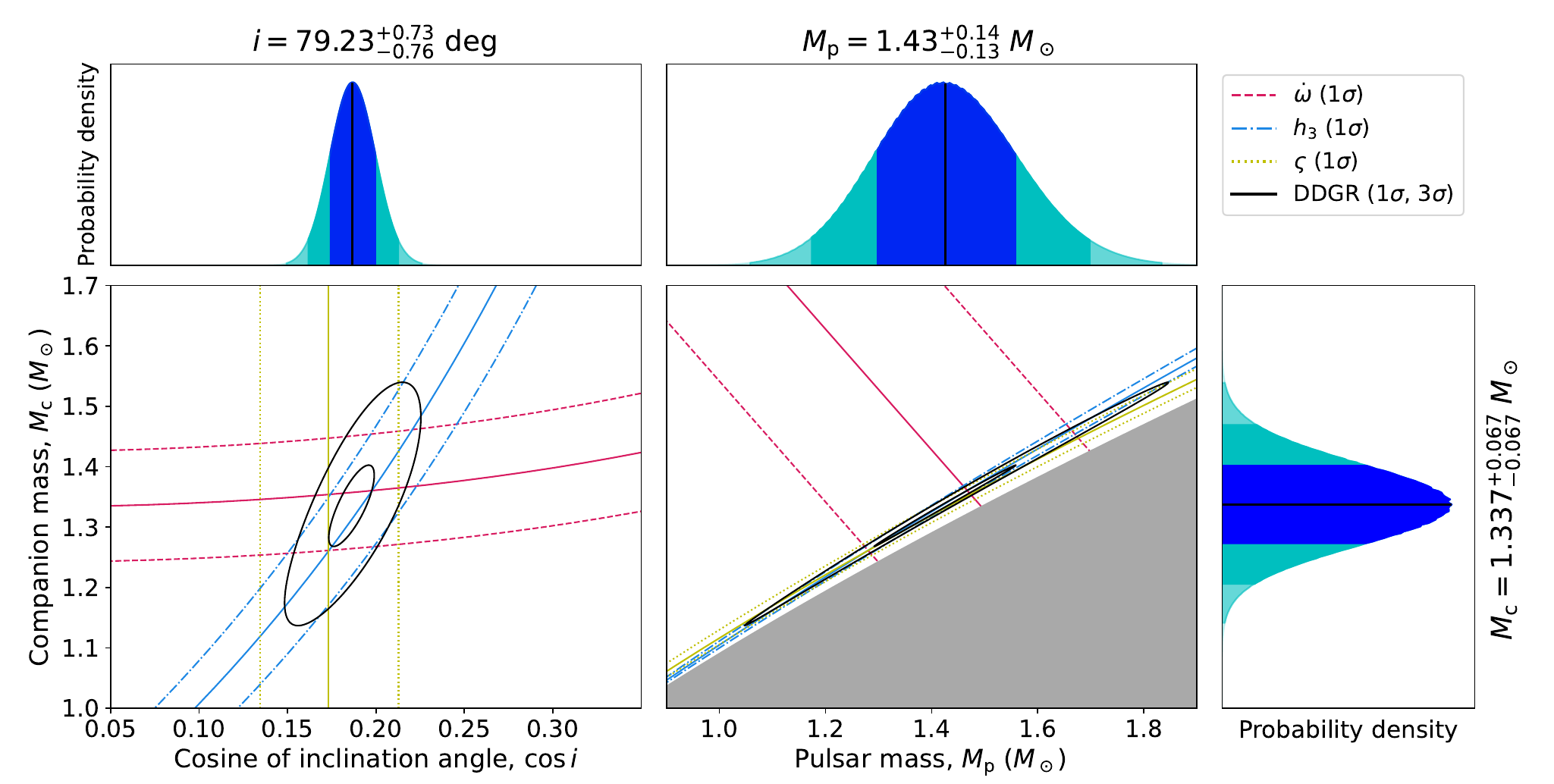}
 \caption{Mass and inclination angle constraints from the DDH PK measurements and the $\chi^2$ mapping with the DDGR model, both from the MeerKAT+UWL fit and under the assumption of the \textbf{Ls} noise model (Table~\ref{fits}). Contours have been drawn following the same principles as in Fig.~\ref{mass_diagram}.}
 \label{DM100_R500_mkt+uwl}
\end{figure}

\begin{figure}[h]
\centering
 \includegraphics[width=1\columnwidth]{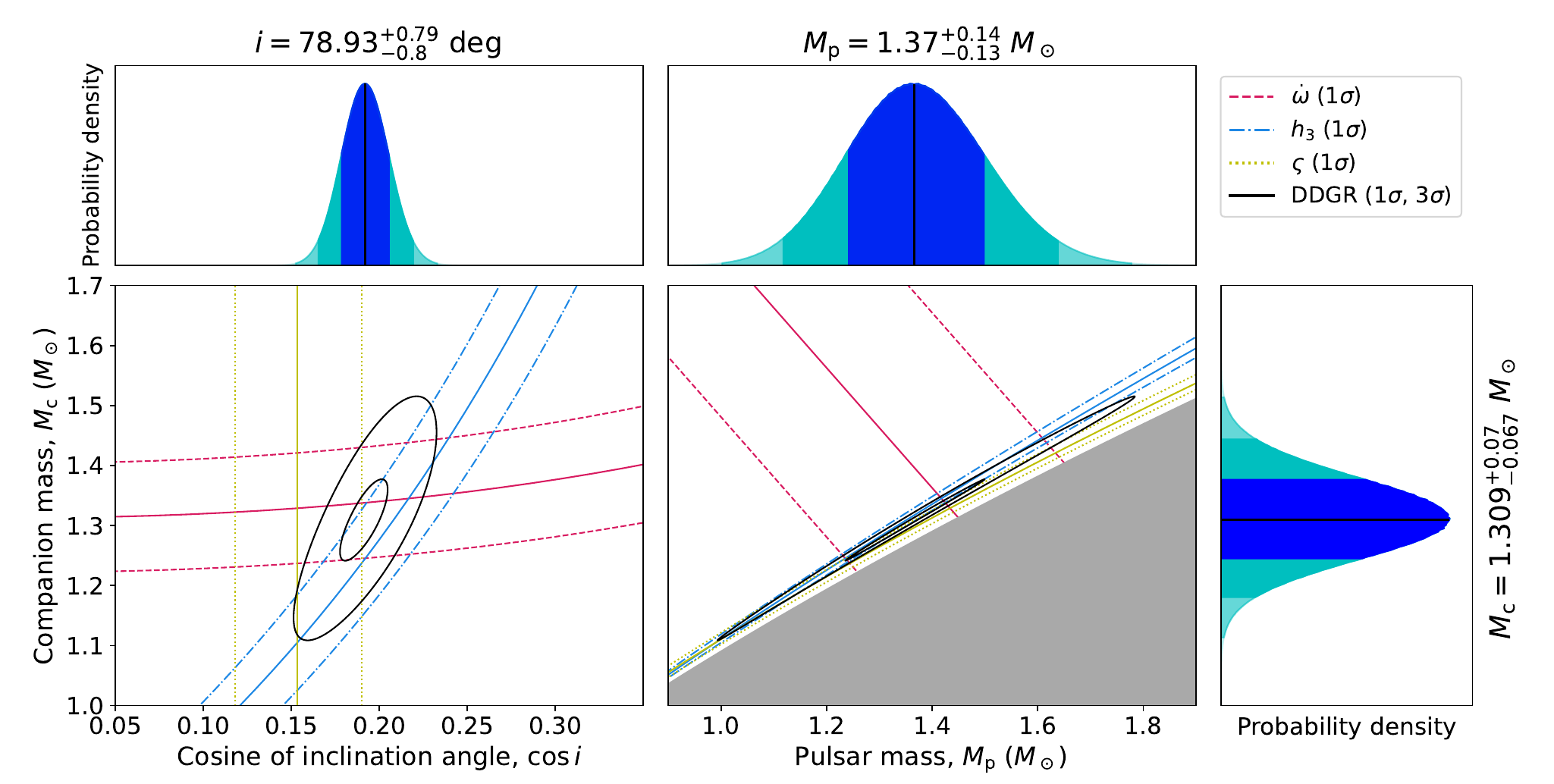}
 \caption{Mass and inclination angle constraints from the DDH PK measurements and the $\chi^2$ mapping with the DDGR model, both from the MeerKAT+UWL fit and under the assumption of the \textbf{Cs} noise model (Table~\ref{fits}). Contours have been drawn following the same principles as in Fig.~\ref{mass_diagram}.}
  \label{DM50_R100_mkt+uwl}
\end{figure}

\end{appendix}

\end{document}